\documentclass[12pt]{article}
\usepackage{epsfig}
\usepackage{amsmath}
\usepackage{hhline}
\usepackage{amssymb} 
\usepackage{times}
\usepackage{cite,./mcite}
\usepackage{xspace}

\newlength{\dinwidth}
\newlength{\dinmargin}
\def\gsim{\,\lower.25ex\hbox{$\scriptstyle\sim$}\kern-1.30ex%
\raise 0.55ex\hbox{$\scriptstyle >$}\,}
\def\lsim{\,\lower.25ex\hbox{$\scriptstyle\sim$}\kern-1.30ex%
\raise 0.55ex\hbox{$\scriptstyle <$}\,}

\newcommand{\dEdx}{\ensuremath{{\rm d}E\!/\!{\rm d}x}}
\newcommand{\eV}{\mbox{e\hspace{-0.08em}V}}
\newcommand{\ie}{i.\,e.\ }
\newcommand{\eg}{e.g.\ }
\newcommand{\dsdx}[1]{$d\sigma\!/\!d #1$\xspace}

\newcommand{\dstar}{\ensuremath{D^*}}
\newcommand{\dstarp}{\ensuremath{D^{*+}}}
\newcommand{\dstarm}{\ensuremath{D^{*-}}}
\newcommand{\dstarpm}{\ensuremath{D^{*\pm}}}

\newcommand{\dstarpj}{\ensuremath{\dstar\! \mbox{+jet}}}
\newcommand{\dstarmj}{\ensuremath{\dstar\! \mbox{-jet}}}

\newcommand{\dstardj}{{\it \dstar\ tagged dijet}}
\newcommand{\dstarDj}{\dstar\ tagged Dijet}
\newcommand{\dstarpmdj}{\ensuremath{\dstarpm\! \mbox{ + 2 jet }}}
\newcommand{\dstarpotherj}{{\it \ensuremath{\dstar\! \mbox{ + other }\;\mbox{jet}}}}
\newcommand{\dstarpotherJ}{\ensuremath{\dstar\! \mbox{ + other }\;\mbox{Jet}}}
\newcommand{\dstarpmpotherj}{\ensuremath{\dstarpm\! \mbox{ + other }\;\mbox{jet}}}

\newcommand{\zDs}{\ensuremath{z(\dstar )}}
\newcommand{\Wgp}{\ensuremath{W_{\gamma p}}}
\newcommand{\ptds}{\ensuremath{p_t(\dstar )}}
\newcommand{\etads}{\ensuremath{\eta(\dstar )}}
\newcommand{\ptj}{\ensuremath{p_t(\mbox{jet})}}
\newcommand{\ptjn}[1]{\ensuremath{p_t(\mbox{jet$_{#1}$})}}
\newcommand{\etaj}{\ensuremath{\eta(\mbox{jet})}}
 
\newcommand{\dphidsj}{\ensuremath{\Delta \phi(\dstar \! , \mbox{jet})}}
\newcommand{\detadsj}{\ensuremath{\eta(\dstar )\, \mbox{-}\, \etaj}}

\newcommand{\xgjj}{\ensuremath{x_\gamma^{\mbox{\footnotesize obs}}}}

\newcommand{\as}{\ensuremath{\alpha_s}}
\newcommand{\pt}{\ensuremath{p_t}}
\newcommand{\kt}{\ensuremath{k_t}}
\def\pythia{{\sc Pythia}}
\def\cascade{{\sc Cascade}}

\newcommand{\grad}{\ensuremath{^\circ\!}} 
\newcommand{\Lineup}{\rule[-1.1ex]{0pt}{3.3ex}}
\newcommand{\Linedown}{\rule[-1.8ex]{0pt}{3.3ex}}

\newcommand{\Mycaption}[1]{\caption{\em #1}}

\begin{document}  
\begin{titlepage}

\noindent
\begin{flushleft}
DESY 06-110\hfill ISSN 0418-9833\\
July 2006
\end{flushleft}

\vspace{2cm}

\begin{center}
\begin{Large}

{\bf \boldmath Inclusive \dstarpm\ Meson Cross Sections \\
and \dstarpm -Jet  Correlations \\
 in Photoproduction   at HERA\\
}
\vspace{2cm}

H1 Collaboration

\end{Large}
\end{center}

\vspace{2cm}

\begin{abstract}
\noindent

Differential
photoproduction cross sections are measured for events containing \dstarpm\ 
mesons. The data were taken with the H1 detector at the $ep$ collider HERA
and correspond to an integrated luminosity of $51.1$~pb$^{-1}$.
The kinematic region covers small photon virtualities $Q^2 < 0.01$~G\eV$^2$
and photon-proton centre-of-mass energies of $171 < W_{\gamma\,p} < 256$~G\eV.
The details of the heavy quark production process are further investigated in
events with one or two jets in addition to the \dstarpm\ meson.
Differential cross sections for \dstarpj\ production are determined and
the correlations between the \dstarpm\ meson and the jet(s)
are studied.
The results are compared with perturbative QCD predictions applying collinear-
or \kt -factorisation. 
\end{abstract}

\vspace{1.5cm}

\begin{center}
Submitted to \emph{Eur.~Phys.~J.} \textbf{C}
\end{center}

\end{titlepage}

\begin{flushleft}
A.~Aktas$^{9}$,                
V.~Andreev$^{25}$,             
T.~Anthonis$^{3}$,             
B.~Antunovic$^{26}$,           
S.~Aplin$^{9}$,                
A.~Asmone$^{33}$,              
A.~Astvatsatourov$^{3}$,       
A.~Babaev$^{24, \dagger}$,     
S.~Backovic$^{30}$,            
A.~Baghdasaryan$^{37}$
% ...

\bigskip{\it
 $ ^{1}$ I. Physikalisches Institut der RWTH, Aachen, Germany$^{ a}$ \\
 $ ^{2}$ School of Physics and Astronomy, University of Birmingham,
          Birmingham, UK$^{ b}$ \\
 $ ^{3}$ Inter-University Institute for High Energies ULB-VUB, Brussels;
          Universiteit Antwerpen, Antwerpen; Belgium$^{ c}$ \\
% ...
$ ^{9}$ DESY, Hamburg, Germany \\
% ...
$ ^{24}$ Institute for Theoretical and Experimental Physics,
          Moscow, Russia$^{ k}$ \\
 $ ^{25}$ Lebedev Physical Institute, Moscow, Russia$^{ e}$ \\
 $ ^{26}$ Max-Planck-Institut f\"ur Physik, M\"unchen, Germany \\
% ...
$ ^{30}$ Faculty of Science, University of Montenegro,
          Podgorica, Serbia and Montenegro$^{ e}$ \\
% ...
$ ^{33}$ Dipartimento di Fisica Universit\`a di Roma Tre
          and INFN Roma~3, Roma, Italy \\
% ...
$ ^{37}$ Yerevan Physics Institute, Yerevan, Armenia \\
% ...
$ ^{\dagger}$ Deceased \\

\bigskip
 $ ^a$ Supported by the Bundesministerium f\"ur Bildung und Forschung, FRG,
      under contract numbers 05 H1 1GUA /1, 05 H1 1PAA /1, 05 H1 1PAB /9,
      05 H1 1PEA /6, 05 H1 1VHA /7 and 05 H1 1VHB /5 \\
 $ ^b$ Supported by the UK Particle Physics and Astronomy Research
      Council, and formerly by the UK Science and Engineering Research
      Council \\
 $ ^c$ Supported by FNRS-FWO-Vlaanderen, IISN-IIKW and IWT
      and  by Interuniversity
Attraction Poles Programme,
      Belgian Science Policy \\
% ...
$ ^e$ Supported by the Deutsche Forschungsgemeinschaft \\
% ...
$ ^k$ Partially Supported by Russian Foundation
      for Basic Research,  grants  03-02-17291
      and  04-02-16445 \\
% ...
}
\end{flushleft}

\newpage

\section{Introduction}
\label{sec:intro}
\begin{figure}[htbp]
  \begin{center}
    \setlength{\unitlength}{\textwidth}
    \begin{picture}(1,0.28)
      \put(0.,0.06){\includegraphics[width=.21\textwidth]{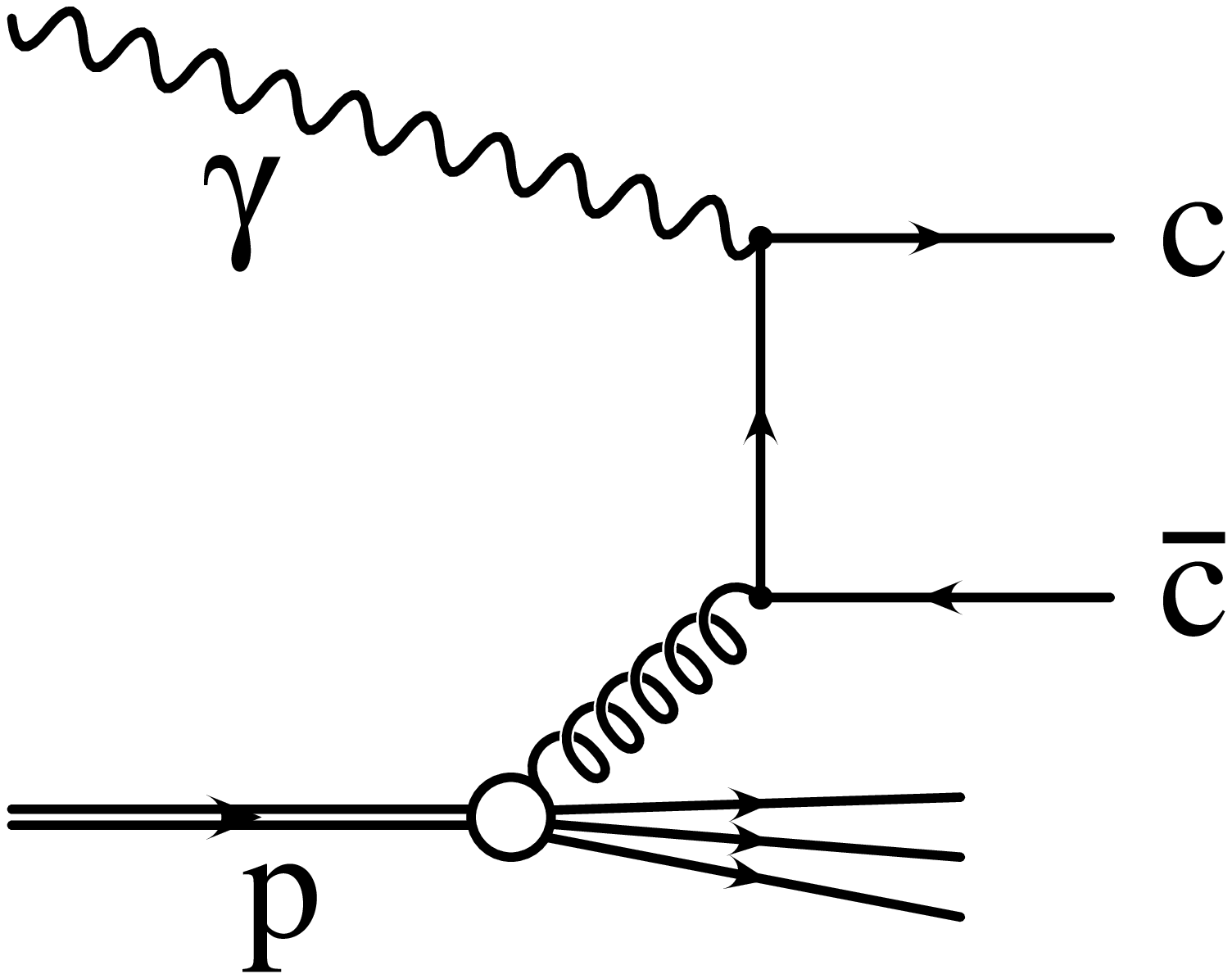}}
      \put(0.25,0.){\includegraphics[width=.21\textwidth]{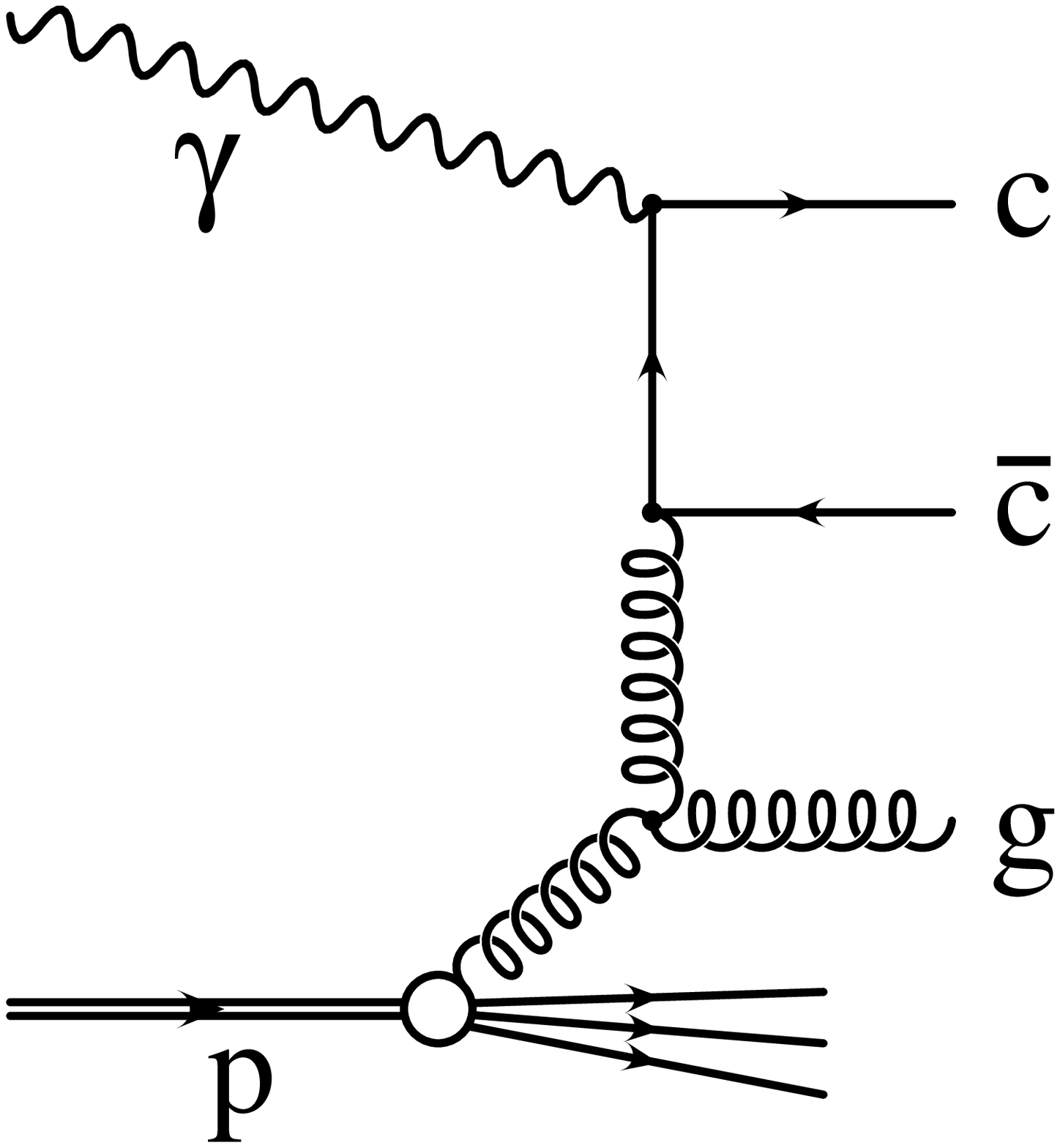}}
      \put(0.5,0.06){\includegraphics[width=.21\textwidth]{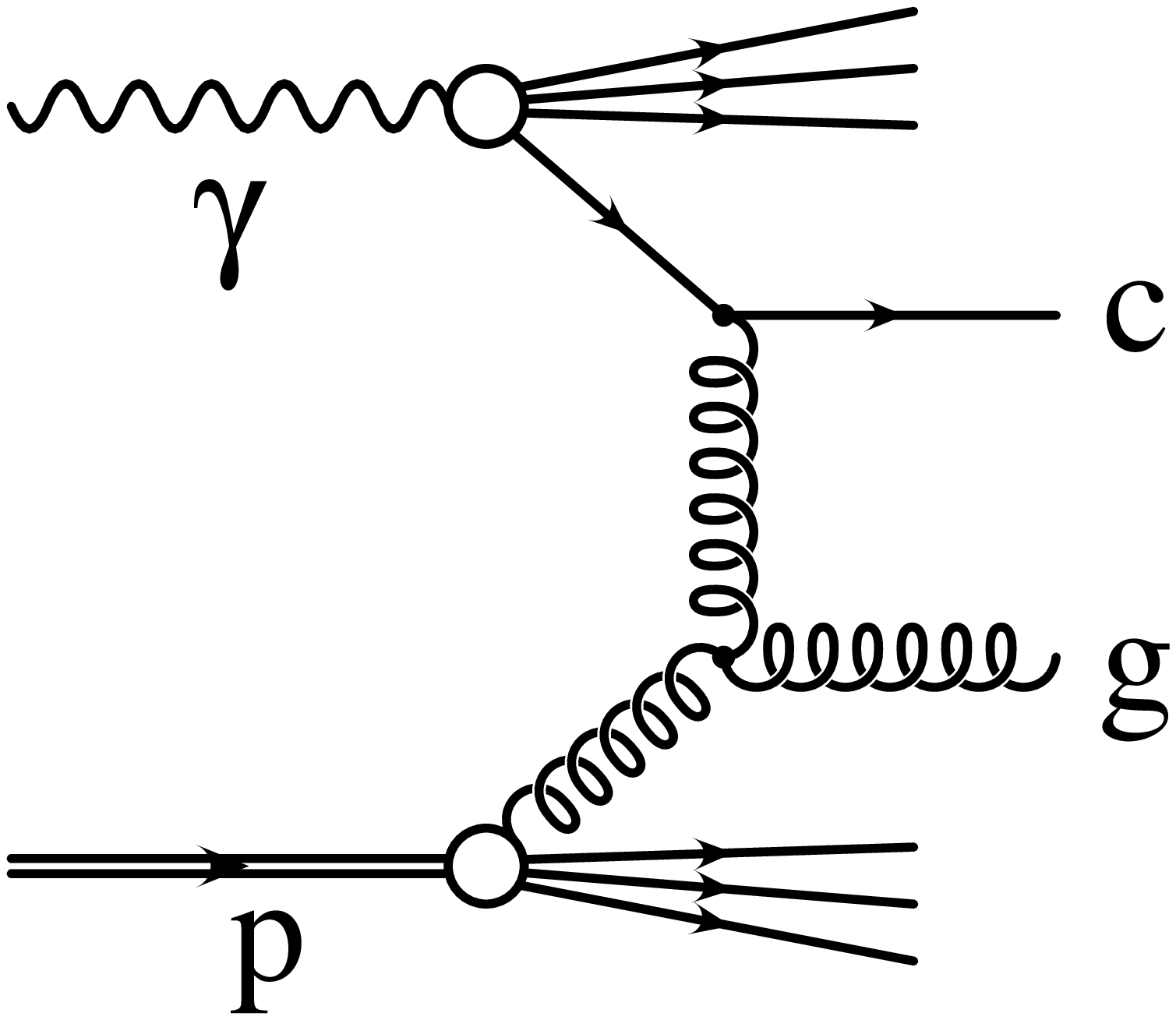}}
      \put(0.75,0.06){\includegraphics[width=.21\textwidth]{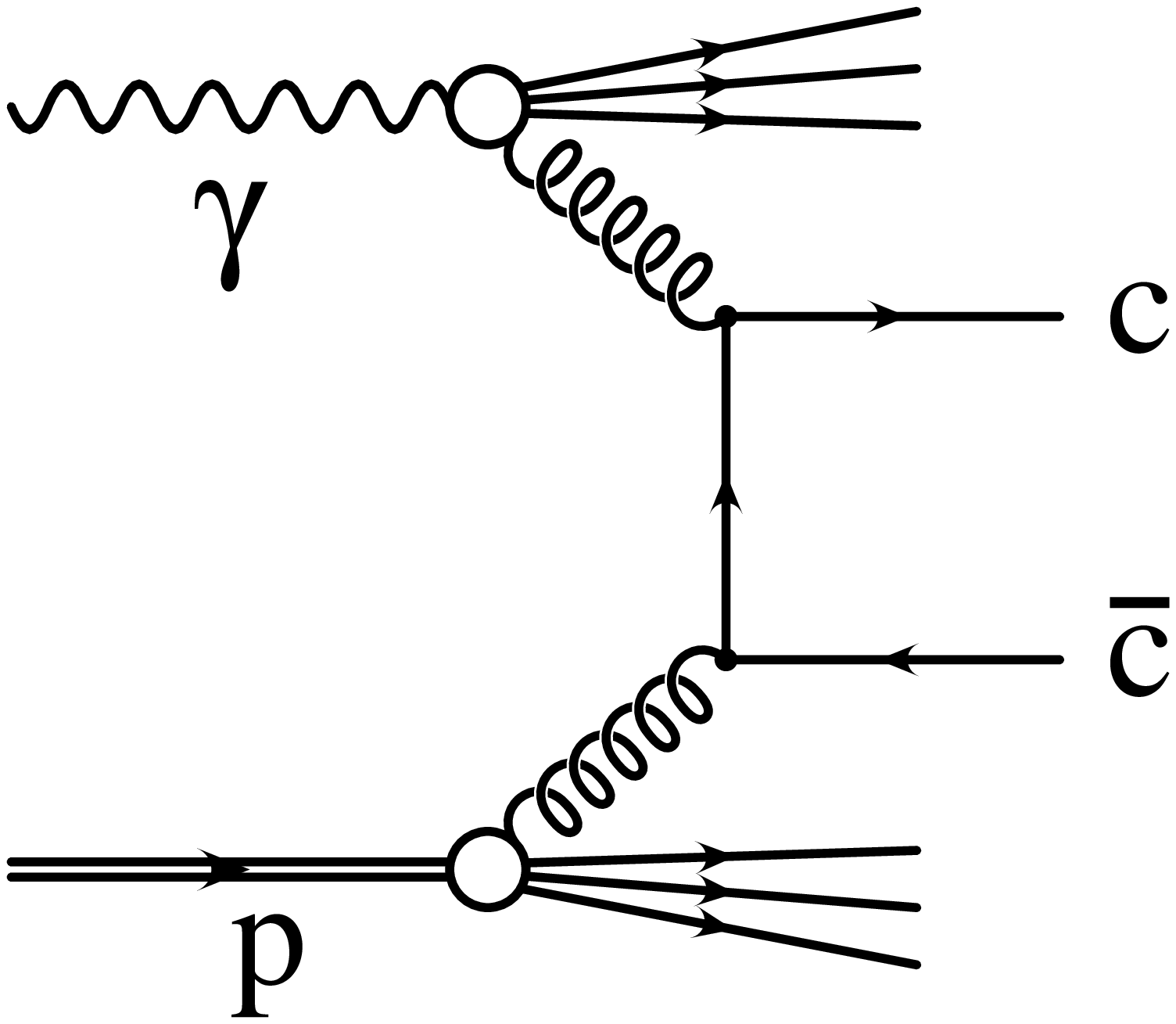}}
      \put(0.03,0.25){a) Direct $\gamma$}
      \put(0.22,.25){b) Real Gluon Emission}
      \put(0.53,.25){c) Excitation}
      \put(0.77,.25){d) Hadron - like}
    \end{picture}
    \Mycaption{Diagrams for charm photoproduction: direct photon processes (a) and (b),  and resolved photon
    processes with (c) charm excitation 
    and (d) the hadronic manifestation of the real photon. 
    }
    \label{fig:processes}
  \end{center}
\end{figure}

Charm photoproduction in $ep$ collisions at HERA proceeds 
predominantly via photon-gluon fusion
as shown in Fig.~\ref{fig:processes}, where the quasi real photon
(virtuality $Q^2 \simeq 0$~G\eV$^2$) is emitted from the beam lepton.
The charm quark mass provides a hard scale which allows perturbative QCD (pQCD) 
to be applied over the full phase space.
Therefore, charm photoproduction is particularly well suited to test
perturbative calculations and the underlying theoretical approaches. 

Previous measurements have focused on inclusive  
\dstarpm\ meson production~\cite{H1:DstarGluonDens99}, \dstarpm\ meson with associated dijet 
production~\cite{ZEUS:dstarjetgammap99,ZEUS:dijetDstar03,ZEUS:dstarGammapJetCorr05} and 
heavy quark pair production
using events with a \dstarpm\ meson and a muon~\cite{H1D*mu:05}.
Here \dstarpm\ photoproduction is considered,
using a data sample five times larger than in the previous
H1 analysis~\cite{H1:DstarGluonDens99}.
Single and double differential cross sections for inclusive \dstarpm\ 
production are presented and compared with theoretical calculations
using collinear-factorisation~\cite{DGLAP:72,DGLAP:75,DGLAP77,DGLAP77_2} 
or \kt -factorisation~\cite{GribLevRys:SemihardQcd84,LevRys:HqSemiHard91,CatCiaf:HqSmallX91,CollEll:HqInHighE91}.
Details of the heavy quark production process are investigated further by
studying events either with an additional jet 
not containing the \dstar\ meson (``\dstarpotherj '') or with  two jets  (``\dstardj '').
The jets are measured down to a transverse momentum of $\ptj = 3$~G\eV,
which extends the region explored in~\cite{ZEUS:dstarjetgammap99,ZEUS:dijetDstar03,ZEUS:dstarGammapJetCorr05}
to significantly smaller values.
Whereas the \dstar\ always originates from a charm or anticharm quark produced in the hard subprocess, the
non-\dstar -tagged jet can result from either the other heavy quark (Fig.~\ref{fig:processes}a,b,d) or a
light parton (\eg a gluon, Fig.~\ref{fig:processes}b,c) coming from higher order processes.
Measurements of correlations between the \dstar\ and the jet
are performed which are sensitive to these higher order effects and to the
longitudinal and transverse momenta of the partons
entering the hard scattering process.

The paper is organised as follows. In section 2 a brief description of
the H1 detector is given, followed by the details of the event selection and
the reconstruction of the \dstar\ mesons.
In section~\ref{sec:QCDcalc} the theoretical 
calculations are described. In section~\ref{sec:xSecDeterm} the
determination of the cross sections and the systematic uncertainties are
presented
and the measured cross sections are compared with theoretical calculations.


\section{Detector Description}
\label{sec:det}

The H1 detector is described in detail 
in~\cite{H1:Detektor97,H1:Detektor97TrackCalMuon} and
only the components most relevant for this analysis
are briefly mentioned here.
The origin of the H1 
coordinate system is the nominal $ep$ interaction point,  
the direction of the proton beam defining the positive 
$z$-axis (forward region). The transverse momenta are measured 
in the $xy$ plane. 

Charged particles
are measured in the Central Tracking Detector (CTD) which covers the
range in pseudo-rapidity\footnote{The pseudo-rapidity
$\eta$ corresponding to a polar angle $\theta$ 
(measured with respect to the positive $z$-axis)
is given by $\eta = -\ln \; \tan (\theta/2)$.} 
$1.74 >\eta >-1.74$.
The CTD comprises two large
cylindrical Central Jet Chambers (CJCs) 
arranged concentrically around the beam-line
within a solenoidal magnetic field of 1.15 T.
Two additional drift chambers improve the $z$-coordinate reconstruction.
The track resolution is further improved using hit information from
the central silicon track detector (CST)~\cite{CST2000}.
The CTD also provides trigger information which is based on measurements
in the $r$-$\phi$ plane of the CJCs and the $z$-position of the
interaction vertex obtained from a double layer of multiwire proportional
chambers.

The track detectors are surrounded in the forward and
central directions by a fine grained Liquid Argon
calorimeter (LAr, $ 3.4>\eta> -1.5$) and in the backward region 
by a lead-scintillating fibre calorimeter
(SpaCal, $-1.4>\eta >-4.0 $)~\cite{H1SPACAL97} both of which contain electromagnetic and hadronic sections.
For the present analysis the hadronic final state (HFS) is reconstructed from combined
objects, built from calorimeter clusters and tracks, 
using an algorithm which
ensures that no double counting of energy occurs.
Compared to the case of clusters alone the use of combined objects improves the reconstruction of
low momentum particles.  
\par
A crystal \v{C}erenkov calorimeter (electron tagger) located close to the beam pipe at
$z = -33.4$~m detects positrons scattered through a very
small angle $(\pi - \theta_{e'}) < 5$~mrad and is used to trigger photoproduction events.
The $ep$ luminosity is determined from the QED
bremsstrahlung $(ep \to ep\gamma)$ event rate by detecting the
radiated photon in another \v{C}erenkov calorimeter located at $z = -103$~m
(photon detector).

\section{Event Selection and Reconstruction}
\label{sec:sel}
The data were recorded  in $e^+p$ scattering at HERA in 1999 and 2000 and
correspond to an integrated luminosity of \mbox{$\mathcal{L} = 51.1$~pb$^{-1}$}.
The energy of the positrons was 
 $E_e = 27.6$~G\eV\ and that of the protons $E_p = 920$~G\eV .
 
The events were triggered by requiring signals from the central drift
chambers and the multi-wire proportional chambers in coincidence 
with a signal
in the electron tagger. 
In addition, an on-line  software
filter selected events with candidates for charmed hadron decays by
calculating invariant masses of track combinations.

The analysis of photoproduction events 
is restricted to $0.29 < y < 0.65$,
 where the inelasticity $y = 1 - E_{e'}/E_e$ is calculated from
the reconstructed positron energy $E_{e'}$ in the tagger.
In this $y$-range the average
tagger acceptance is almost 40\%. This range
corresponds to a photon-proton centre-of-mass energy
$171 \lsim W_{\gamma p} \lsim 256$~G\eV .
For the small scattering angles considered
the photon virtuality $Q^2$ is below $0.01$~G\eV$^2$.

\subsection{\boldmath \dstar\ Selection and Fit}
The \dstarpm\ meson is detected via the decay channel\footnote{
  Charge conjugate states are implicitly implied.
}
$\dstarp \to D^0 \pi_s^+ \to K^- \pi^+ \pi_s^+$, 
which has a branching ratio 
$\mathcal{BR}(\dstar \to K \pi \pi_s) = (2.57\pm0.06)\%
$~\cite{PDG2004}.
Here $\pi_s$ refers to the low momentum pion from the \dstar\ decay. 
In each event, tracks with opposite charges, fitted to the event vertex,
are combined in pairs
and both invariant masses $m(K^\pm\pi^\mp)$ are calculated where one track is assigned
the kaon mass and the other the pion mass.
If the result is consistent with the 
nominal $D^0$ mass~\cite{PDG2004}, any remaining track with a charge opposite
to that of the kaon candidate is assigned the pion mass and combined with the
$D^0$ candidate to form a \dstar\ candidate.
The measured specific energy loss per path length \dEdx\ of each track 
has to be consistent with the respective particle hypothesis~\cite{diss:geroflucke05}.

The \dstar\ candidate is accepted if it fulfils the selection cuts listed in
Tab.~\ref{tab:dstarcuts}.
\begin{table}[tb]
  \begin{center}
    \begin{tabular}{|l|c|c|}
\hline
\textbf{Selection} of  & \multicolumn{2}{c|}{$\pt(K,\pi) > 0.3$~G\eV} \\
\hspace{0.5em}{\small $\dstarp \to D^0 \pi_s^+ \to K^- \pi^+ \pi_s^+$}
                         & \multicolumn{2}{c|}{$\pt(\pi_s) > 0.12$~G\eV} \\
                         & \multicolumn{2}{c|}{$20\grad < \theta(K,\pi,\pi_s) < 160\grad$} \\
                         & \multicolumn{2}{c|}{$\dEdx(K,\pi,\pi_s)$ consistent with particle hypothesis} \\
\cline{2-3} 
                         & \multicolumn{2}{c|}{\rule[-4.ex]{0pt}{9.ex}
                           $ \small m(K\pi) - m_{D^0} \le \left\{
                               \begin{array}{rcl}
                                  80\mbox{ M\eV} & \mbox{for} & \ptds < 6.0\mbox{ G\eV} \\ 
                                 100\mbox{ M\eV} & \mbox{for} & 6.0\le\ptds <8.5\mbox{ G\eV}\\
                                 140\mbox{ M\eV} & \mbox{for} & \ptds \ge 8.5\mbox{ G\eV}
                               \end{array}\right.
                           $} \\
                         & \multicolumn{2}{c|}{$\Delta m = m(K\pi\pi_s) - m(K\pi) < 167.5$~M\eV}  \\ 
\cline{2-3}
\Lineup                  & \multicolumn{2}{c|}{$\pt(K) + \pt(\pi) > 2.2$~G\eV} \\
\Lineup\Linedown     & \multicolumn{2}{c|}{$\frac{\pt(\dstar)}{\sum_{HFS}^{\theta > 10\grad} E_{t,i}} >0.10$}\\
\hline\hline
\textbf{Visible kinematic region} & \multicolumn{2}{c|}{{\it inclusive \dstar}} \\\hline
                         & \multicolumn{2}{c|}{$\ptds \geq 2.0$~G\eV }\\
                          & \multicolumn{2}{c|}{$|\etads | < 1.5$}\\
                         & \multicolumn{2}{c|}{$Q^2 < 0.01$~G\eV$^2$} \\
                         & \multicolumn{2}{c|}{$0.29 < y < 0.65$ }\\
\hline
                         & {   \dstarpotherj}  & {    \dstardj\ } \\\cline{2-3}
                         & $|\etaj | < 1.5$                   & $|\etaj | < 1.5$\\
                         & \hspace*{2.5em}$\ptj \geq 3$~G\eV\hspace*{2.5em}&$\ptjn{1}\geq 4$~G\eV\\
                         & $\dstar \notin \mbox{ jet}$                     &$\ptjn{2}\geq 3$~G\eV\\
\hline
    \end{tabular}
    \Mycaption{\dstar\ selection cuts and definition of the {\it ``visible''} kinematic regions
      for which the {\it inclusive \dstar}, the \dstarpotherj\ and the  
      \dstardj\  cross sections are measured.
    }
    \label{tab:dstarcuts}
  \end{center}
\end{table}
The restrictions 
in the transverse momentum \ptds\ and in the pseudorapidity \etads\ ensure
good detector acceptance.
To reduce the combinatorial background, cuts are applied to the scalar sum of
$\pt(K)$ and $\pt(\pi)$ and to the fraction of the transverse momentum carried by the
\dstar\ with respect to the scalar sum of transverse energies   
of the full hadronic final state, excluding the forward region ($\theta < 10\grad$).

In Fig.~\ref{fig:dmsignal}a the distribution of the mass difference
$\Delta m = m(K\pi\pi_s) - m(K\pi)$ of the
final \dstar\ candidates is shown.
A clear peak is observed around the nominal value of
$\Delta m = (145.421\pm 0.010)$~M\eV~\cite{PDG2004}.
The number of reconstructed \dstar\ mesons $N(\dstar)$ is extracted from
a likelihood fit to the $\Delta m$-distribution with a function which is 
a superposition of a Gaussian for the signal and a phase space threshold
function with a quadratic correction term, 
$F^b(\Delta m) = 
u_n(\Delta m -m_\pi)^{u_{e}}\cdot (1-u_s(\Delta m)^2)$, for the background.
The fit yields a signal of $1166\pm 82$ \dstar\ mesons.
Separate fits are performed in each bin of the cross section measurement, in which
the mean and the width of the Gaussian,
the background parameters $u_s$ and, in cases where statistics are very low, $u_{e}$
are fixed to the values obtained from the fit to the {\it inclusive \dstar\ } sample.
The background normalisation $u_n$ is a free parameter.

\subsection{Jet Selection}
In the {\it inclusive \dstar\ } sample jets are defined by the inclusive
$k_t$-algorithm~\cite{jetKT93} in the \pt-re\-com\-bina\-tion
and $\Delta R$-dis\-tance scheme (with $R_0 = 1$, \cite{Kt++:02}).
The jet algorithm is applied in the laboratory frame to all reconstructed particles
of the HFS.
To prevent the decay particles of the \dstar\ candidate from being ascribed to
different jets,
the \dstar\ candidate is used as a single particle in the jet algorithm,
replacing its decay products.
In events which contain more than one \dstar\ candidate, the jet algorithm is run separately
for each candidate. 
The transverse momentum is required to satisfy $\ptj > 3$~G\eV.
For jets originating from charm 
the requirement of $\ptj > 3$~G\eV\ matches approximately the requirement of $\ptds > 2$~G\eV , since
the charm fragmentation function to a \dstar\ peaks where the \dstar\ meson  takes
$\sim 70 \% 
$ of the charm quark momentum \cite{Artuso:2004pj}.

To ensure a good jet reconstruction at these low values
of $\pt$, jets are restricted to the central detector region $|\etaj | < 1.5$
where precise track information is available.
The jet transverse momentum resolution is around 30\% over the whole momentum
range considered here.

\begin{figure}[htbp]
  \begin{center}
    \includegraphics[width=0.48\textwidth]{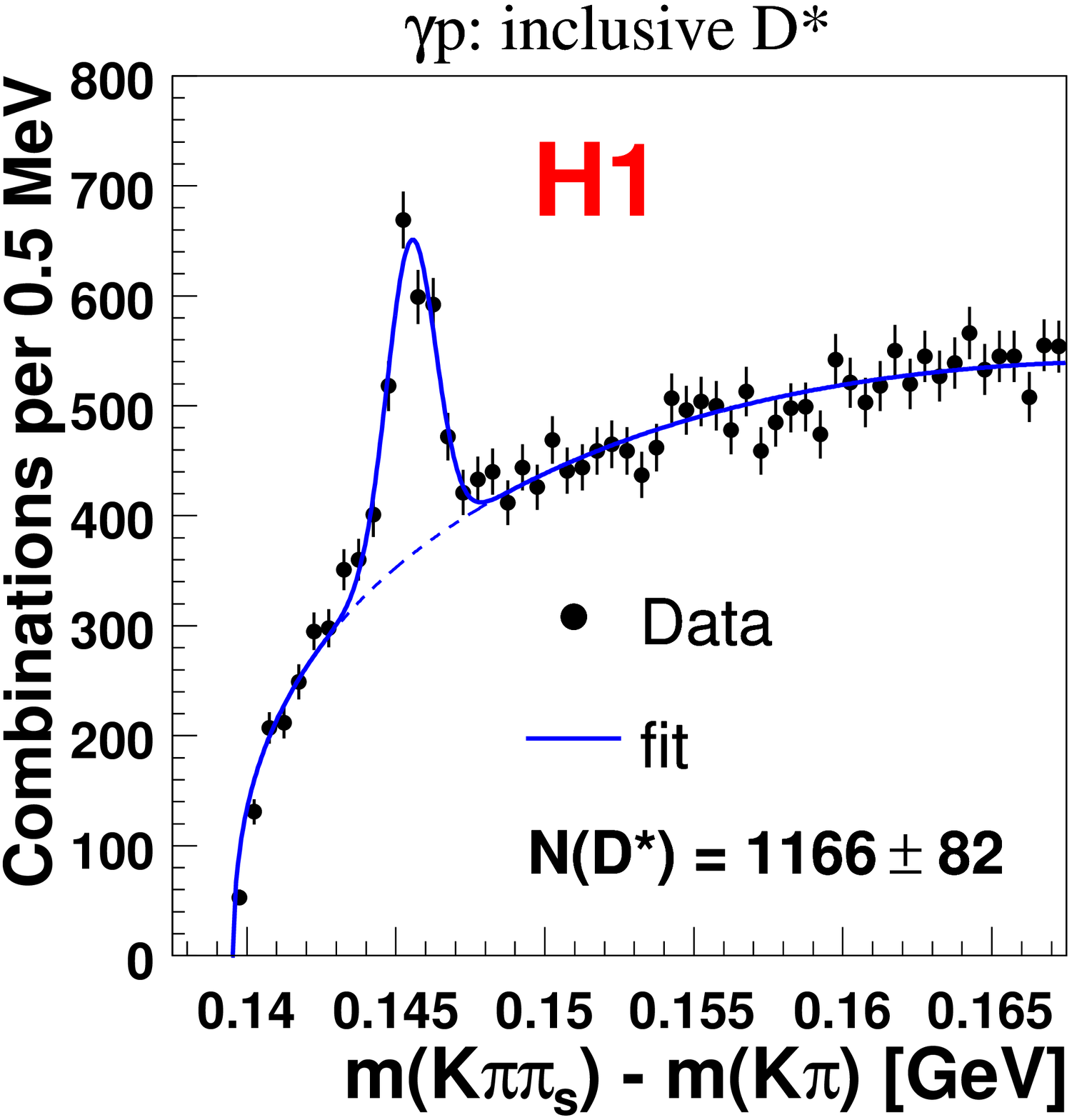}
    \includegraphics[width=0.48\textwidth]{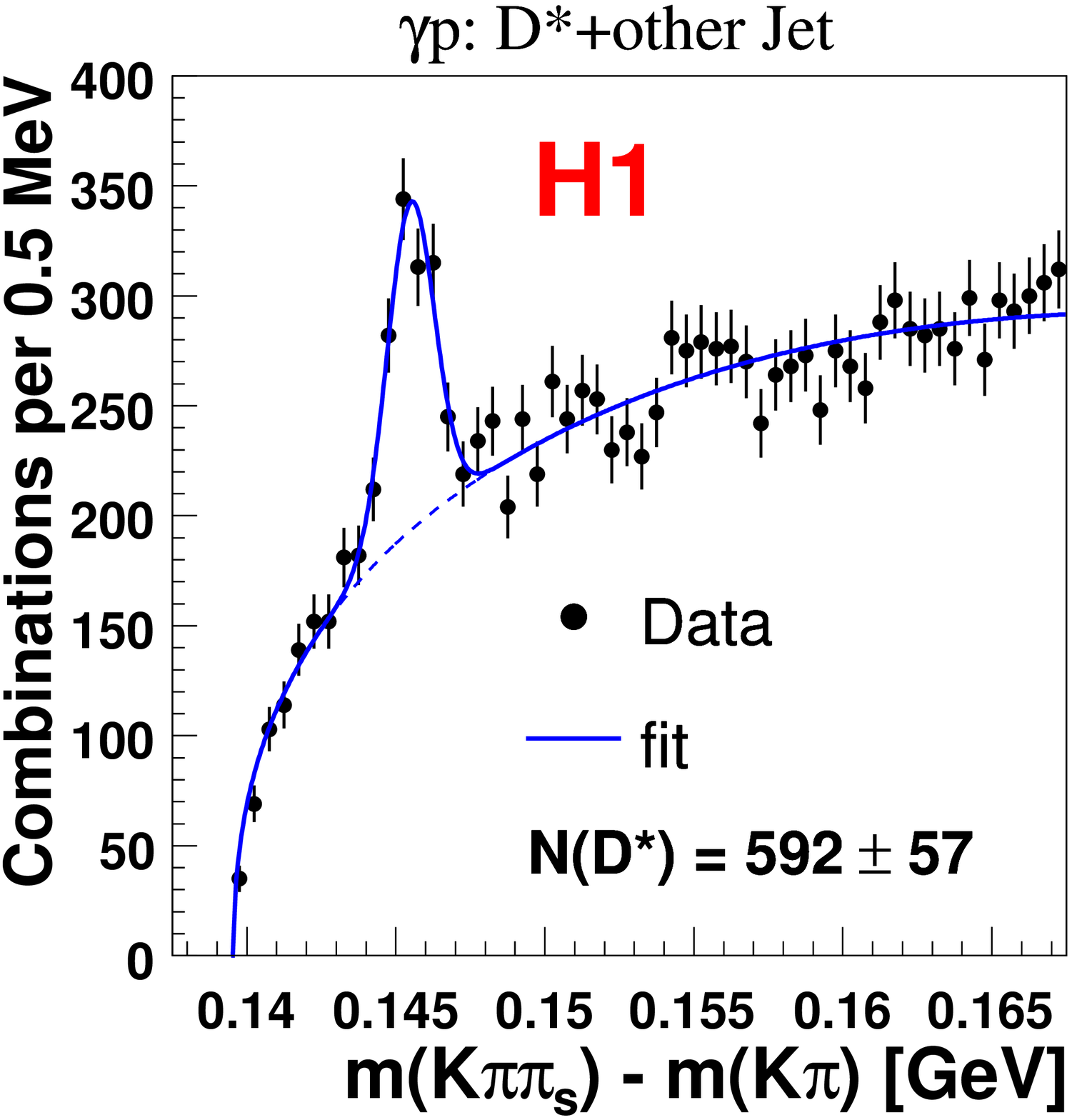}
    \setlength{\unitlength}{\textwidth}
    \begin{picture}(0,0)
      \put(-0.88,0.4){\bfseries a)}
      \put(-0.39,0.4){\bfseries b)}
    \end{picture}

    \Mycaption{\dstar\ signal in the distribution of the mass difference
      $\Delta m = m(K\pi\pi_s) - m(K\pi)$ of the 
      {\it inclusive \dstar\ } (a) and the \dstarpotherj\ (b) selection.
      The solid lines represent the fits to determine the number of \dstar\ mesons
      in the signal and the dashed lines indicate the resulting background parametrisations.
    }
    \label{fig:dmsignal}
  \end{center}
\end{figure}

The jet with the highest \ptj\ not containing the \dstar\ meson is considered
together with the \dstar\ meson in the ``\dstarpotherj '' analysis.
The $\Delta m$-distribution of the \dstar\ candidates 
of this sample is shown in Fig.~\ref{fig:dmsignal}b.
A signal of $592 \pm 57$ 
\dstar\ mesons is found. In about 10\% 
of the \dstar\ events a second (non-\dstar -)jet is observed.

In addition to the \dstarpotherj\ sample a  ``\dstardj '' sample
is selected. 
At least two jets are required with $\ptj > 4$~G\eV\ for the highest \pt\ jet
and $\ptj > 3$~G\eV\ for the second jet (see Tab.~\ref{tab:dstarcuts}),
irrespective of whether the \dstar\ meson is attributed to either of the jets.
A signal of $496 \pm 53$ \dstardj\ events is observed. 
The fraction of events where the \dstar\ meson is contained in neither of 
the two jets is negligible.%

\section{QCD Calculations}
\label{sec:QCDcalc}
Figure~\ref{fig:processes} shows examples of diagrams for charm photoproduction
in leading order ${\cal O}(\as)$
and next-to-leading order ${\cal O}(\as^2)$.
In direct-photon processes (Figs.~\ref{fig:processes}a,b),
the photon emitted from the beam lepton enters the hard process directly.
In resolved-photon processes (Figs.~\ref{fig:processes}c,d),
the photon acts as a source of incoming partons, one of
which takes part in the hard interaction.
The distinction between these two classes of processes depends on the
factorisation scheme and the order in which the calculation is
performed.

The data presented in this analysis are compared with 
leading order (LO) calculations supplemented by parton showers as well as
with next-to-leading order (NLO) calculations.
The calculations are performed using either the collinear factorisation
or the \kt -factorisation approach.
The collinear factorisation makes use of the \mbox{DGLAP~\cite{DGLAP:72,DGLAP:75,DGLAP77,DGLAP77_2}} 
evolution equations. In this approach   
transverse momenta obtained through the initial state QCD evolution are 
neglected and all the transverse momenta are generated in the hard scattering process, \ie 
the incoming partons are  
collinear with the proton, resulting at 
lowest order in a back-to-back configuration of the heavy
quark pair in the transverse plane. 
Effects from the finite transverse momentum of the 
gluons enter only at the NLO level (for example in a process like 
$\gamma g \to c \bar{c} g$ as shown in Fig.~\ref{fig:processes}b).
In the \kt -factorisation ansatz the transverse momentum (\kt) of incoming gluons 
are already included at leading order both in the \kt -dependent off-shell matrix element
and the 
\kt -dependent unintegrated gluon density~\cite{CollJungNeedUnintPdf:05}.
Therefore, higher order corrections, \ie hard parton emissions, are partially
considered.

Heavy quark production is calculated either in the
massive scheme~\cite{Frixione:diffHQ95},
where heavy quarks are produced only perturbatively via boson gluon fusion, 
or in the massless scheme~\cite{Kniehl:massless02},
where heavy quarks are treated as massless partons.
These two schemes should be appropriate in different regions of
phase space~\cite{heavyQuarkConcept:01}: the massive scheme should be reliable
when the transverse momentum $\pt$ of the heavy quarks 
 is of similar size compared to $m_c$, whereas the massless scheme is expected to be valid for
$\pt \gg m_c$. Recently, new calculations combining and matching the two
approaches in the photoproduction regime 
have become available~\cite{Kniehl:2004fy,Kniehl:2005mk}. 
 
The uncertainties on the calculations are estimated by varying
the charm quark mass, the renormalisation scale and where possible the
factorisation scales.
The uncertainty in each bin is obtained by taking the maximal deviations from
the central value resulting from the separate variations.
The detailed settings are summarised in Tab.~\ref{tab:calculations}. 
\begin{table}[tb]
\begin{center}
{\footnotesize
\begin{tabular}{l@{\hspace{.36cm}}c@{\hspace{.36cm}}c@{\hspace{.36cm}}c@{\hspace{.36cm}}c@{\hspace{.36cm}}c@{\hspace{.36cm}}}
  & {\bf PYTHIA}
  & {\bf CASCADE}
  & {\bf FMNR}
  & {\bf ZMVFNS}
  & {\bf GMVFNS}
  \rule[-2mm]{0mm}{5mm}
  \\
\hline
\hline
Version       &   6.224
              &   1.2010
              &
              &   
              &
              \rule[-1mm]{0mm}{5mm}
              \\
\hline
Proton PDF
              &  CTEQ6L~\cite{CTEQ6:02}
              &  A0($\pm$)~\cite{unintGlu_Jung:04}
              &  CTEQ6M~\cite{CTEQ6:02}
              &  CTEQ6M 
              &  CTEQ6M 
              \rule[-1mm]{0mm}{5mm}
              \\

Photon PDF
              &  GRV-G LO~\cite{GRV:92}
              &  {--}
              &  GRV-G HO~\cite{GRV:92}
              &  GRV-G HO
              &  GRV-G HO
              \rule[-2mm]{0mm}{5mm}
              \\

\hline
Renorm. scale  %
               & $m_t$ %
               & $\left.  \begin{array}{r} 2 \\ \bf 1 \\ 0.5 \end{array} \!\!\right\}
               \cdot m'_t$
               &  $\left. \begin{array}{r} 2 \\ \bf 1 \\ 0.5 \end{array} \!\!\right\}
               \cdot  m_t $ 
               &   $\left. \begin{array}{r} 2 \\ \bf 1 \\ - \end{array} \!\!\right\}
               \cdot m_{t,\dstar}$ 
               &   $\left. \begin{array}{r} 2 \\ \bf 1 \\ 0.6 \end{array} \!\!\right\}
               \cdot m_{t,\dstar}$
               \\ \hline  
Factor. scale  
               &  $m_t$ 
               &  $\sqrt{\hat{s} + Q_t^2}$
               &   $\left. \begin{array}{r} 1 \\ {\bf 2} \\ 4 \end{array} \!\!\right\}
               \cdot m_t $ 
               & $\left. \begin{array}{r} 1/1.5 \\ \bf 2 \\ 4 \end{array} \!\!\right\}
               \cdot m_{t,\dstar}$ %
               & $\left. \begin{array}{r} 0.6 \\ \bf 1 \\ 2 \end{array} \!\!\right\}
               \cdot m_{t,\dstar}$
               \\ \hline
\Lineup $m_c \ [ {\rm G\eV} ]$
               & 1.5
               & $1.5^{+0.2}_{-0.2}$
               & $1.5^{+0.2}_{-0.2}$
               & 1.5
               & 1.5
               \\
\hline
Fragmentation
                & $\epsilon_{pet} = 0.04$
                & $\epsilon_{pet} = 0.04$
                & $\epsilon_{pet} = 0.035$~\cite{Nason:99}
                & BKK O~\cite{BKKfrag:98}
                & KKSS~\cite{KKSSfrag:05}
               \\
\hline
\hline
\end{tabular}
\Mycaption{Parameters used in the pQCD calculations where $m_c$
denotes %
the charm quark mass.
The ``transverse mass'' variables are defined as
{$m_t^2 = m_c^2 + (p_{t,c}^2 + p_{t,\bar{c}}^2)/2\,$,
$m^{'2}_t = 4m_c^2 +  p_{t,c}^2$} 
and
{ $m_{t,\dstar}^2  = m_c^2 + p_{t,\dstar}^2$}
where $p_{t,c}$ ($p_{t,\bar{c}}$ ) is the transverse momentum of the 
charm quark (antiquark)
and $p_{t,\dstar}$
is the transverse momentum of the
\dstar\ meson. The squared invariant mass and the transverse momentum squared of the  $c \bar{c}$ pair  
are denoted by
$\hat{s}$ and  $Q_t^2$, respectively.
$\epsilon_{pet}$ is the Peterson fragmentation parameter.
If a parameter is varied to determine the uncertainty of the prediction, the central line
gives the value used for the main prediction.
}
\label{tab:calculations}}
\end{center}
\end{table}
All predictions (except where scale dependent fragmentation functions are used)
are based on a \dstar\ fragmentation ratio of
$f(c\to\dstar) = 0.235$~\cite{Gladilin:c->D*99}. The beauty contribution is included in
all predictions and amounts to a few percent.

\paragraph{PYTHIA}
In \pythia\ ~\cite{pythia61}
three different processes are generated separately using
leading order matrix-elements:
direct photon-gluon fusion (Fig.~\ref{fig:processes}a), resolved photon
processes in which a charm quark 
(charm excitation, Fig.~\ref{fig:processes}c) or a gluon (Fig.~\ref{fig:processes}d)   
from the photon enters the hard scattering. 
In the excitation processes the charm quark is treated as a
massless parton, whereas in the other processes the charm
mass is accounted for in all steps of the calculation.
Higher order contributions are simulated with leading log parton showers in the 
collinear approach.
The Lund string fragmentation model~\cite{Lund83} is used for the simulation of the
hadronisation process.
For the longitudinal
fragmentation of the charm quark into the \dstar\ meson the Peterson
parametrisation~\cite{Peterson83} is used. 
No uncertainties are calculated.

\paragraph{CASCADE}
The \cascade\ ~\cite{cascade,cascadeII,CASCADE12:manual} program is used 
for leading order 
calculations in the \kt-factorisation approach. In the 
$\gamma g^* \to c \bar{c}$ matrix element,
which takes the charm mass into account, the incoming gluon is
treated off mass-shell and can have a finite transverse momentum.
Higher order QCD corrections are simulated with initial
state parton showers applying the 
CCFM evolution~\cite{CCFM:Ciaf88,CCFM:CatFioMarch90a,CCFM:CatFioMarch90b,CCFM:March95} 
equations with an unintegrated parton density function
(uPDF) including angular ordering constraints for the emitted partons.
The uPDF has been obtained from an analysis of the inclusive structure function
$F_2$ in the CCFM approach~\cite{unintGlu_Jung:04}. For the variation of the  
renormalisation scale, dedicated unintegrated gluon density parameterisations have been 
used~\cite{unintGlu_Jung:04}. The factorisation scale cannot be varied.
The final state radiation off the heavy quarks and the fragmentation is performed 
with  \pythia.

\paragraph{FMNR}
The FMNR program~\cite{Frixione:totHQ95,Frixione:diffHQ95} implements a 
massive scheme
next-to-leading order calculation (${\cal O}(\alpha_s^2)$) in the collinear 
factorisation approach.
It provides weighted parton level events 
with two or three outgoing partons, \ie a $c\bar{c}$
quark pair and possibly one additional light parton.
The fragmentation of the charm quarks into \dstar\ mesons is treated by a 
downscaling of their three-momenta according to the Peterson fragmentation 
function in a frame where the quark-antiquark pair is back-to-back. 

\paragraph{ZMVFNS}
The zero-mass vari\-able-flavour-number scheme
(ZMVFNS)~\cite{KniehlHein:D*+jet04,Kniehl:massless02} provides a next-to-leading order 
calculation (${\cal O}(\alpha_s^2)$) for \dstarpotherj\ cross sections 
in the collinear approach, neglecting the charm mass. 
The transition from the charm quark to 
the \dstar\ meson is treated using the scale dependent
fragmentation functions determined in~\cite{BKKfrag:98}.
In the determination of the uncertainty 
the initial and final state factorisation scales are varied
simultaneously (with a lowest value of $m_{t,\dstar}$ and
1.5~$m_{t,\dstar}$, respectively, see  Tab.~\ref{tab:calculations}). 

\paragraph{GMVFNS}
The general-mass vari\-able-flavour-number scheme  (GMVFNS) combines
the massless  with the massive scheme~\cite{Kniehl:2005mk,Kniehl:2004fy}.
Scale dependent fragmentation functions  as
determined in~\cite{KKSSfrag:05} are used. The calculation is only available
for inclusive \dstar\ production. 
The initial and final state factorisation scales are 
varied separately in the uncertainty determination.

For the NLO calculations which are compared to the 
\dstarpotherj\ and \dstardj\ measurements (FMNR and ZMVFNS), 
additional corrections for hadronisation effects 
(transition form partons to jets) are applied.
These hadronisation  corrections are calculated using \pythia .
In \pythia\ parton level jets are constructed from the generated quarks and gluons after
the parton showers have been simulated. 
The ratio of hadron and parton level cross sections in each bin is applied
to the NLO calculations as a hadronisation correction factor.

\section{Results}
\label{sec:xSecDeterm}
\subsection{Cross Section Measurement}

The bin averaged visible differential cross section with respect to a variable $Y$ 
(with bin width $\Delta Y$) 
is calculated according to
\begin{equation} 
\label{eq:diffcrosssec}
\frac{{\rm d}\sigma_{vis}(ep \to e' \dstarpm\mbox{(+jet)} X)}{{\rm d}Y} = 
\frac{N(\dstarpm\mbox{(+jet)}) \cdot (1-r)}
     {\Delta Y \cdot \mathcal{BR}(\dstar \to K \pi \pi_s) \cdot \mathcal{L} \cdot 
       \epsilon }
\end{equation}
where a 
correction $r=0.035$~\cite{H1:DstarGluonDens99} is applied to account  for
reflections from other $D^0$ decays within the $D^0$ mass window.
$\mathcal{BR}(\dstar \to K \pi \pi_s)$ is the 
branching ratio of the analysed \dstar\ decay chain  
and $\mathcal{L}$  
is the integrated luminosity.
The correction factor $\epsilon$ takes into account detector acceptances, trigger
and reconstruction efficiencies and migrations between bins.
The cross section is defined as the sum of the \dstarp\ and \dstarm\ cross sections 
and includes \dstar\ mesons  
from $b$-quark decays.

The average acceptance of the electron tagger is calculated as a convolution of the 
predicted $y$ - distribution 
with the $y$ - dependent tagger acceptance which is
determined as in~\cite{H1:PhotonProton95}. 
The reconstruction and trigger efficiencies as well as the acceptance
of the selection criteria are determined using a GE\-ANT~3.15~\cite{geant3} based simulation of
the detector response to events generated with \pythia.
The efficiency of
the particle identification using \dEdx\ and of the software filter
are deduced from the data.

The following sources of systematic uncertainties 
(summarised and quantified in Tab.~\ref{tab:systematics} for the integrated cross section)
have been studied:
\begin{itemize}
\itemsep0pt
\item The simulation of the trigger signals from the CJCs and the
  multi-wire proportional chambers have been verified (using data) to within an
  uncertainty of 4\%. Including
  the efficiency of the software filter this amounts to a
  4.5\% uncertainty on the cross section measurement.
\item The positron beam parameters and the absolute energy scale of the electron tagger have
  been varied within their uncertainties. The resulting
  average uncertainty on the tagger acceptance is 5.8\%, depending on \Wgp.
\item The uncertainty on the track reconstruction efficiency
  amounts to 6\% per \dstar\ meson.
\item The uncertainty on the efficiency of the particle identification using \dEdx\ is 
  estimated to be 2\% per \dstar\ meson~\cite{diss:geroflucke05}.
\item The uncertainty on the \dstar\ branching ratio is 2.5\%~\cite{PDG2004}.
\item The uncertainty on the correction for reflections from other $D^0$ decays is 
   1.5\%~\cite{H1:DstarGluonDens99}.
\item The uncertainty on the extraction of the \dstar\ signal from the
  $\Delta m$ distributions has been determined
  in~\cite{diss:geroflucke05} and amounts to 3\%.
\item The model dependence of the correction factors is estimated by using 
  \cascade\ instead of \pythia. Half of the resulting deviation is taken
  as the uncertainty on the signal extraction procedure, \ie 1.3\% for the total
  {\it inclusive \dstar\ }
  sample and up to 17\% for the differential distributions.
\item The luminosity is determined with a precision of 1.5\%.
\item The uncertainty on the energy of the HFS objects 
  leads to a systematic uncertainty on the \dstarmj\ cross sections of 2.8\%.
\end{itemize}
\begin{table}[bt]
  \begin{center}
    \begin{tabular}{|l|c|c|}
  \hline
  sources  & \multicolumn{2}{c|}{\begin{tabular}{c|c}
              {\it inclusive \dstar } & \dstar\ with jets \\
              \hline
               [\%]  & [\%]  
              \end{tabular}}
               \\
  \hline\hline
  trigger efficiency & \multicolumn{2}{c|}{{4.5}} \\ \hline 
  electron tagger acceptance & \multicolumn{2}{c|}{{5.8}} \\\hline 
  track reconstruction    &\multicolumn{2}{c|}{{6}}\\\hline
  particle identification & \multicolumn{2}{c|}{2} \\ \hline
  $\mathcal{BR}(\dstar \to K \pi \pi_s)$& \multicolumn{2}{c|}{2.5} \\\hline
  reflections       & \multicolumn{2}{c|}{1.5}\\ \hline
  signal extraction & \multicolumn{2}{c|}{3}\\\hline
  model dependence  & \multicolumn{2}{c|}{{1.3}} \\\hline
  luminosity $\mathcal{L}$ & \multicolumn{2}{c|}{1.5} \\ \hline
  HFS objects energy scale & \multicolumn{2}{c|}{\begin{tabular}{c|c}
                             --- & 2.8 \end{tabular}} \\
  \hline\hline
  {\bf in total} & \multicolumn{2}{c|}{\begin{tabular}{c|c} {\bf 10.7} & {\bf 11.1}\end{tabular}} \\
  \hline
\end{tabular}
\Mycaption{Systematic uncertainties of the integrated cross section measurements.
}
\label{tab:systematics}
\end{center}
\end{table}
The total
systematic uncertainty is obtained by adding each uncertainty in quadrature
and amounts to 11\% for the integrated cross sections and up to 20\% for differential
distributions.

\subsection{\boldmath Inclusive \dstar\ Cross Sections}
\label{sec:xSecD}
The integrated \dstar\  photoproduction cross section 
is measured to be:
\begin{displaymath}
  \label{eq:totXsec}
  \sigma_{vis}(ep \to e' \dstarpm\ X) = 
  6.45 \pm 0.46\mbox{ (stat.)} \pm 0.69\mbox{ (sys.)} \mbox{ nb},
\end{displaymath}
in the visible range given by 
 $0.29 < y <0.65,$  $Q^2 < 0.01$~G\eV$^2$, 
$\ptds > 2$~G\eV\ and $|\etads| < 1.5$. 
In Tab.~\ref{tab:totCrossSec} the predictions from QCD calculations are listed and
compared with this measurement.
\begin{table}[tbp]
  \centering
\begin{tabular}{l|c|c|c}
  $\sigma_{vis} $ [nb] & inclusive \dstar &\dstarpotherj 
  &  \dstardj \\
  \hline
  \Lineup {\bf Data} & {\boldmath $6.45 \pm 0.46 \pm 0.69$}& {\boldmath $3.01 \pm 0.29 \pm 0.33$ }
  & {\boldmath $2.32 \pm 0.25 \pm 0.26$}\\ \hline\hline
  \Lineup {FMNR}     & { $5.9^{+2.8}_{-1.3}$}    &{\small$\left(2.65^{+0.78}_{-0.42}\right)$}
  & {\small$\left(2.44^{+0.97}_{-0.52}\right)$}\\ 
  \Lineup \hspace{1em} $\otimes$ had. corr. &         & {$2.35^{+0.69}_{-0.37}$}
  & {$2.09^{+0.83}_{-0.44}$}\\ \hline%
  \Lineup {ZMVFNS}      & -- & {\small$\left(3.05^{+0.62}_{-0.47}\right)$ } & --\\
  \Lineup \hspace{1em}  $\otimes$ had. corr.&        & {$2.71^{+0.55}_{-0.42}$}& --\\ \hline
  \Lineup {GMVFNS}      &{ $8.2^{+5.3}_{-4.0}$}&  -- & --\\ \hline
  {\pythia}  & { $8.9$}             & {$3.8$} & {$2.8$}\\ \hline
  \Lineup {\cascade} & { $5.38^{+0.54}_{-0.62}$ } & {$3.08^{+0.22}_{-0.28}$} 
  & {$2.48^{+0.17}_{-0.20}$}\\%
\end{tabular}

  \Mycaption{Integrated cross section in the visible range for 
  {\it inclusive \dstar },
    \dstarpotherj\ and \dstardj\ photoproduction.
    For the cross sections for processes involving jets the result of the NLO calculations before
    the correction for hadronisation effects is given in brackets.
  }
  \label{tab:totCrossSec}
\end{table}
The central values from FMNR and \cascade\ are slightly lower than the measured result, whereas those of 
\pythia\ and GMVFNS are 
higher. 

\begin{figure}[htbp]
  \begin{center}
    \includegraphics[width=0.48\textwidth]{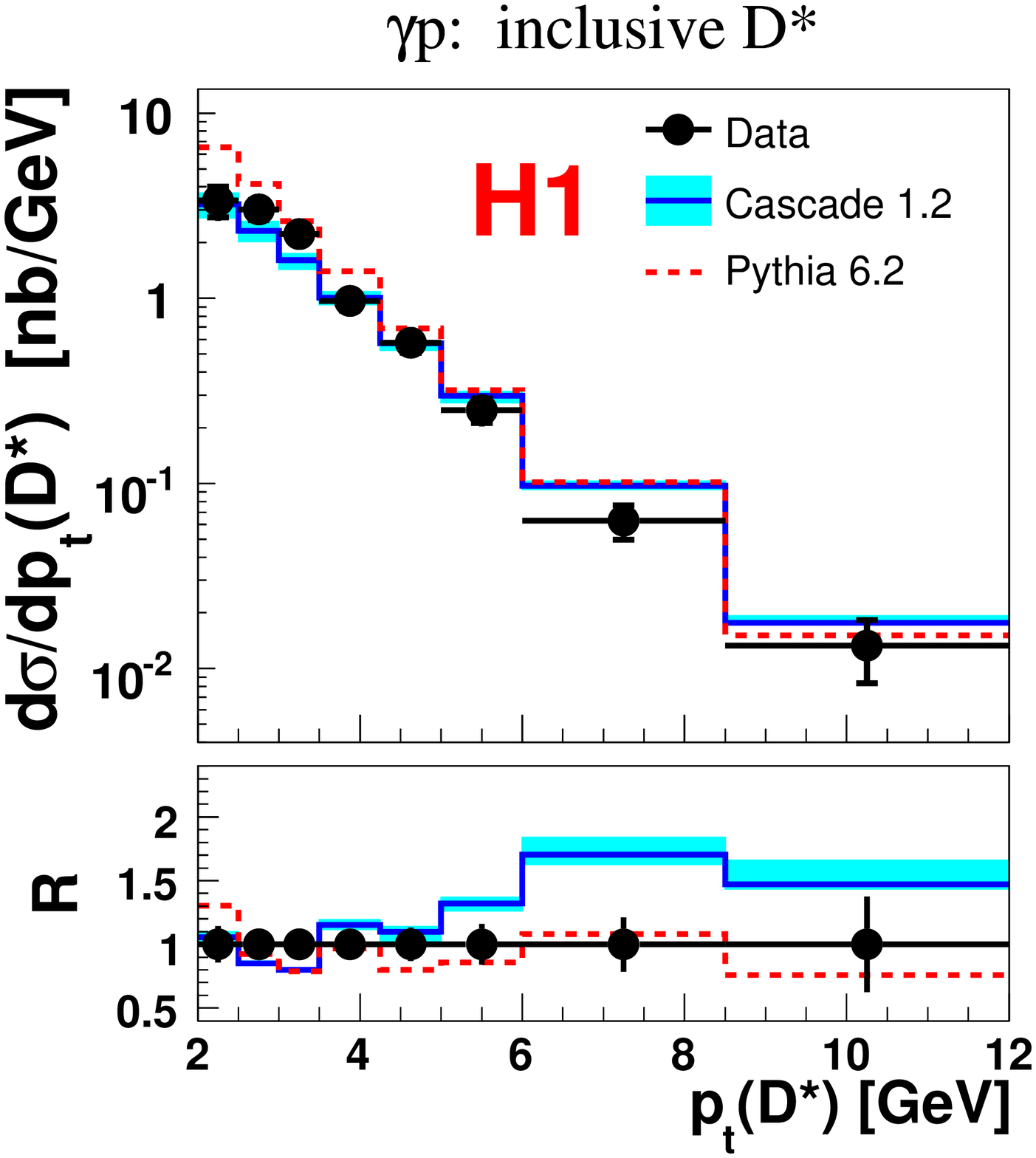}
    \includegraphics[width=0.48\textwidth]{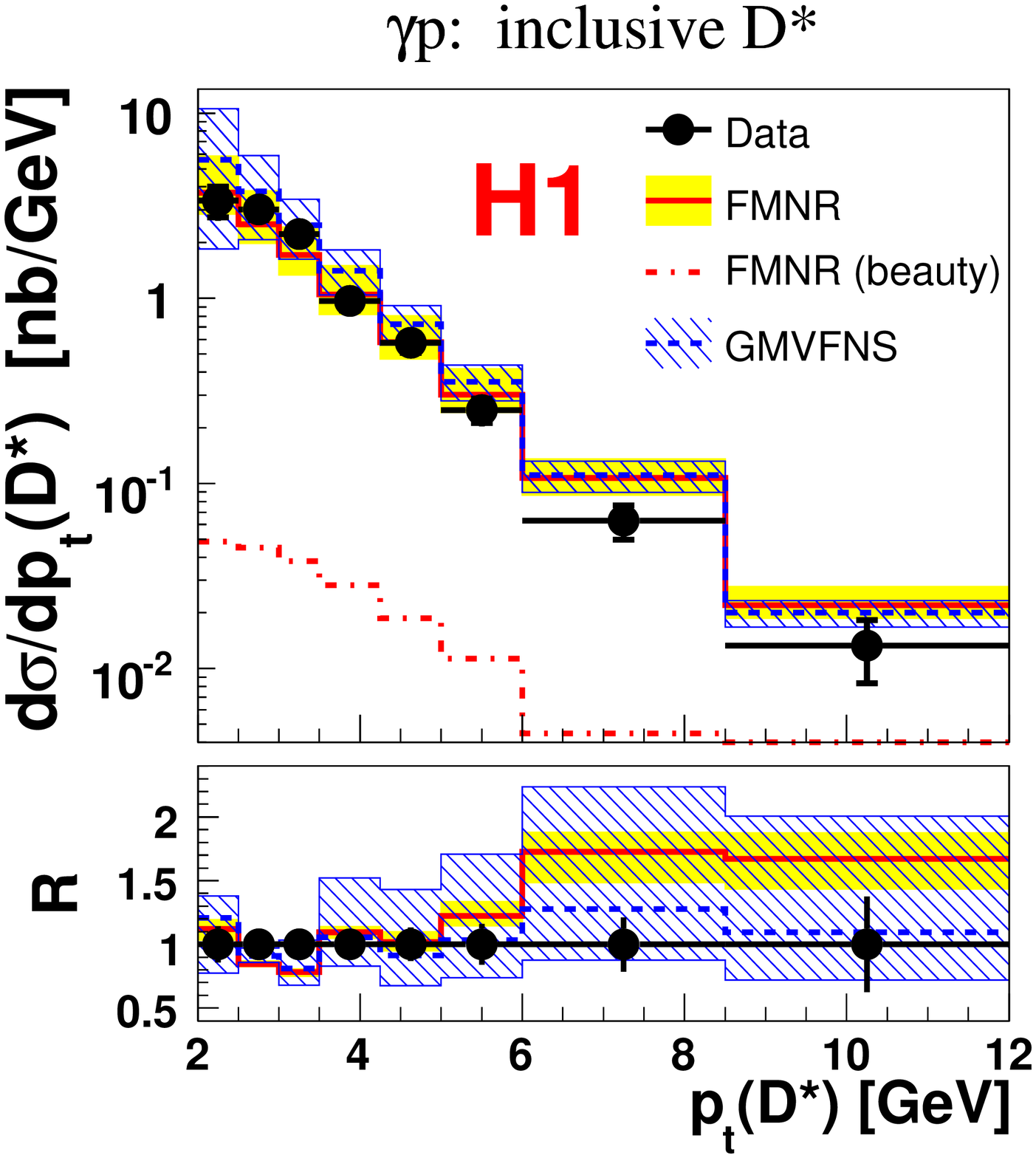}
    \setlength{\unitlength}{\textwidth}
    \begin{picture}(0,0)
      \put(-0.85,0.22){\bfseries a)}
      \put(-0.37,0.22){\bfseries b)}
    \end{picture}
    \includegraphics[width=0.48\textwidth]{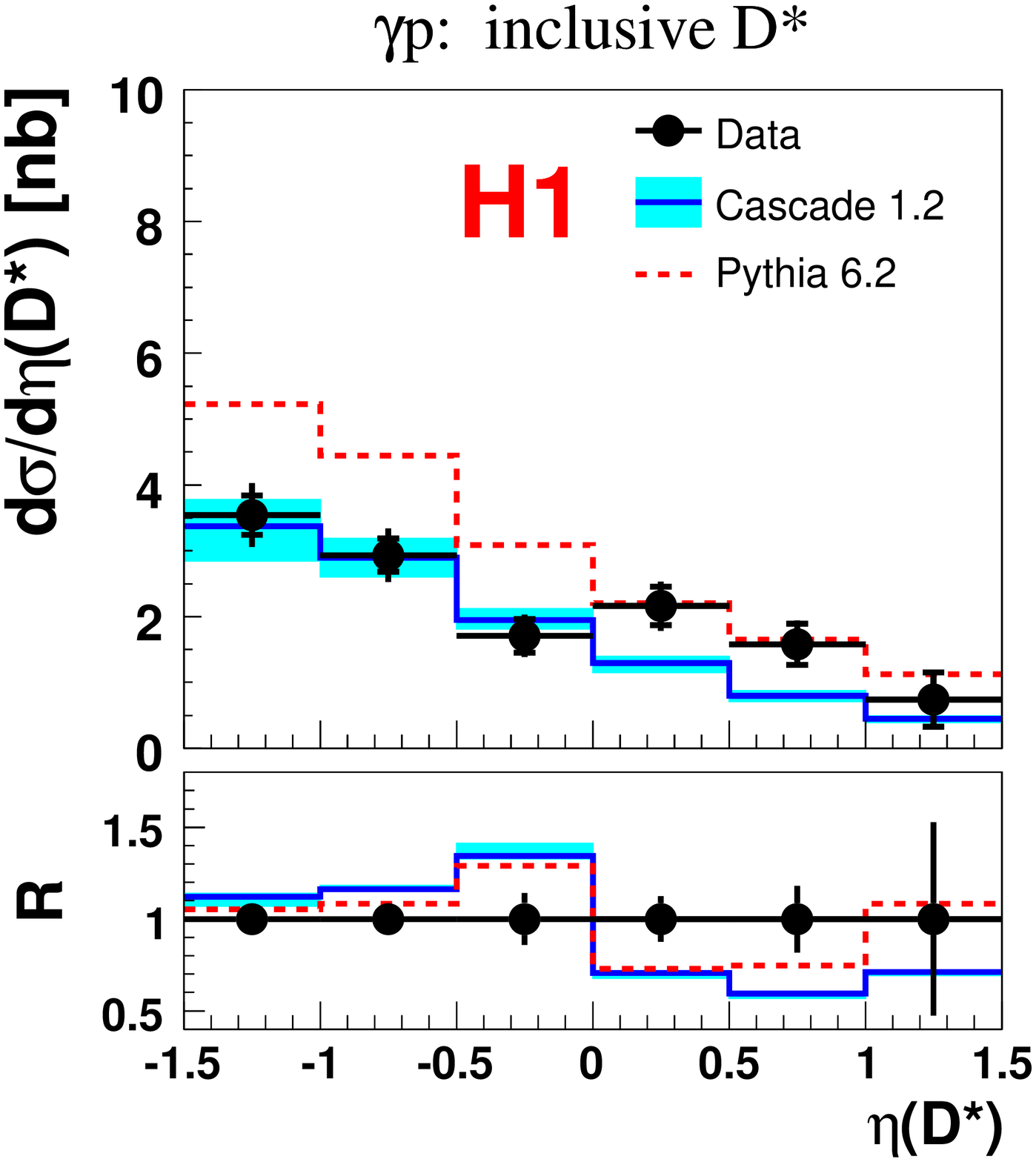}
    \includegraphics[width=0.48\textwidth]{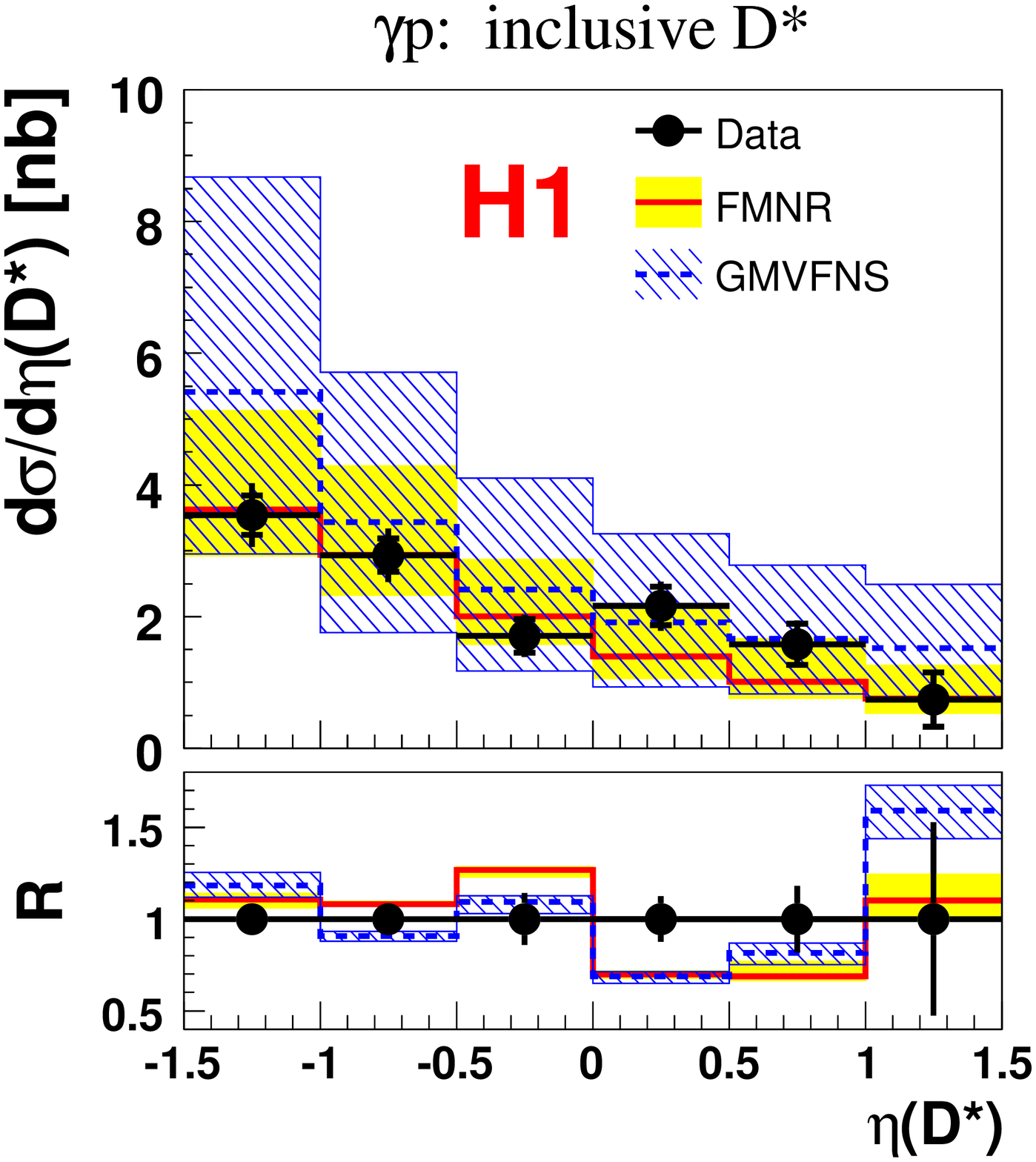}
    \begin{picture}(0,0)
      \put(-0.85,0.22){\bfseries c)}
      \put(-0.37,0.22){\bfseries d)}
    \end{picture}
    \Mycaption{Inclusive \dstar\ cross sections as a function of \ptds\ (a,b) and
      \etads\ (c,d) 
      compared with the predictions of PYTHIA and CASCADE on the left and of the
      next-to-leading order calculations  FMNR and GMVFNS on the right.
      For FMNR the beauty contribution is shown separately for \ptds.
      Here and in the following figures the inner error bars indicate the statistical errors and
      the outer error bars show the statistical and systematic uncertainties added in quadrature.
      The normalised ratio $R$ (Eq.~\protect\ref{eq:Rdef}) is also shown.
     }
    \label{fig:xsecIncl_pteta}
  \end{center}
\end{figure}

\begin{figure}[htbp]
  \begin{center}
    \includegraphics[width=0.4\textwidth]{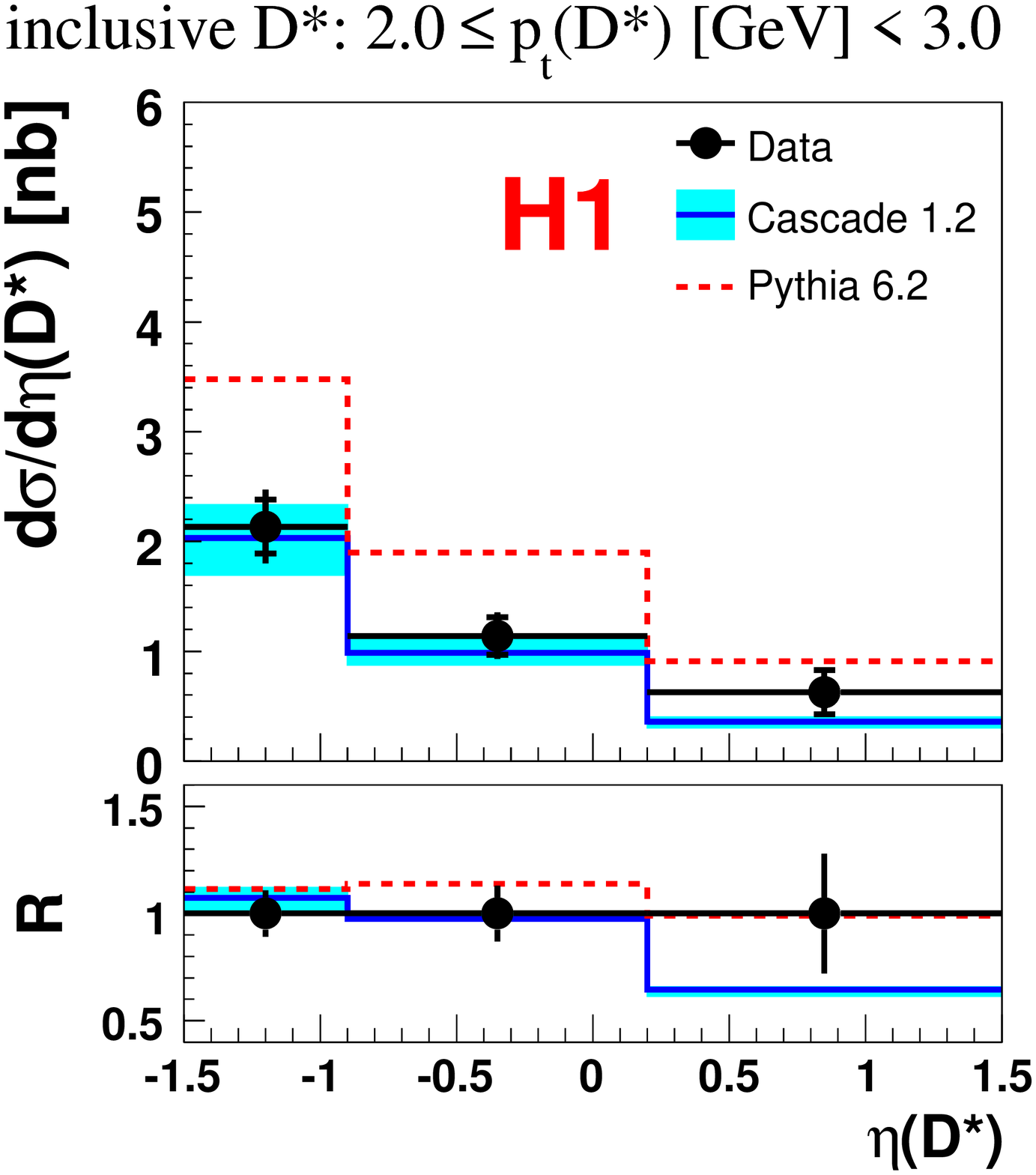} \hspace{0.03\textwidth}
    \includegraphics[width=0.4\textwidth]{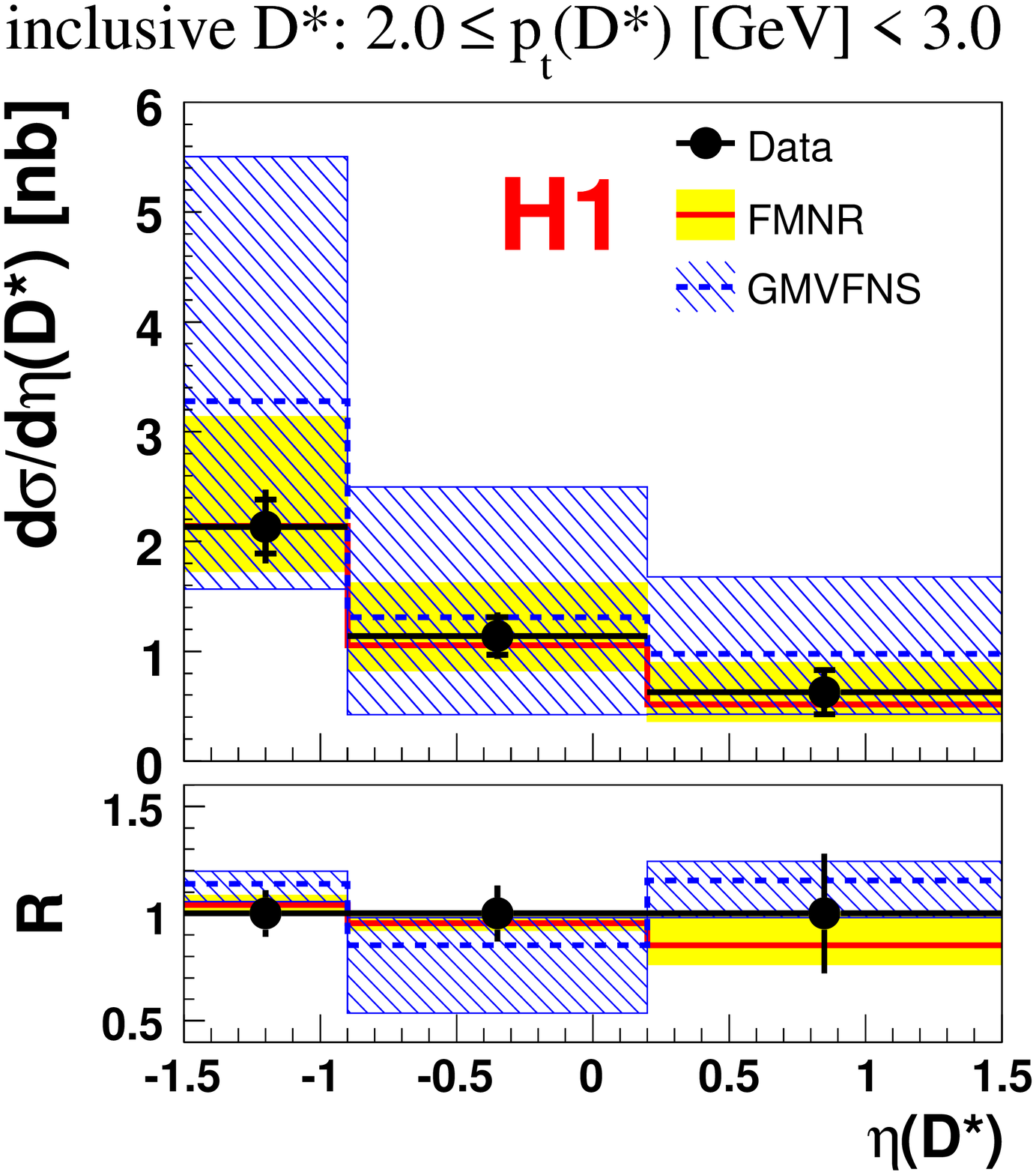}
    \setlength{\unitlength}{\textwidth}
    \begin{picture}(0,0)
      \put(-0.745,0.37){\bfseries a)}
      \put(-0.25,0.37){\bfseries b)}
    \end{picture}
    \includegraphics[width=0.4\textwidth]{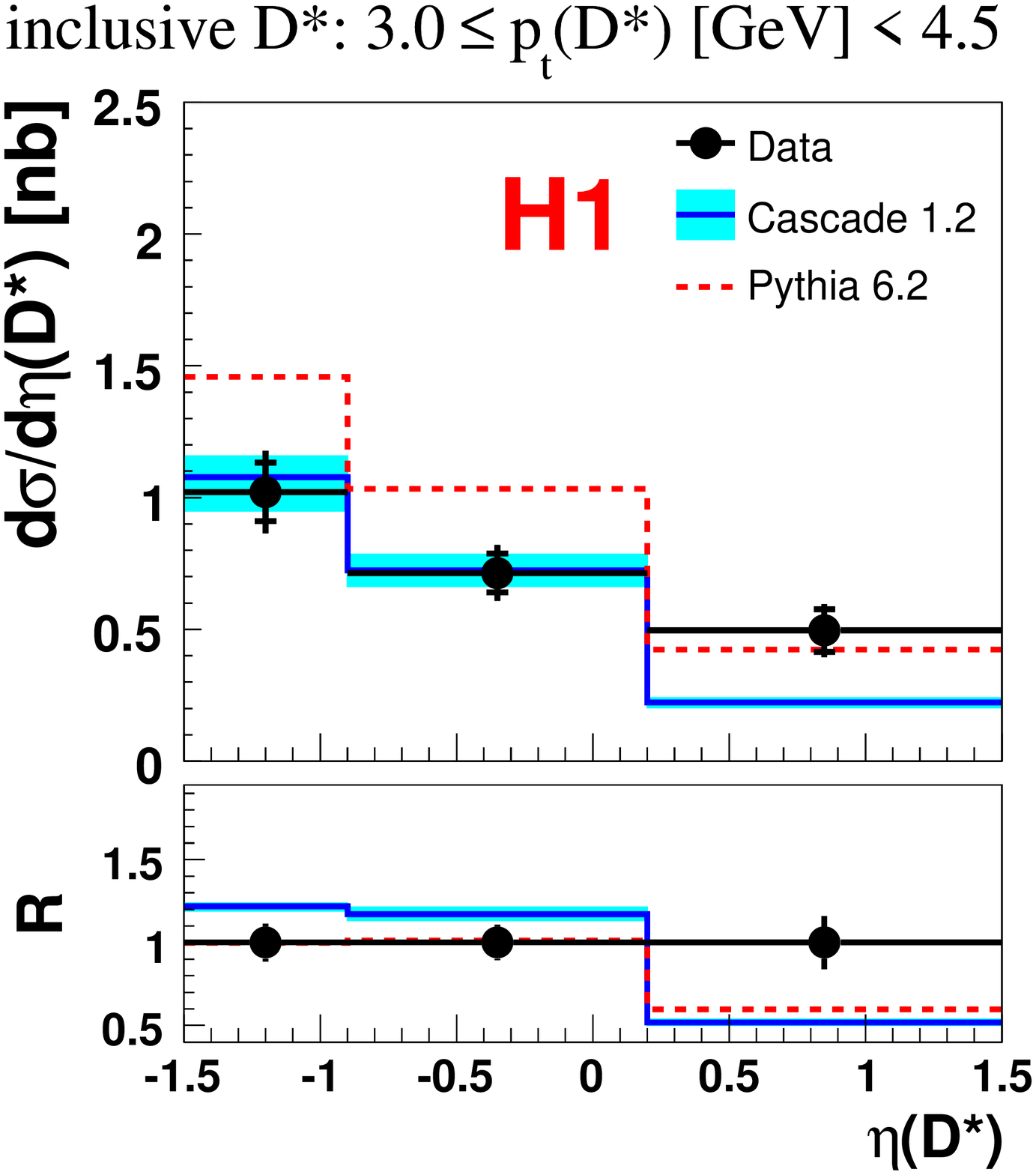} \hspace{0.03\textwidth}
    \includegraphics[width=0.4\textwidth]{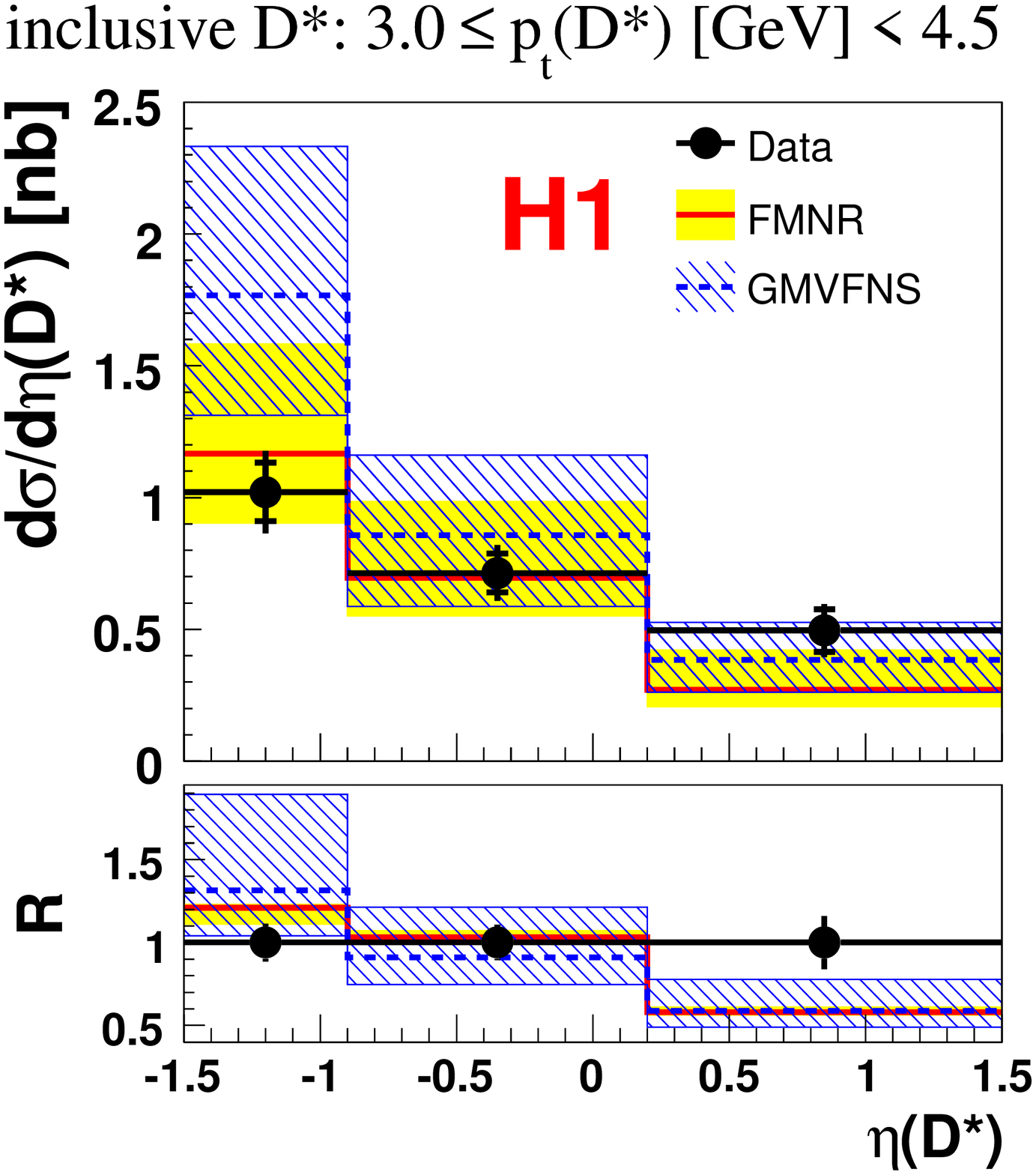}
    \begin{picture}(0,0)
      \put(-0.745,0.37){\bfseries c)}
      \put(-0.25,0.37){\bfseries d)}
    \end{picture}
    \includegraphics[width=0.4\textwidth]{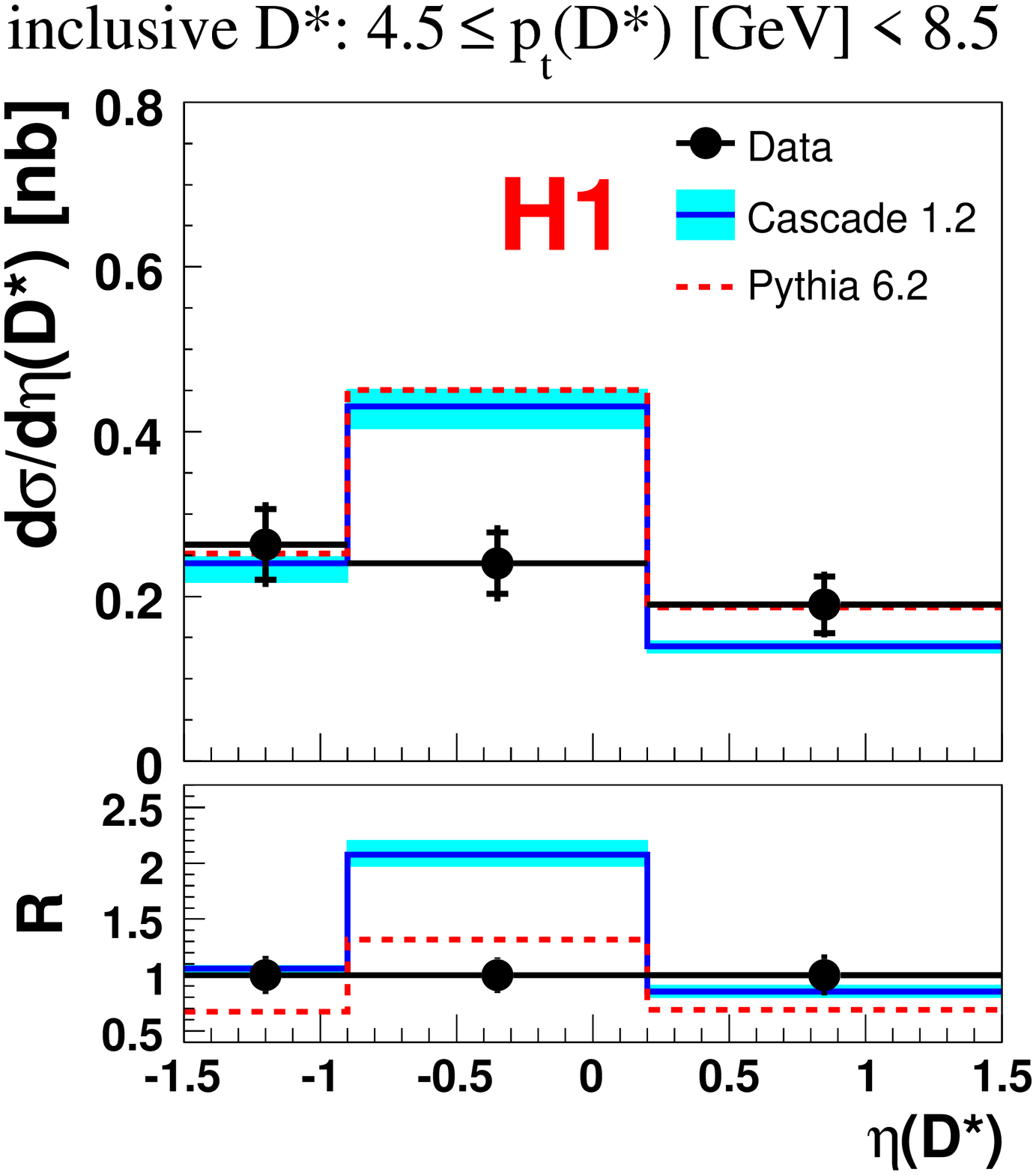} \hspace{0.03\textwidth}
    \includegraphics[width=0.4\textwidth]{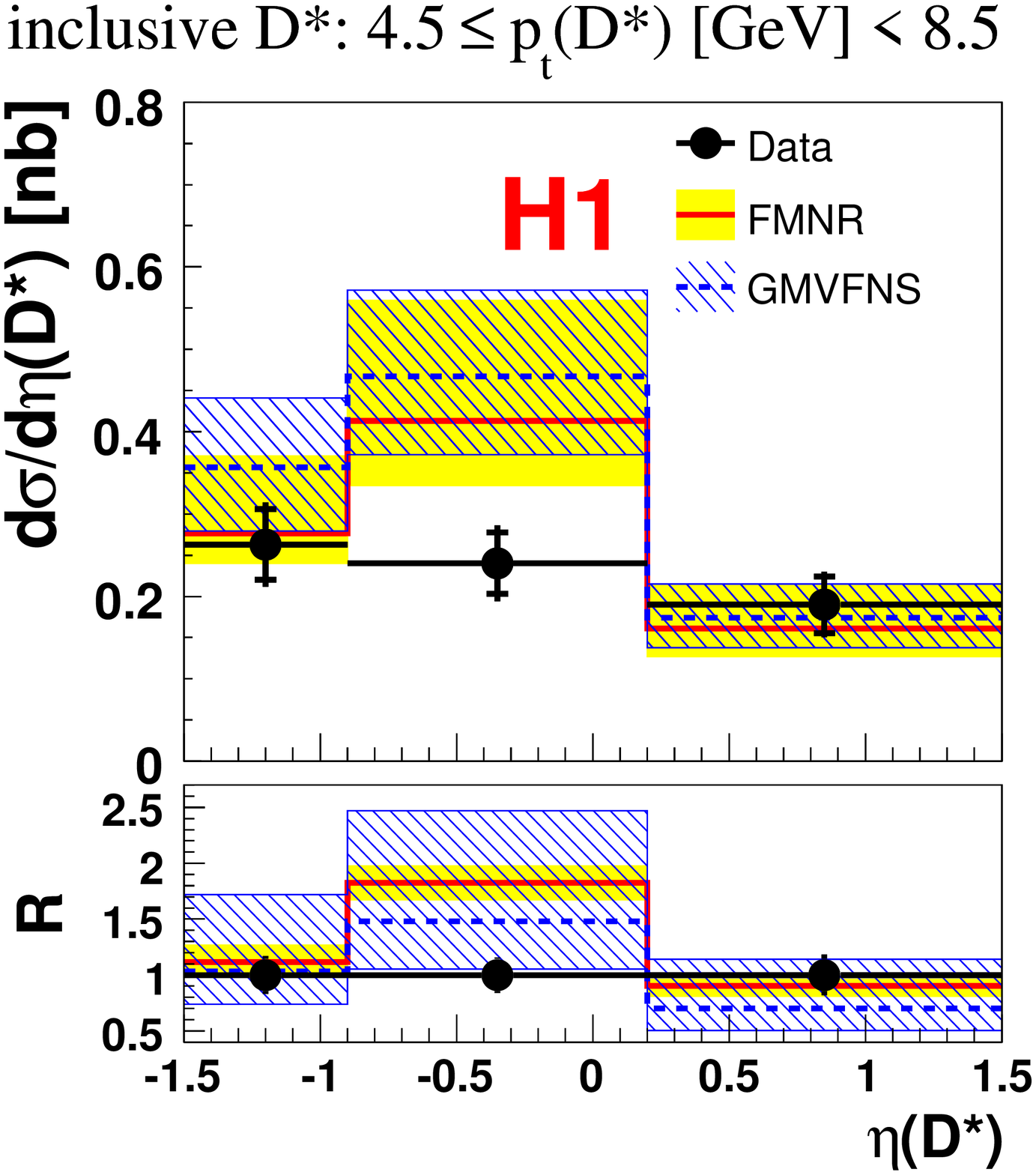}
    \begin{picture}(0,0)
      \put(-0.745,0.37){\bfseries e)}
      \put(-0.30,0.37){\bfseries f)}
    \end{picture}
    \Mycaption{Inclusive \dstar\ cross sections as a function of \etads\ for three bins of \ptds\ 
      compared with the predictions of PYTHIA and CASCADE on the left and of the
      next-to-leading order calculations  FMNR and GMVFNS on the right. 
      }
    \label{fig:xsecInclDoDiff}
  \end{center}
\end{figure}

\begin{figure}[htbp]
  \begin{center}
    \setlength{\unitlength}{\textwidth}
    \includegraphics[width=0.48\textwidth]{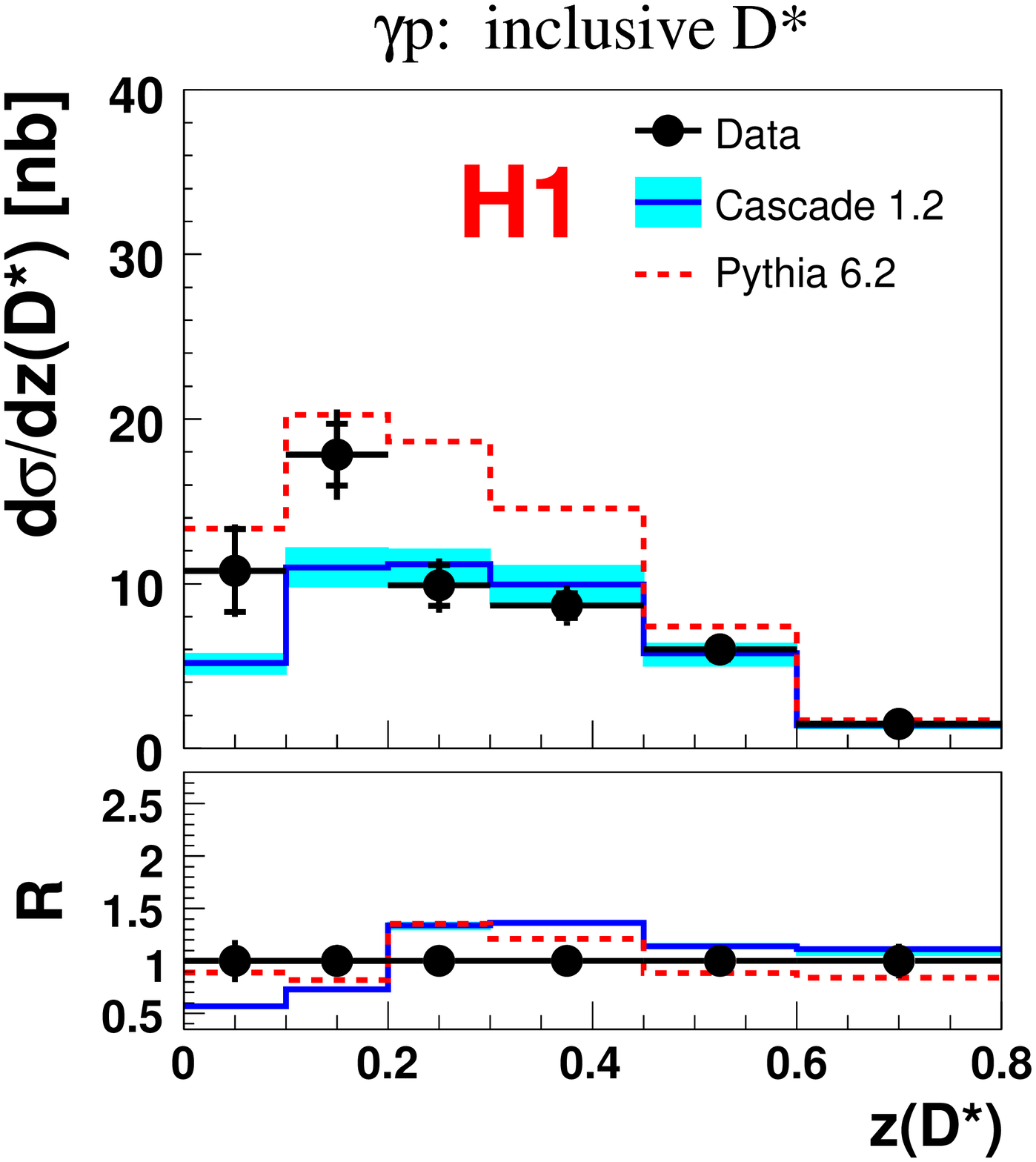}
    \includegraphics[width=0.48\textwidth]{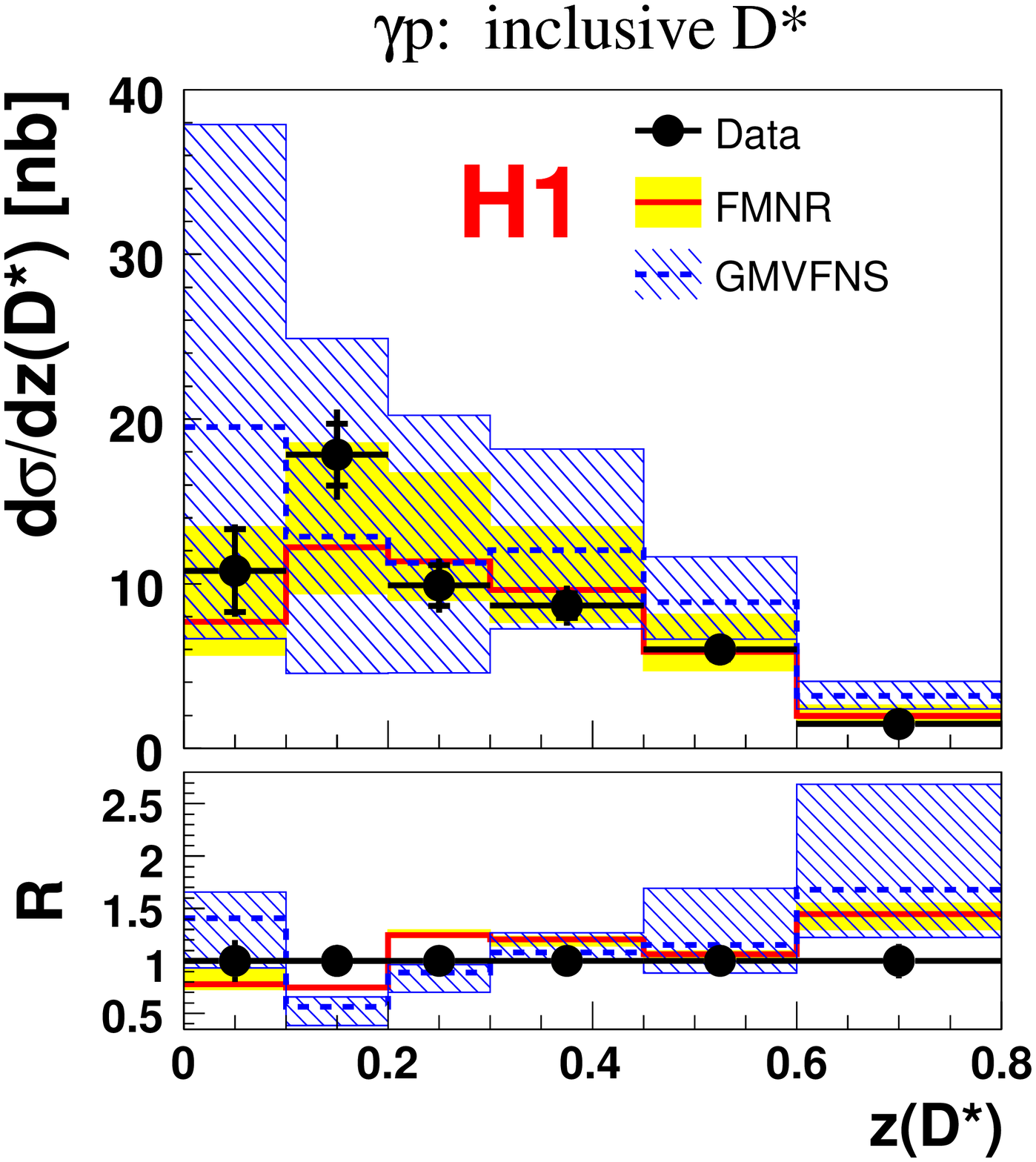}
    \begin{picture}(0,0)
      \put(-0.85,0.45){\bfseries a)}
      \put(-0.33,0.45){\bfseries b)}
    \end{picture}
    \includegraphics[width=0.48\textwidth]{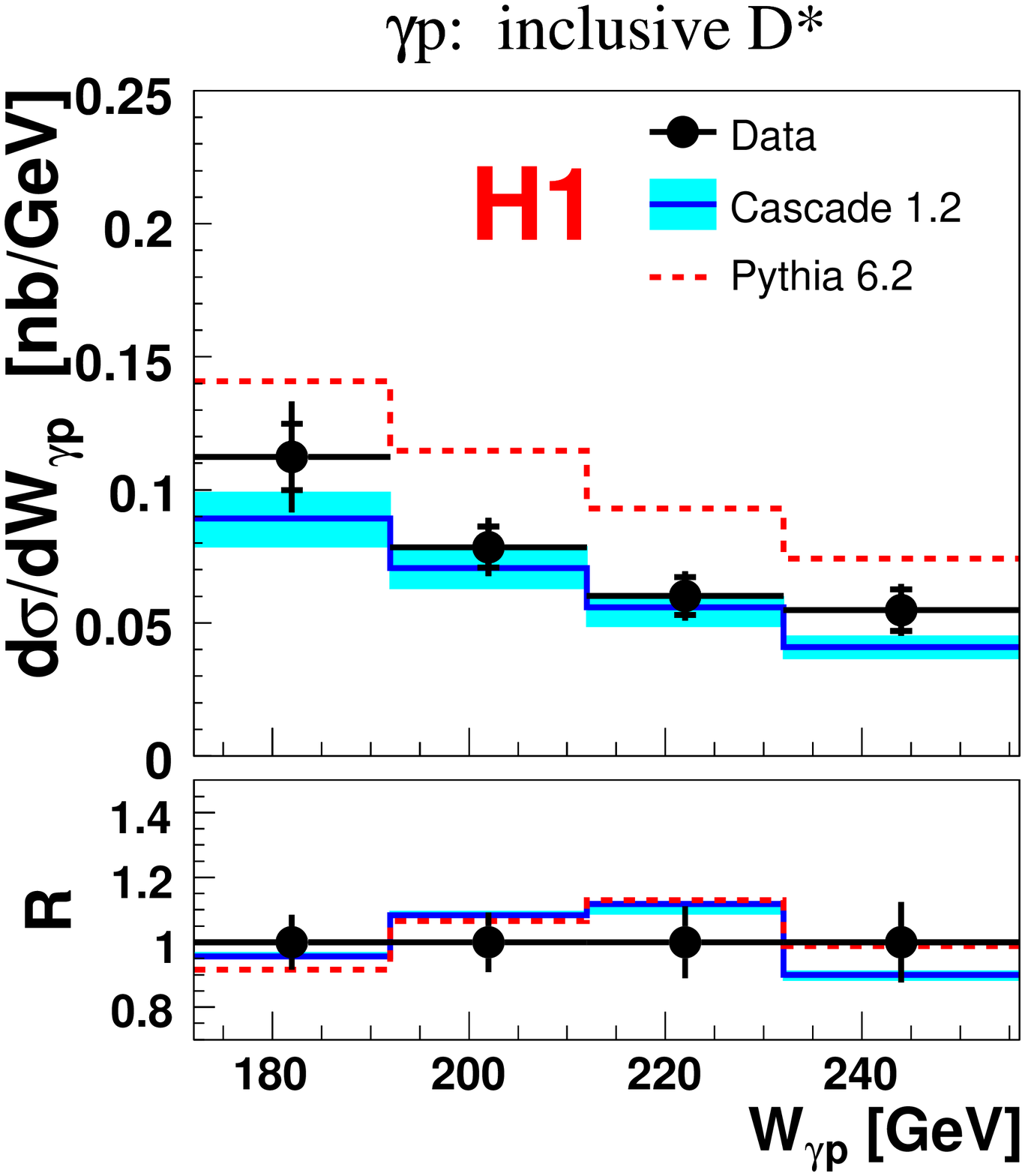}
    \includegraphics[width=0.48\textwidth]{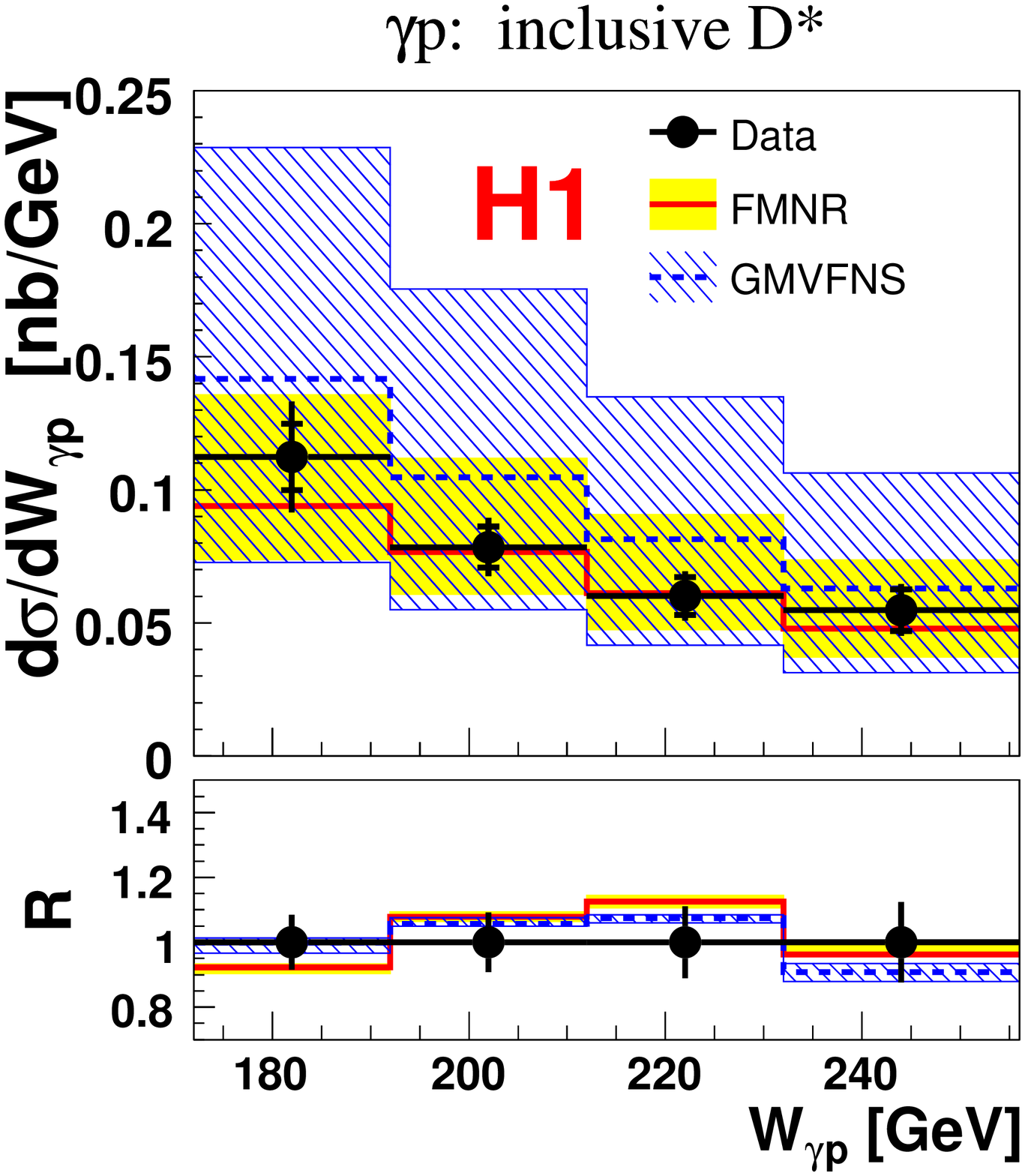}
    \begin{picture}(0,0)
      \put(-0.85,0.45){\bfseries c)}
      \put(-0.36,0.22){\bfseries d)}
    \end{picture}
    \Mycaption{
      Inclusive \dstar\ cross sections as a function of the inelasticity \zDs\ (a,b)
      and the photon-proton centre-of-mass energy \Wgp\ (c,d) 
      compared with the predictions of PYTHIA and CASCADE on the left and of the
      next-to-leading order calculations  FMNR and GMVFNS on the right.
   }
    \label{fig:xsecIncl_zdswgp}
  \end{center}
\end{figure}

The  measured bin averaged differential cross sections are shown in 
Figs.~\ref{fig:xsecIncl_pteta}-\ref{fig:xsecIncl_zdswgp} and given in
Tabs.~\ref{tab:diffXsec1D}-\ref{tab:diffXsec1Ddd}.
The predictions show large uncertainties compared with those on the data.  
These uncertainties partially cancel in the comparison of the shapes of the distributions.
The ratio
\begin{equation}
R = \frac{\frac{1}{\sigma^{\rm calc}_{vis}}\frac{d\sigma^{\rm calc}}{dY}}
{\frac{1}{\sigma^{\rm data}_{vis}}\frac{d\sigma^{\rm data}}{dY}}
\label{eq:Rdef}
\end{equation} 
is therefore also presented, where 
$Y$ denotes any measured variable. Note that here each 
differential cross section is normalised
to its own visible cross section
({\it inclusive \dstar}, \dstarpotherj\ or \dstardj).

In Fig.~\ref{fig:xsecIncl_pteta}a and b the cross section is shown differentially  in 
the transverse momentum of the \dstar .
The cross section falls steeply with increasing \ptds\ as predicted by all
calculations. \cascade\ and FMNR predict a distribution which falls less steeply
at large \ptds\ than is visible in the data.
The \pythia\ and GMVFNS calculations describe the slope well. However, the 
theoretical uncertainty of the GMVFNS calculations is large. 
In Fig.~\ref{fig:xsecIncl_pteta}b the small contribution coming from
$b$-decays as determined from FMNR is shown separately.

In Fig.~\ref{fig:xsecIncl_pteta}c and d the differential cross section  is shown as a function
of the pseudorapidity \etads .
The cross section decreases with increasing $\eta$ (the forward direction $\eta > 0$).
All calculations predict a similar shape, differing from that in the data, which 
shows a larger relative contribution in the forward direction.

In Fig.~\ref{fig:xsecInclDoDiff} the differential cross section  is shown as a function of
\etads\ for three bins in \ptds .
As for  the inclusive \etads\ distribution none of the calculations is able to 
describe  the data completely. 

The inelasticity \zDs , defined by
$\zDs = P \cdot p(\mbox{\dstar})/ (P \cdot q)$ with $P$, $p(\mbox{\dstar})$ 
and $q$ being the
four-momenta of the incoming proton, the \dstar\ meson and the exchanged
photon,
is a measure of the fraction of photon energy transferred to the
\dstar\ meson in the proton rest frame. This
quantity  is sensitive to the production
mechanism and to the $c\to \dstar$ fragmentation function. It is reconstructed as 
$\zDs = (E-p_z)_{\dstar}/(2  y  E_e)$.
The differential cross section as a function of \zDs\ is compared to the predictions in 
Fig.~\ref{fig:xsecIncl_zdswgp}a and b. For $\zDs > 0.2$  the 
 \cascade\ and FMNR predictions agree fairly well with the data.
However, in both cases the predictions (central values) underestimate the
measured cross section in the region $\zDs < 0.2$. A similar finding is
reported for a measurement of \dstar\ production in 
deep inelastic scattering ($Q^2 > 1$~G\eV$^2$)~\cite{H1dstarDIS:01}.
\pythia\  can account for the large cross section at
small \zDs , but overestimates the cross section at medium \zDs .
The central value of the GMVFNS calculation is slightly higher than the measurement,
but the uncertainties cover the whole range.
None of the calculations describes the overall shape of the data.

The cross section as a function of \Wgp\ is shown in Fig.~\ref{fig:xsecIncl_zdswgp}c and d.
It falls weakly with increasing \Wgp.
The behaviour observed in the data is reproduced by all calculations within the uncertainties.

In summary, significant differences are observed in the shape of most 
of the distributions between data and all central theoretical predictions.
A good agreement is only found for the \Wgp\ distribution for which
each bin receives contributions throughout the available phase space of the produced 
\dstarpm\ meson.  
Here all calculations predict very similar shapes. 
For those observables which are especially sensitive to the phase space
distribution of the outgoing charm quark, \ie \ptds , \etads\ and 
\zDs , large deviations in shape between data and theory are observed.
It is interesting to observe that the cross sections
predicted by the calculations which treat  
the heavy quark mass explicitly (\cascade\ and FMNR) are similar. 
  
\subsection{\boldmath \dstarpotherJ\ and \dstarDj\ Cross Sections}
\label{sec:xSecDjet}
A more detailed investigation of the  heavy quark production process is
performed by analysing events with a \dstarpm\ meson with either a jet not containing the 
\dstar\ (\dstarpotherj\ ) or with two jets (\dstardj\ ).
 In this way one can tag a
second outgoing parton from the hard interaction in addition to the
(anti-)charm quark. The requirements for the \dstar\ meson
and the photoproduction selection are the same   
as in the inclusive case. In the \dstarpotherj\ 
analysis the jet is required to have a transverse momentum $\ptj > 3 $~G\eV\ in the range
of $|\etaj| < 1.5$ in the laboratory frame. 
The integrated (\dstarpotherj ) 
cross section in the visible range 
given in Tab.~\ref{tab:dstarcuts} is measured to be
\begin{displaymath}
  \label{eq:totXsecDstarjet}
  \sigma_{vis}(ep \to e' \dstarpmpotherj\ X) = 
  3.01 \pm 0.29\mbox{ (stat.)} \pm 0.33 \mbox{ (sys.)} \mbox{ nb}.
\end{displaymath}
The integrated \dstardj\ cross section ($\ptj > 4(3)$~G\eV, $|\etaj| < 1.5$) is measured to be
(in the visible range given in Tab.~\ref{tab:dstarcuts})
\begin{displaymath}
  \label{eq:totXsecDstarDijet}
  \sigma_{vis}(ep \to e' \dstarpmdj\ X) = 
  2.32 \pm 0.25\mbox{ (stat.)} \pm 0.26 \mbox{ (sys.)} \mbox{ nb}.
\end{displaymath}
In Tab.~\ref{tab:totCrossSec} both values are compared with the predictions from
the QCD calculations. 
All predictions agree with the measurements within the quoted
calculated uncertainties, including ZMVFNS for the \dstarpotherj\ sample.

\begin{figure}[htbp]
  \begin{center}
    \includegraphics[width=0.48\textwidth]{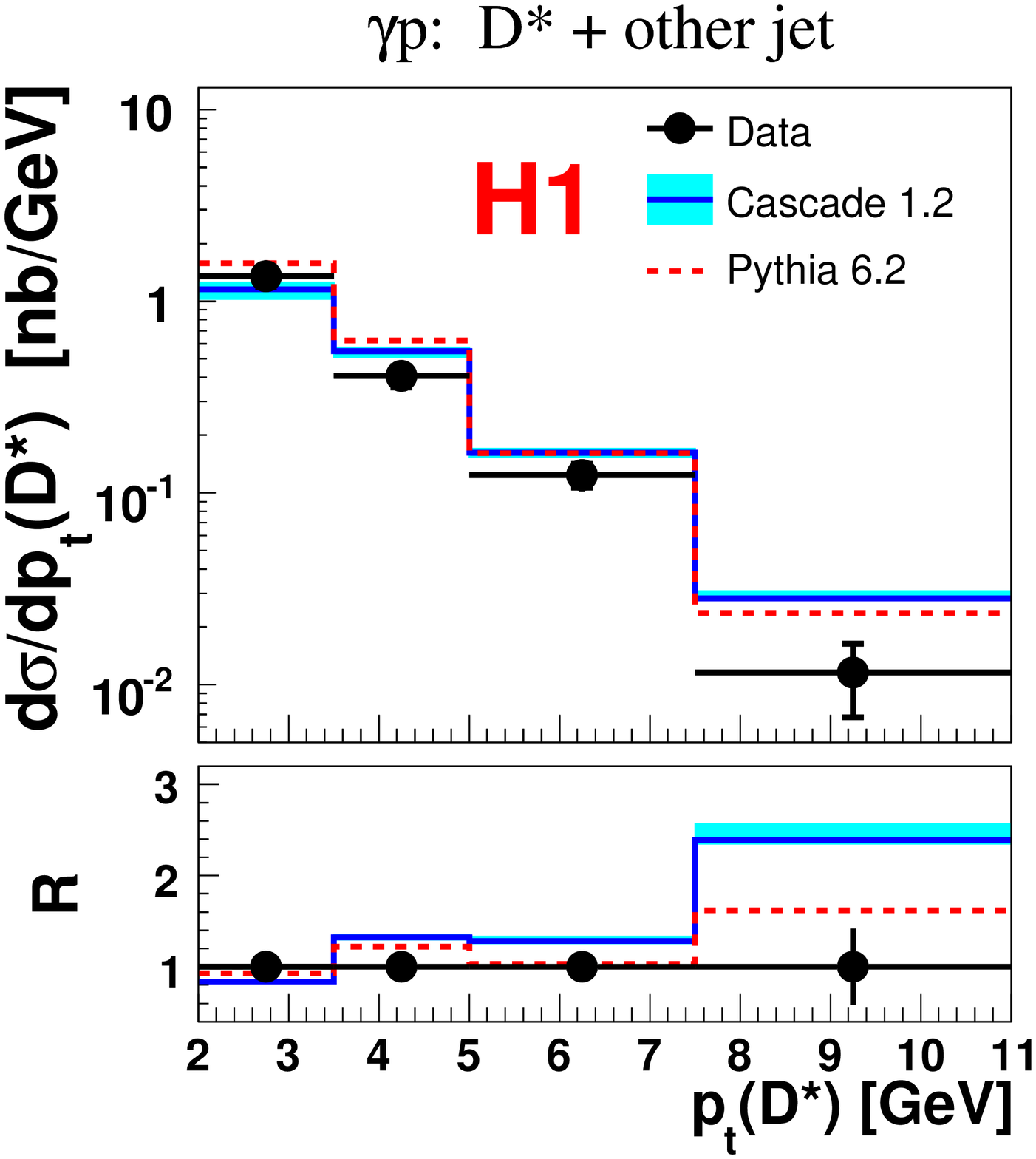}
    \includegraphics[width=0.48\textwidth]{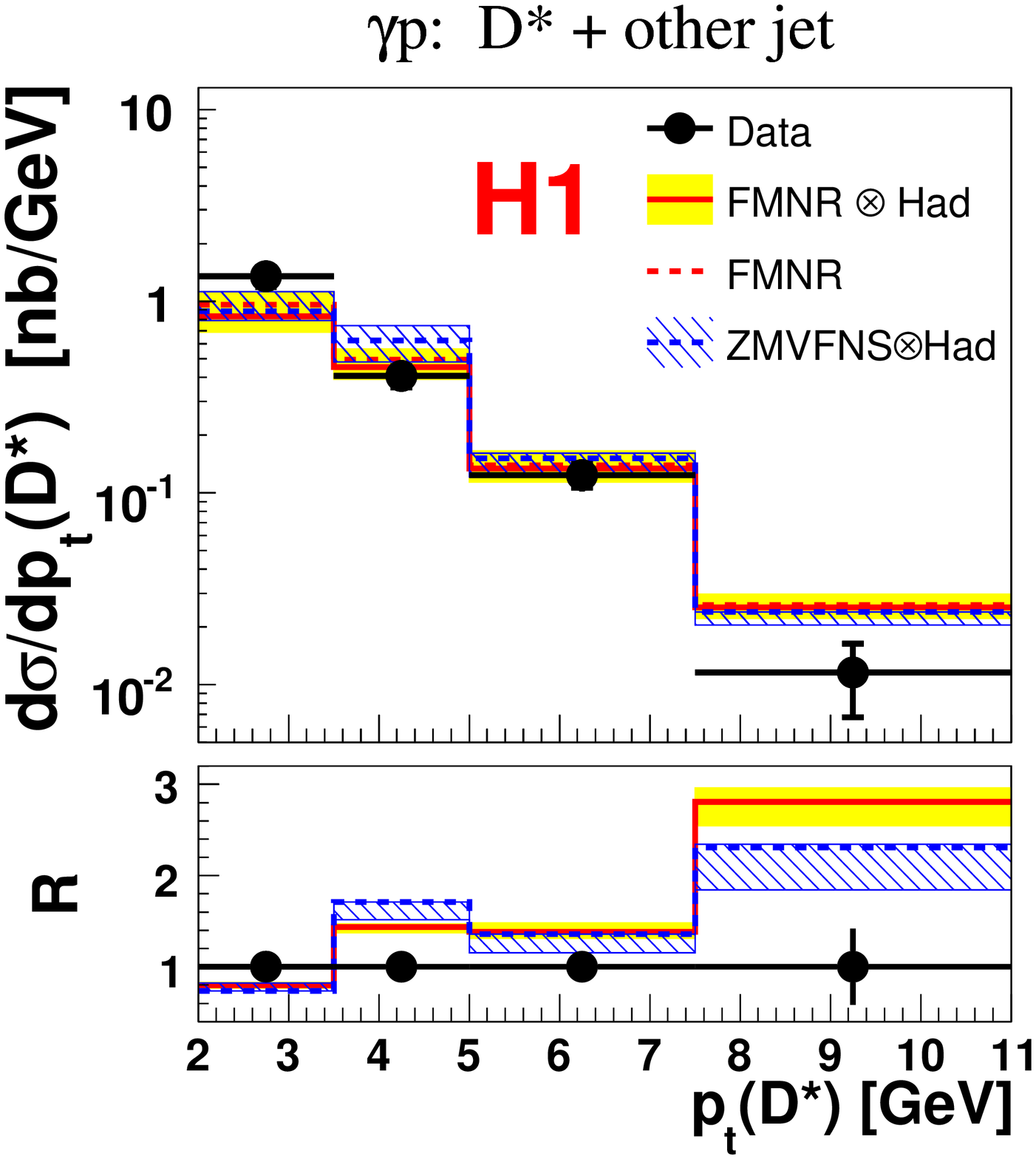}
    \setlength{\unitlength}{\textwidth}
    \begin{picture}(0,0)
      \put(-0.85,0.22){\bfseries a)}
      \put(-0.36,0.22){\bfseries b)}
    \end{picture}
    \includegraphics[width=0.48\textwidth]{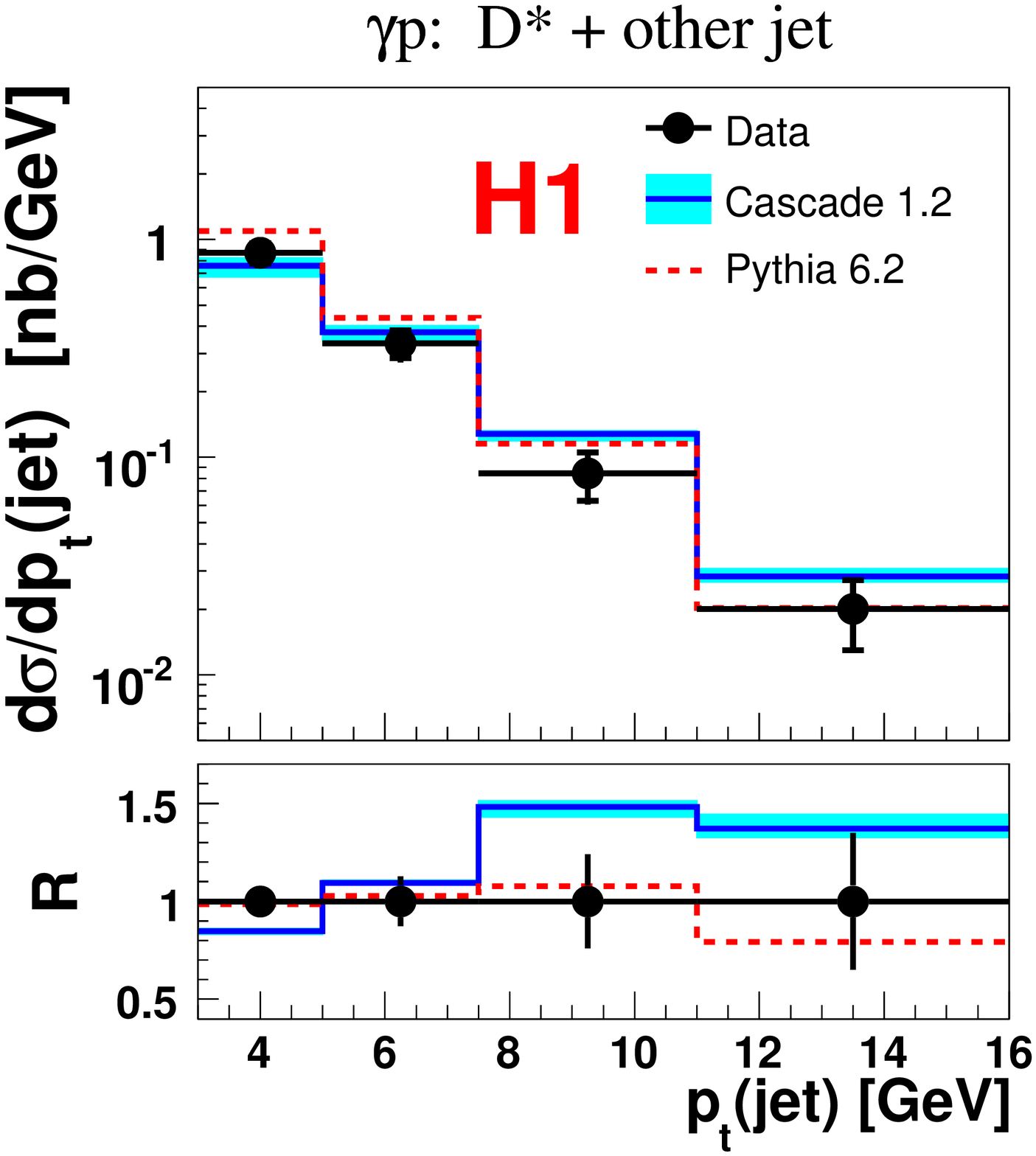}
    \includegraphics[width=0.48\textwidth]{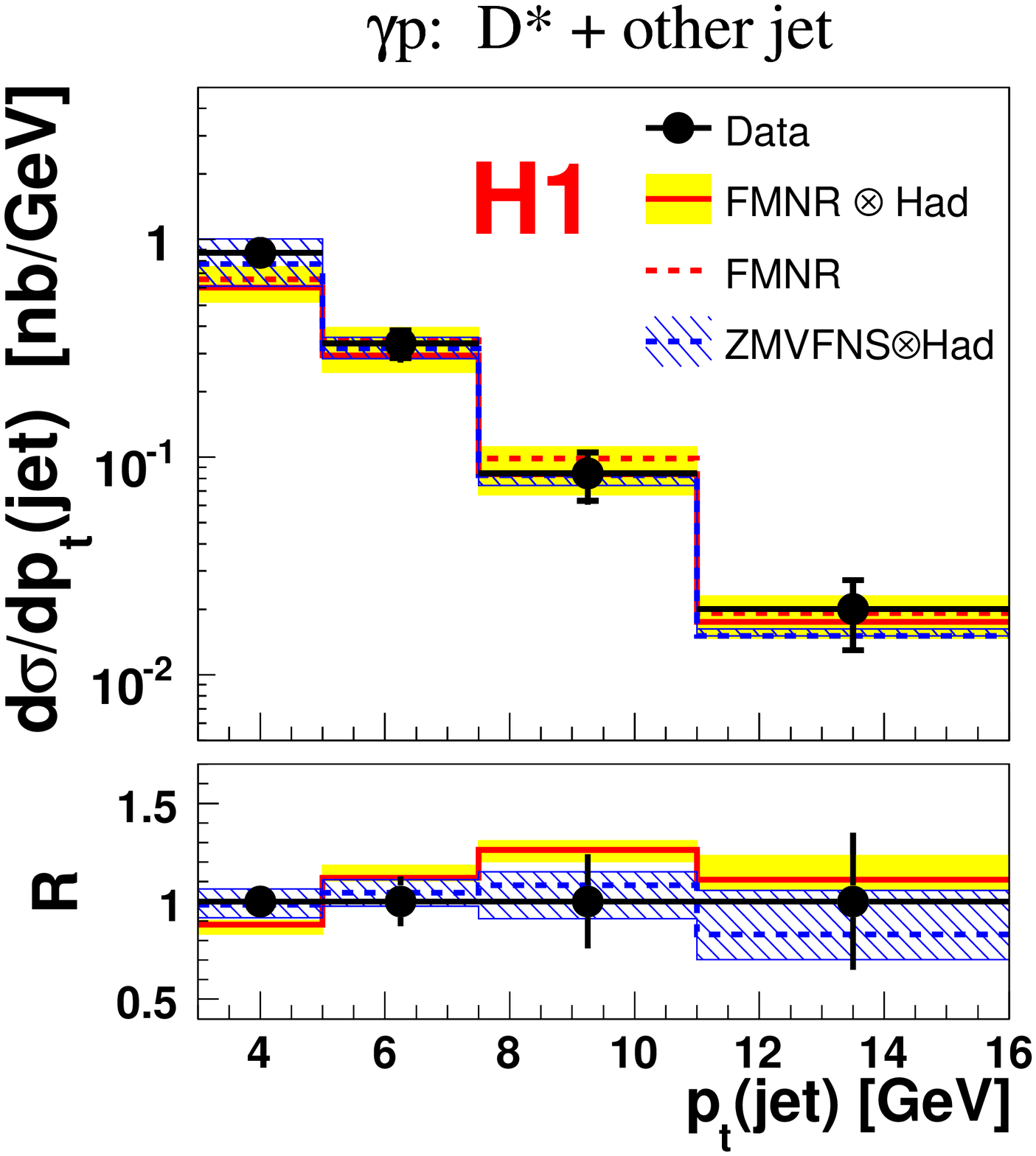}
    \begin{picture}(0,0)
      \put(-0.85,0.22){\bfseries c)}
      \put(-0.36,0.22){\bfseries d)}
    \end{picture}
    \Mycaption{\dstarpotherj\ cross sections as a function of the transverse momenta of the
      \dstar\ and the jet
      compared with the predictions of PYTHIA and CASCADE on the left and of the
      next-to-leading order calculations  FMNR and ZMVFNS on the right. 
      Here and in the following figures the central FMNR prediction is shown
      before and after applying the hadronisation corrections for the jet.
       }
    \label{fig:xsecDsJetPt}
  \end{center}
\end{figure}

\begin{figure}[htbp]
  \begin{center}
    \includegraphics[width=0.395\textwidth]{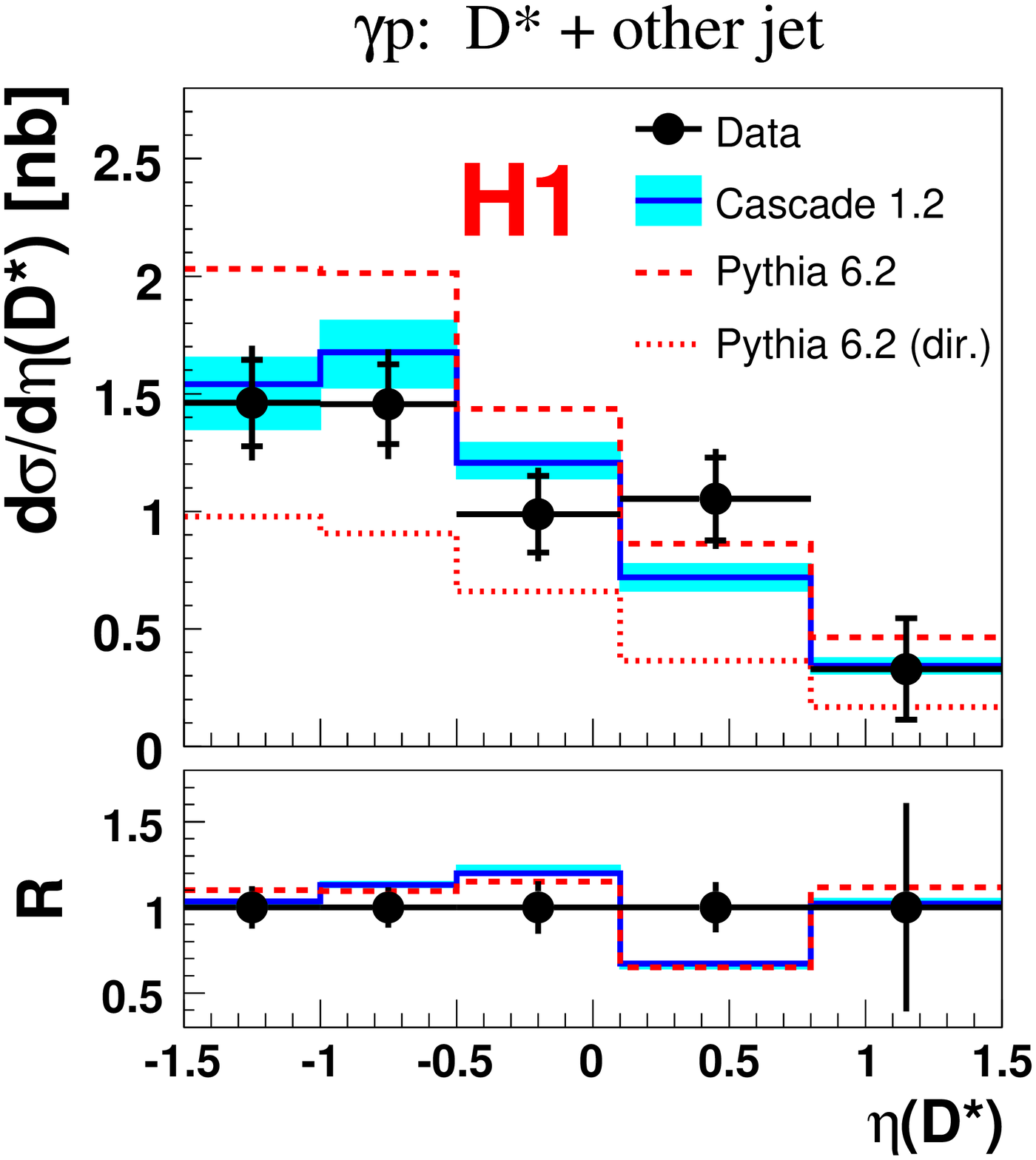} \hspace{0.03\textwidth}
    \includegraphics[width=0.395\textwidth]{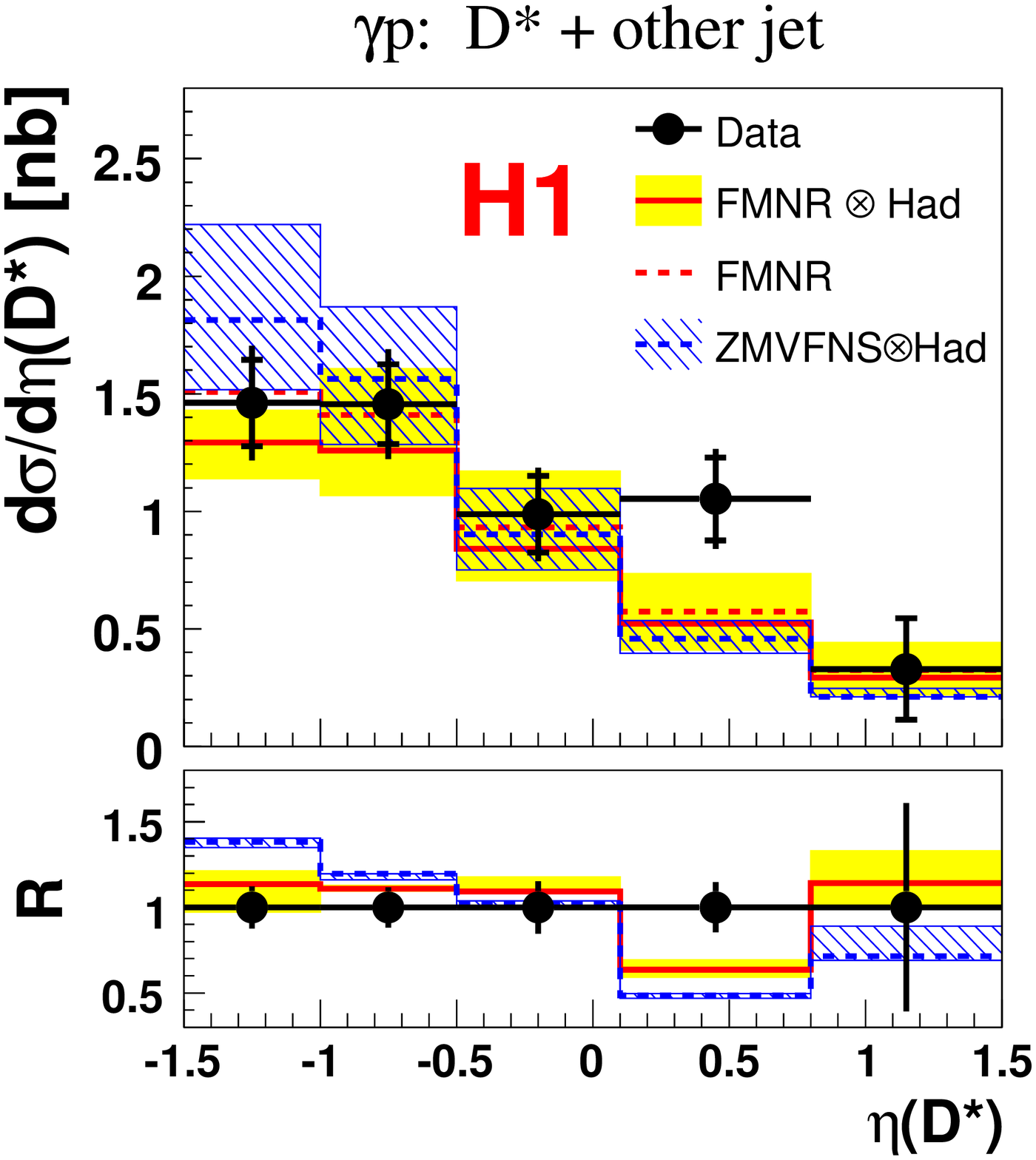}
    \setlength{\unitlength}{\textwidth}
    \begin{picture}(0,0)
      \put(-0.745,0.19){\bfseries a)}
      \put(-0.30,0.19){\bfseries b)}
    \end{picture}
    \includegraphics[width=0.395\textwidth]{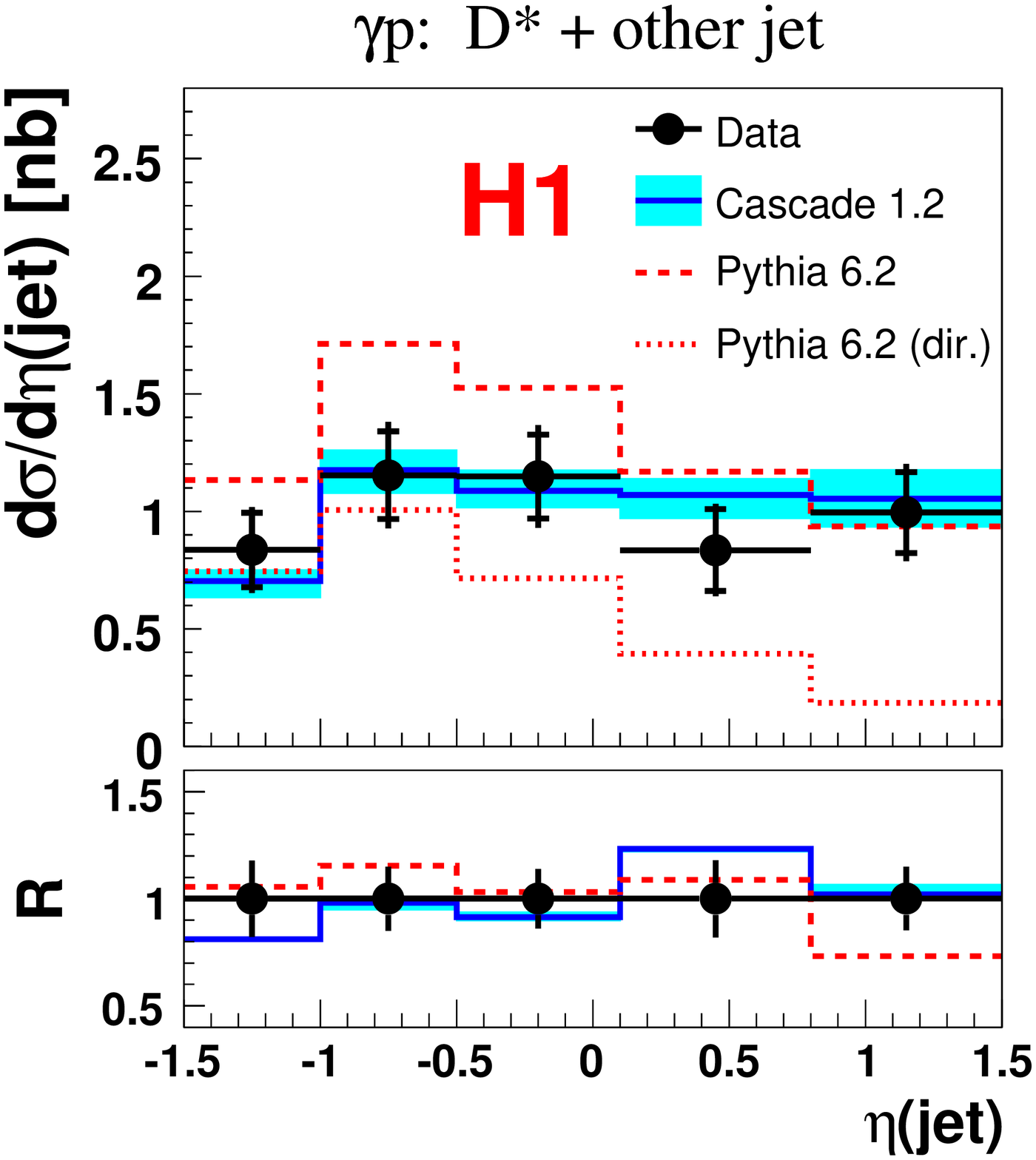} \hspace{0.03\textwidth}
    \includegraphics[width=0.395\textwidth]{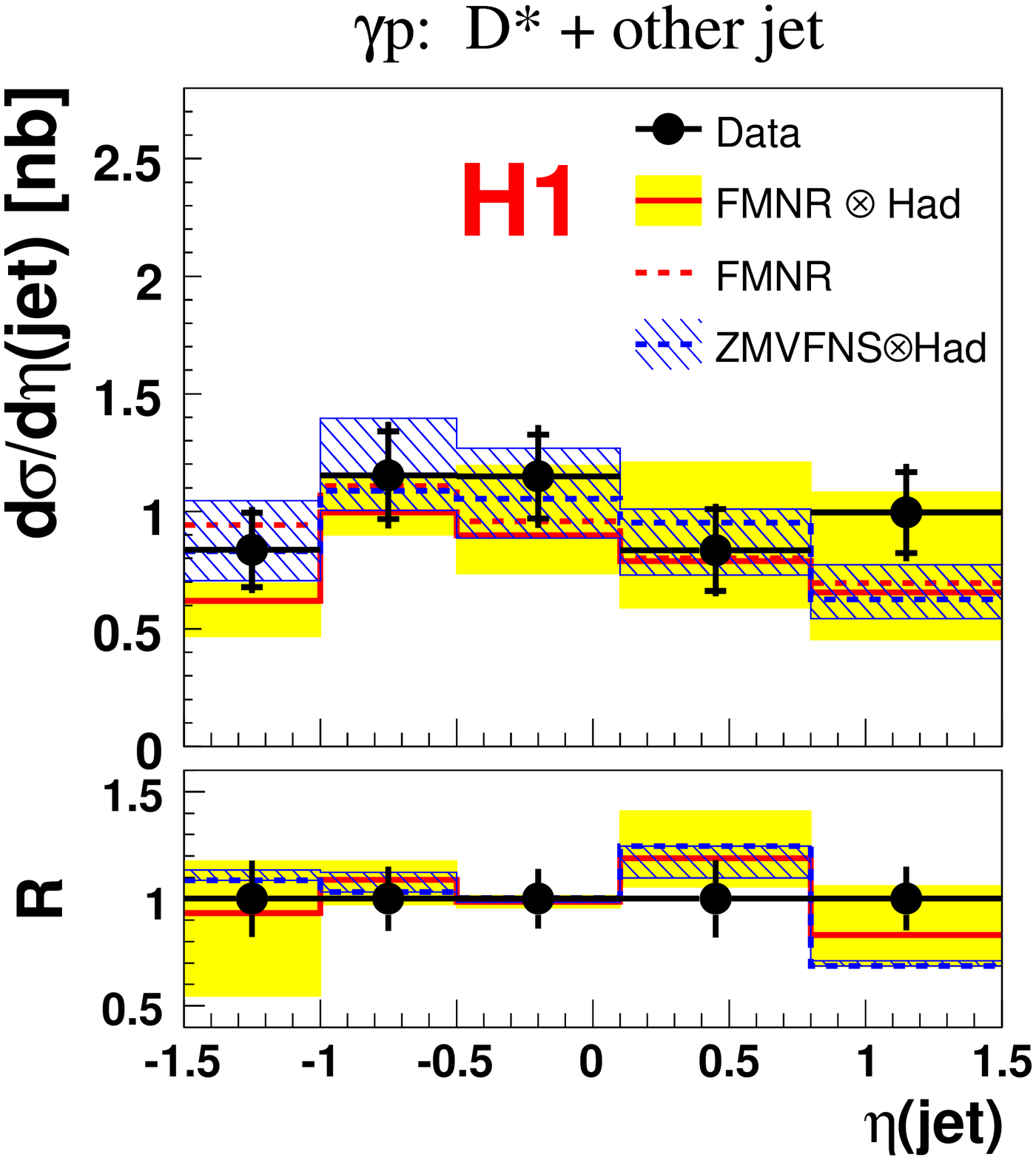}
    \begin{picture}(0,0)
      \put(-0.745,0.37){\bfseries c)}
      \put(-0.30,0.37){\bfseries d)}
    \end{picture}
    \includegraphics[width=0.395\textwidth]{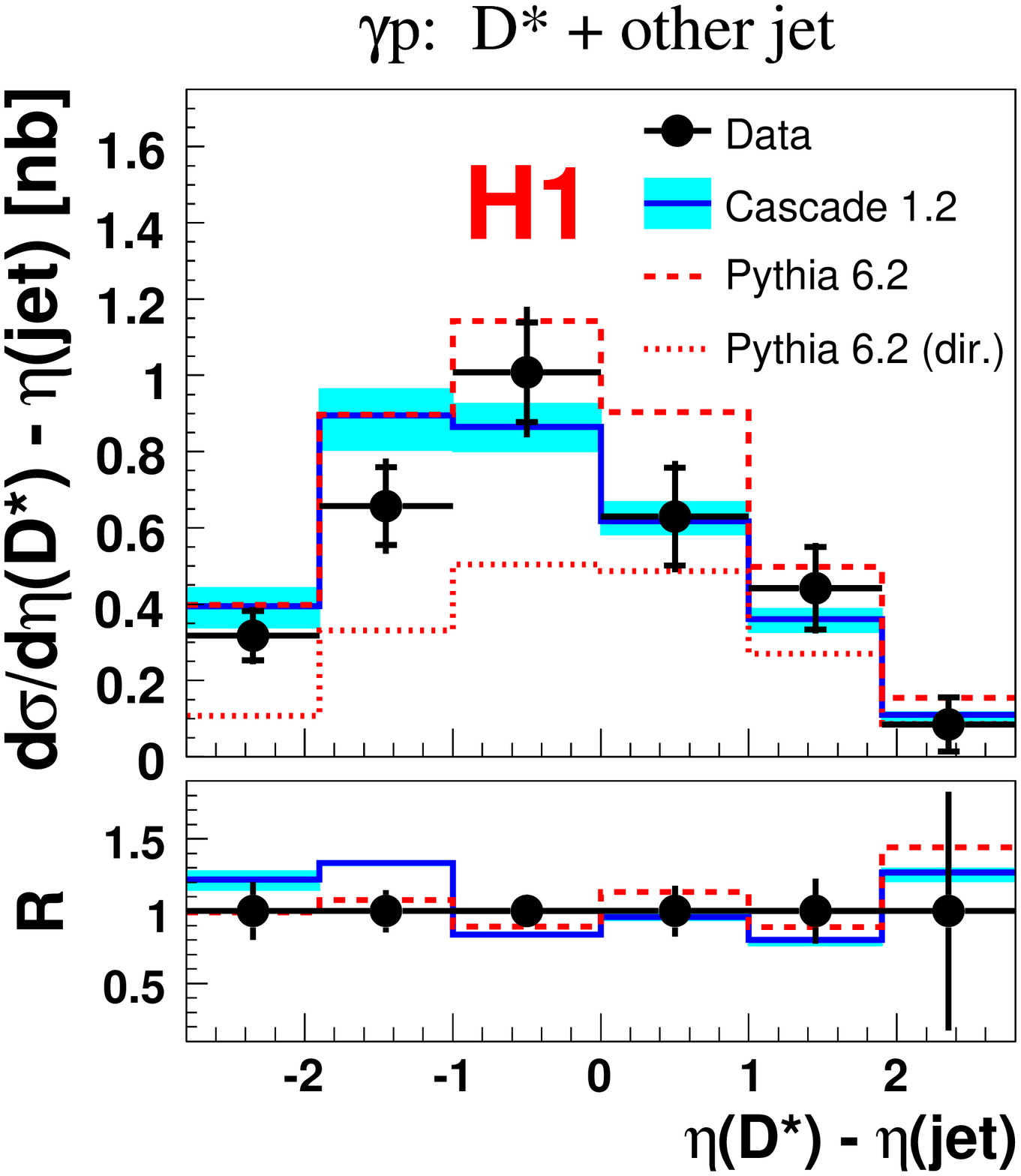} \hspace{0.03\textwidth}
    \includegraphics[width=0.395\textwidth]{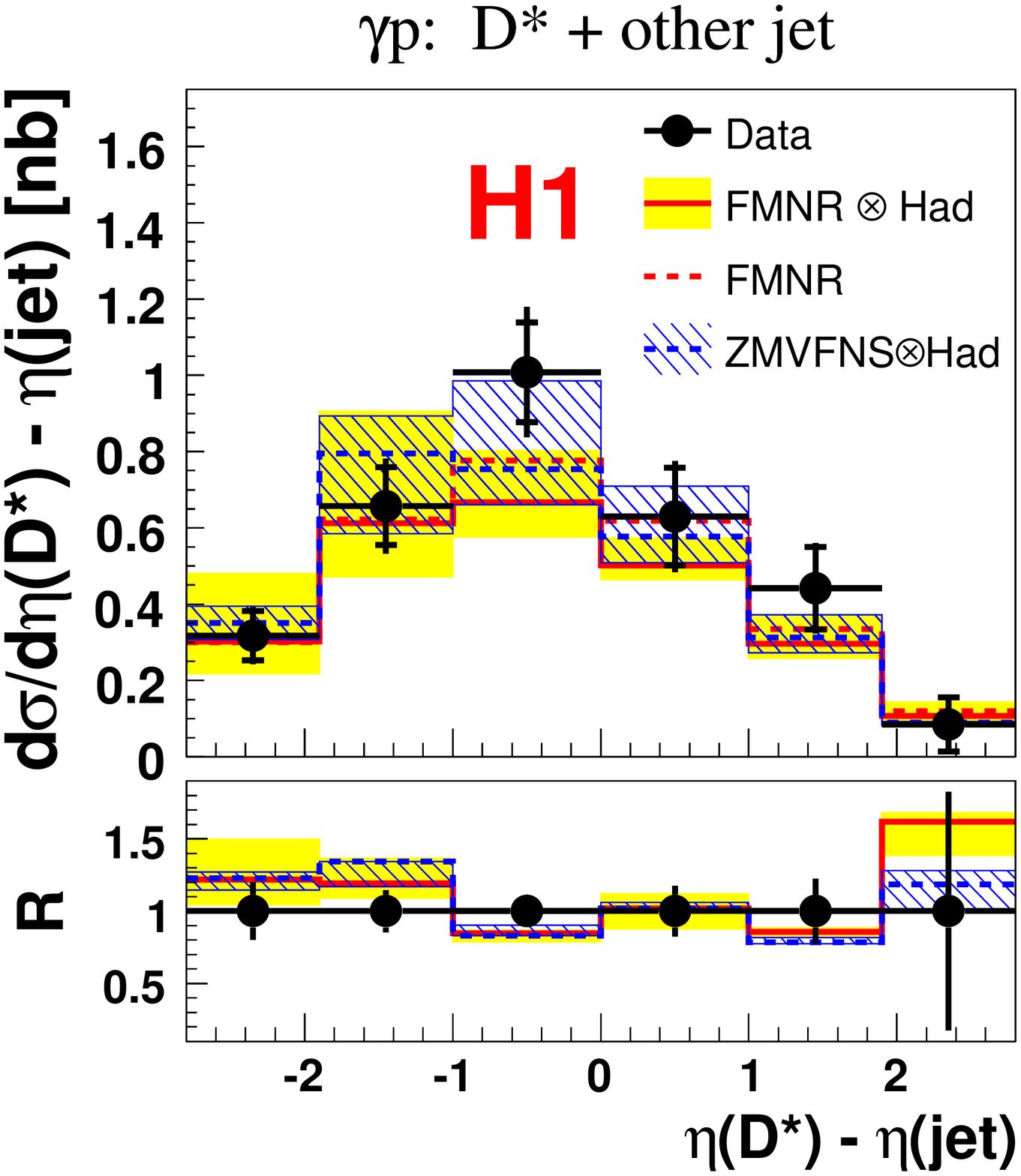}
    \begin{picture}(0,0)
      \put(-0.745,0.37){\bfseries e)}
      \put(-0.30,0.37){\bfseries f)}
    \end{picture}
    \Mycaption{\dstarpotherj\ cross sections as a function of the \etads\ and  \etaj\
      and of their difference, 
      compared with the predictions of PYTHIA and CASCADE (left) and of the
      NLO calculations  FMNR and ZMVFNS (right). 
      Here and in the following figures the direct photon contribution of PYTHIA 
      is shown separately and labelled ``Pythia 6.2 (dir.)''.
}
    \label{fig:xsecDsJetEta}
  \end{center}
\end{figure}

\begin{figure}[htbp]
  \begin{center}
    \includegraphics[width=0.45\textwidth]{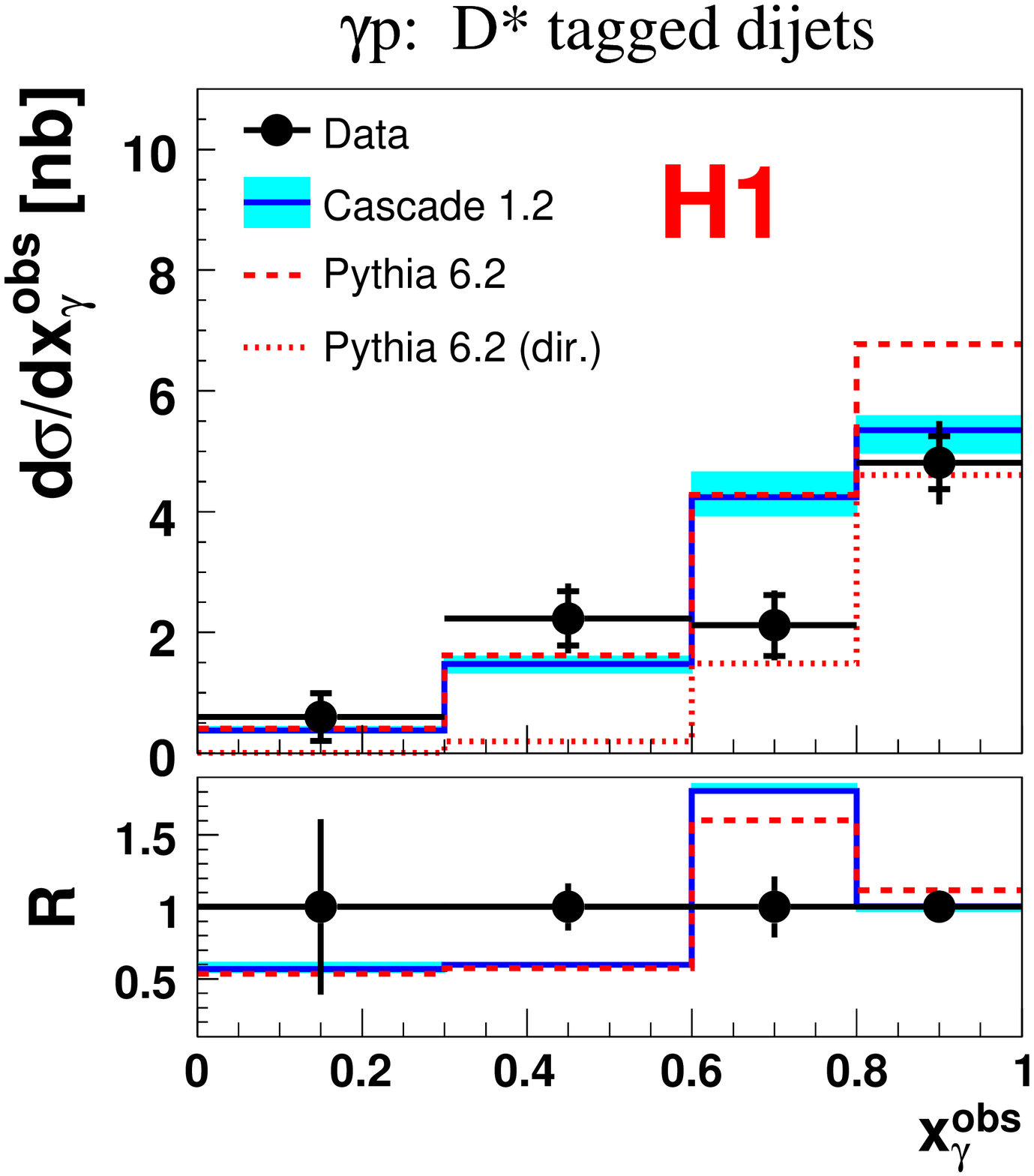} \hspace{0.03\textwidth}
    \includegraphics[width=0.45\textwidth]{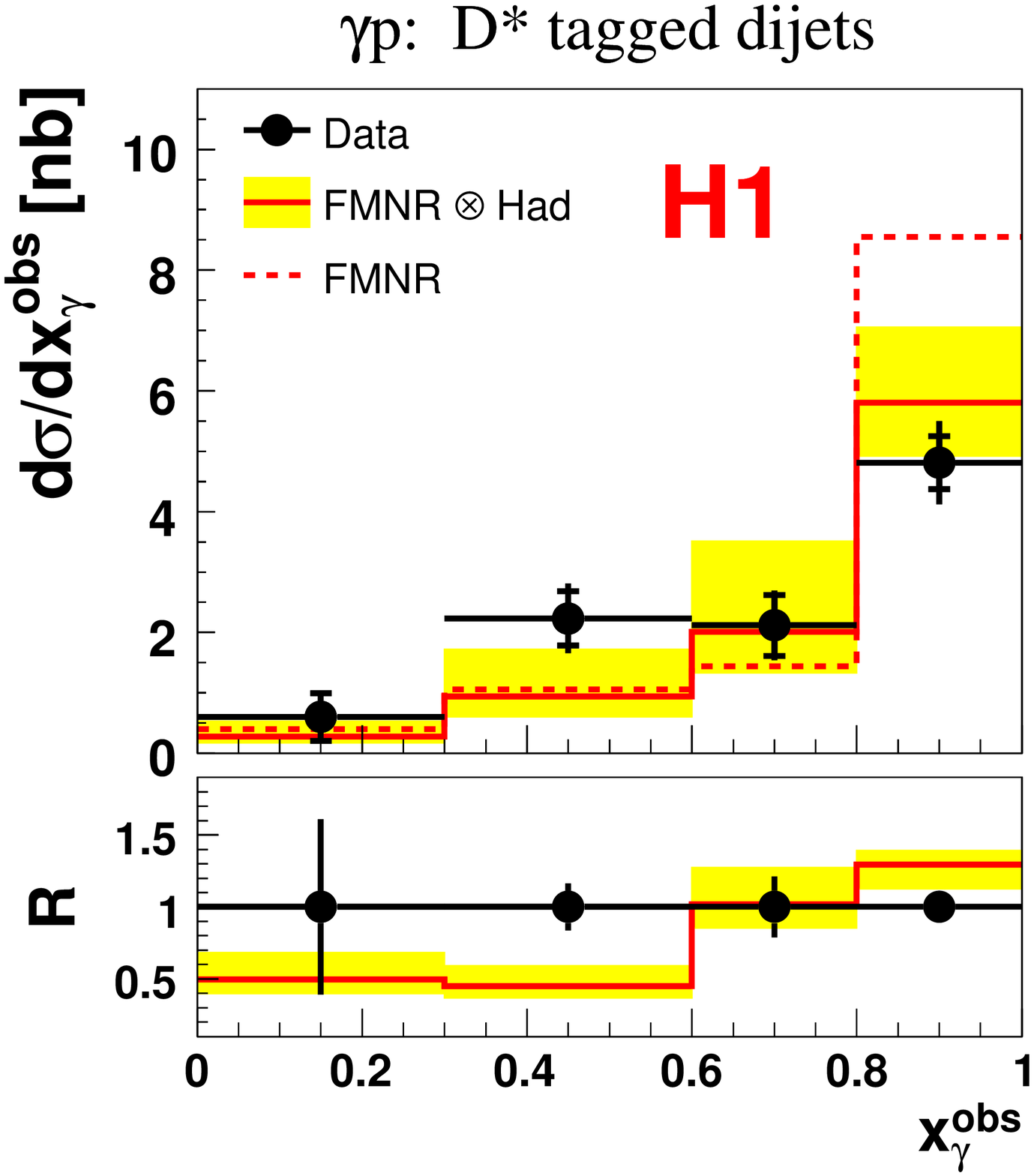}
    \setlength{\unitlength}{\textwidth}
    \begin{picture}(0,0)
      \put(-0.565,0.21){\bfseries a)}
      \put(-0.075,0.21){\bfseries b)}
    \end{picture}
    \Mycaption{\dstardj\ cross sections as a function of \xgjj\
      compared with the predictions of PYTHIA and CASCADE on the left and of
      the next-to-leading order calculation FMNR on the right.
       }
    \label{fig:xsecDsJetxGam}
  \end{center}
\end{figure}

\begin{figure}[htbp]
  \begin{center}
    \includegraphics[width=0.45\textwidth]{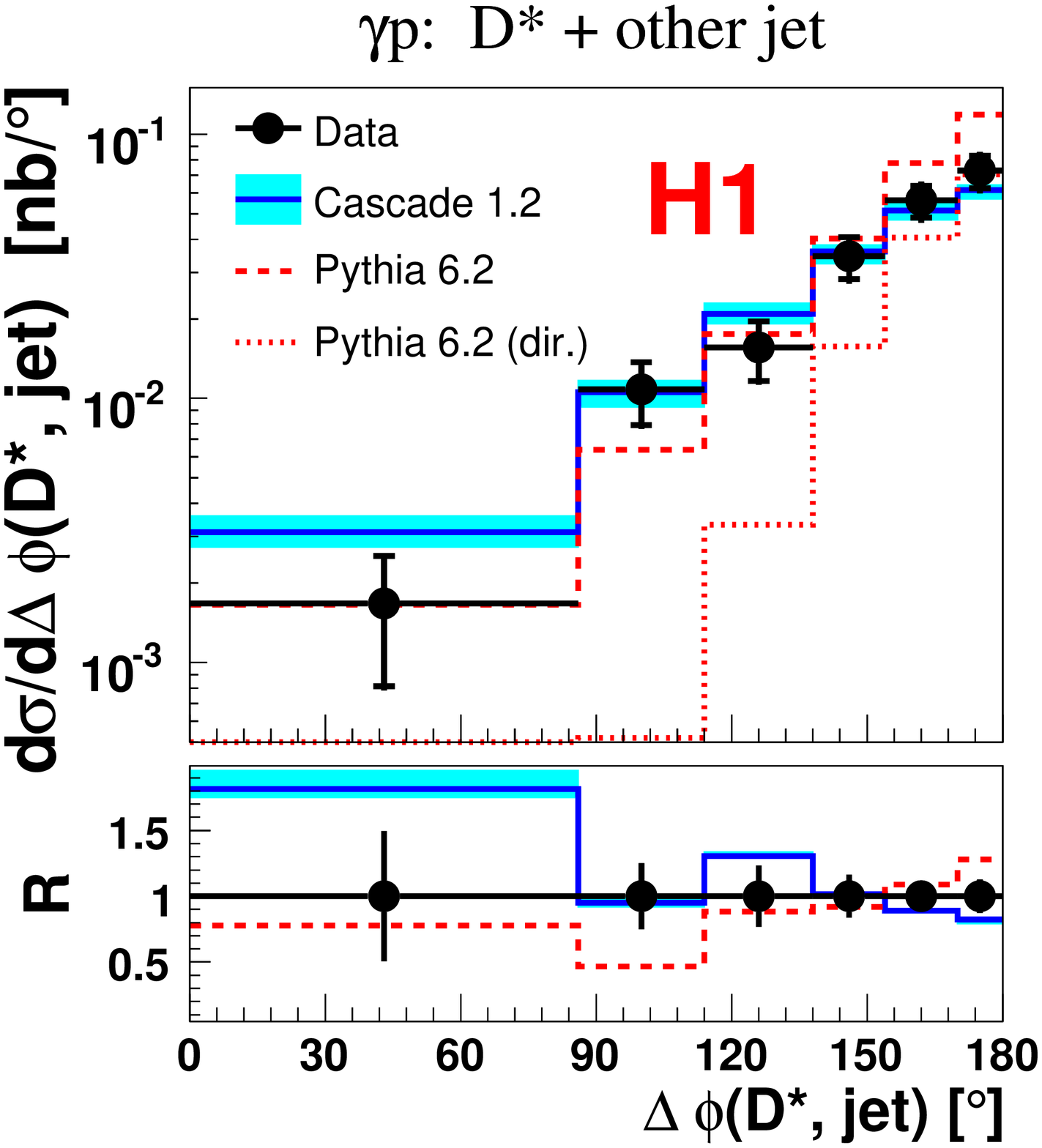} \hspace{0.03\textwidth}
    \includegraphics[width=0.45\textwidth]{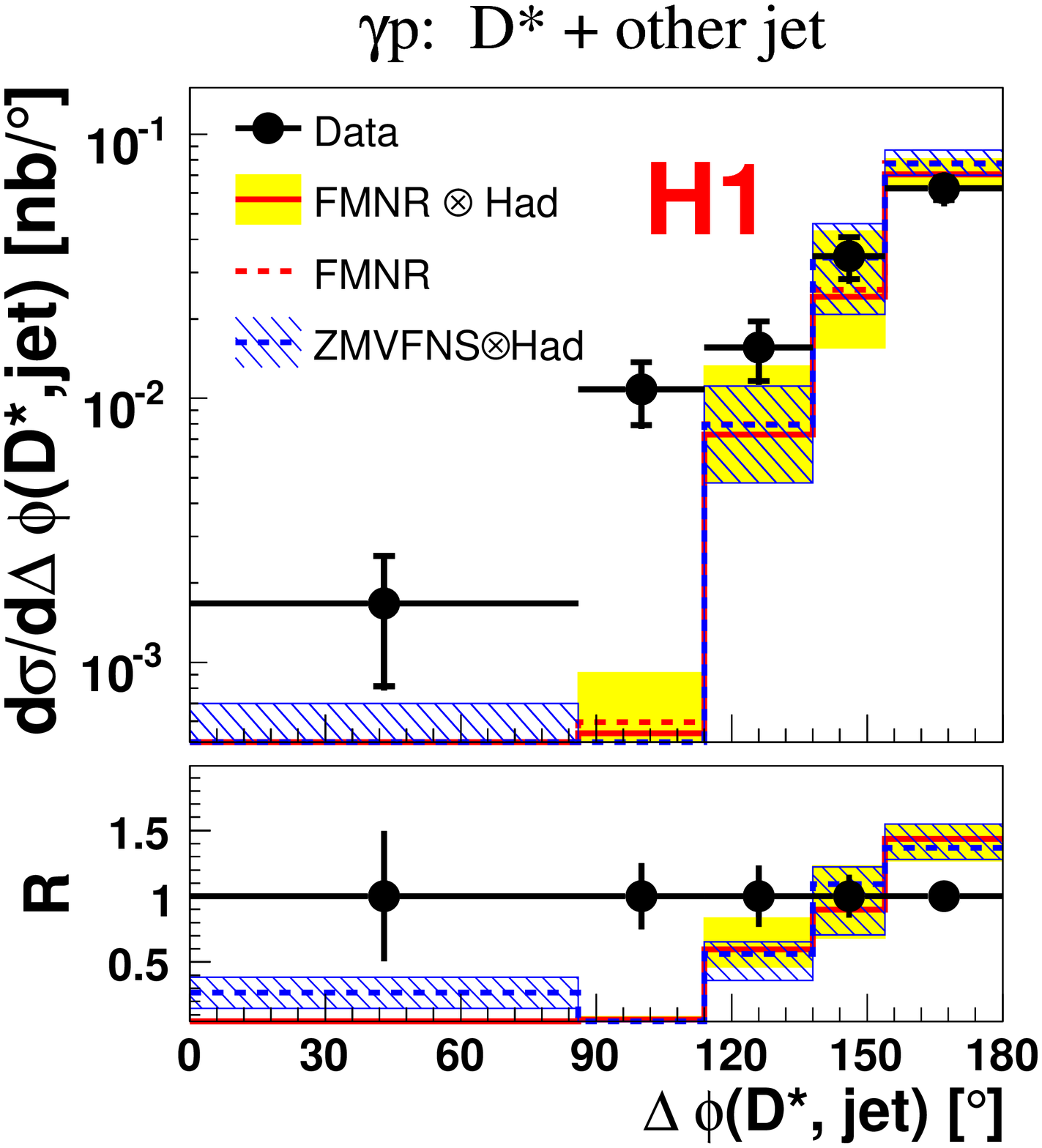}
    \setlength{\unitlength}{\textwidth}
    \begin{picture}(0,0)
      \put(-0.565,0.21){\bfseries a)}
      \put(-0.075,0.21){\bfseries b)}
    \end{picture}
    \Mycaption{\dstarpj\ cross sections as function of  \dphidsj\
      compared with the predictions of PYTHIA and CASCADE on the left and of the
      next-to-leading order calculations  
      FMNR and ZMVFNS on the right.
      Due to divergencies in the  NLO calculations  the last two bins of (a) are merged in (b).
       }
    \label{fig:xsecDsdeltaphi}
  \end{center}
\end{figure}

The bin averaged differential cross section for the \dstarpotherj\  and \dstardj\ are 
listed in Tabs.~\ref{tab:diffXsec2D}-\ref{tab:diffXsecXgammaDphi} and shown in Figs.~\ref{fig:xsecDsJetPt}-\ref{fig:xsecDsdeltaphi}.
The cross sections for the \dstarpotherj\ selection are shown in
Fig.~\ref{fig:xsecDsJetPt} as functions of \ptds\ and \ptj\
together with theoretical predictions.
The calculations overestimate the cross section at large \ptds , while
the cross
section as a function of \ptj\ is well described by the NLO calculations. 
\cascade\  overestimates the cross sections at large \ptj .

Cross sections as a function of \etads\ and \etaj\ 
are shown in Fig.~\ref{fig:xsecDsJetEta}a-d.
They differ noticeably: The
$\etads$ distribution falls steeply with increasing values of $\eta$ 
(similar to the inclusive case), whereas $\etaj$ is almost flat.
For direct photon-gluon fusion processes ($\gamma g \to c\bar{c}$), as shown
separately for \pythia , the shapes of the cross sections as a function of
 $\etads$ and $\etaj$
are found to be similar (dotted lines in Fig.~\ref{fig:xsecDsJetEta}a and c),
indicating that the difference is not caused by the slightly
different kinematic cuts for the \dstar\ and the jet.
The observed difference in shape between the \etads\ and \etaj\ distributions
indicates the presence of hard non-charm partons in the
forward region. In fact, the dominant mechanism for non-charm jets, as predicted
by the calculations, is hard gluon radiation off the gluon from the proton,
either calculated as an NLO correction (Fig.~\ref{fig:processes}b), 
as a (LO) charm excitation (Fig.~\ref{fig:processes}c) or by using an 
unintegrated gluon density. 
All calculations include this diagram and can describe the observed shapes
of \etads\ and \etaj\ in general. 

The \dstarpotherj\ measurement allows the investigation of correlations between the
\dstar\  and the jet. 
The cross section as a function of \detadsj\ is shown in
Fig.~\ref{fig:xsecDsJetEta}e and f.  
The distribution is not symmetric but the
jet is on average more forward than the \dstar.
The cross section is reasonably well described in shape and magnitude by all QCD calculations.

In collinear factorisation at LO a resolved photon process is characterised by
$x_\gamma < 1$, where $x_\gamma$ is the fraction of the longitudinal photon momentum 
entering the hard scattering process.
In the \dstardj\ sample, $x_\gamma$ is experimentally approximated by
\begin{equation}
  \xgjj  = \frac{\sum_{i=1}^{jet1 + jet2} (E -p_z)_i}{\sum_{j=1}^{all} (E-p_z)_j}.
\end{equation}
The sum in the numerator includes the particles in the two selected jets,
whereas the sum in the denominator contains all reconstructed
particles of the hadronic final state.
At NLO \xgjj\ is 
sensitive to ${\cal O}(\as^2)$ contributions. In the
\kt-factorisation approach  the \xgjj\  observable is sensitive to the contribution from 
gluon emission in the initial state.

The \dstardj\ cross section as
a function of \xgjj\ is shown in Fig.~\ref{fig:xsecDsJetxGam}.
The large cross section at small \xgjj\ shows that 
processes beyond direct photon-gluon fusion 
are needed to describe the data in the collinear approach
(as can be seen by comparison with the prediction from
\pythia\ (\emph{dir.}) in Fig.~\ref{fig:xsecDsJetxGam}a.). Both \pythia\ and 
\cascade\ give a poor description of the \xgjj\ distribution. 
All predictions underestimate the region of low $\xgjj < 0.6$ which can be 
clearly seen in the normalised shape $R$.
The hadronisation corrections applied to the FMNR calculation are
large.

 The correlation in the transverse plane is experimentally accessed by the 
difference in the azimuthal angle \dphidsj\ between the \dstar\ and the 
jet in the \dstarpotherj\ sample. 
The cross section as a function of 
\dphidsj\ is shown in Fig.~\ref{fig:xsecDsdeltaphi}. 
Only $\sim$ 25\% 
of the measured cross section originates from a back-to-back configuration in the
transverse plane with $\dphidsj > 170\grad$. Such a
configuration is expected for the process
$\gamma g \to c\bar{c}$ in the collinear approximation. 
The large fraction of events where the \dstar\ and the jet are not back-to-back
can only be described by models which include significant contributions 
from higher order QCD radiation.

In \pythia\ higher order contributions are simulated by leading log 
parton showers and the charm excitation process.
In \cascade\  higher order contributions lead to a significant \kt\ of the
gluon entering the hard subprocess as simulated by the unintegrated gluon density.
The large deviations from the back-to-back topology are well described by both 
\pythia\ and \cascade . 
The back-to-back configuration is overestimated by \pythia ,  while 
with the given parameterisation of the unintegrated gluon density \cascade\
tends to underestimate the 
back-to-back region and overestimates the small $\dphidsj$ configuration which
corresponds to large \kt. It is interesting to note that \cascade\ 
also predicts significantly harder $p_t$ spectra than that observed in the data
for the {\it inclusive \dstar\ } cross section in Fig.~\ref{fig:xsecIncl_pteta}a as well as for 
the \dstar\ and jet cross sections in Fig.~\ref{fig:xsecDsJetPt}a and c for the 
 \dstarpotherj\
sample. These quantities are also sensitive to the \kt\ of the gluon
which suggests that
the current unintegrated gluon density is too hard in \kt.
In the NLO calculations $\dphidsj \neq 180\grad$ comes entirely from real gluon emission
($\gamma g \to c \bar{c}g$) and from processes initiated by light quarks from the proton ($\gamma q \to c \bar{c}q$).
The NLO calculations, although ${\cal O}(\alpha_s^2)$,
include only the lowest order contribution to this region.
The NLO calculations are in reasonable agreement with the measurement in the region
of $\dphidsj > 120\grad$, but show large discrepancies for $\dphidsj < 120\grad$,
suggesting the presence of  higher order 
contributions. A similar behaviour was found in the
measurement of \dstar\ dijet data by the ZEUS collaboration~\protect\cite{ZEUS:dstarGammapJetCorr05}
where significantly higher transverse momenta
of the jets were required.

\section{Conclusions}
The photoproduction of \dstar\ mesons is investigated
with the H1 detector at HERA 
using a data sample five times larger than 
in a previous publication.
Differential cross sections are determined for events with a 
\dstar\ meson ({\it inclusive \dstar}) 
and for events with a \dstar\ meson and  jets 
(\dstarpotherj\ and  \dstardj ).
Jets are selected with transverse momenta down to 3~G\eV.
The results are compared with 
QCD calculations based on different approaches in leading and next-to-leading order pQCD
employing either  collinear or \kt -factorisation.

The precision of the {\it inclusive \dstar\ }  cross section
measurements presented here is much higher than the
accuracy of the NLO calculations.
The comparison of normalised cross sections, for which these theoretical
uncertainties are significantly reduced, reveals sizable differences between
data and theoretical predictions for variables sensitive to the phase
space distribution of the outgoing charm quark. 
However, a good agreement is found for the \Wgp\ distribution for which
each bin receives contributions throughout the available phase space of the produced 
\dstarpm\ meson.  

In the  \dstarpotherj\  analysis 
a jet is required in addition
which does not contain the \dstar\ meson.
In general the predictions agree better with the measurements
than is the case for the inclusive \dstar\ analysis.
The cross section as a function of \etaj\ is almost flat
in contrast to the cross section as a function of \etads\  which   
falls towards increasing \etads.
This indicates the presence of jets originating from non-charmed
partons, most likely from gluons.
All calculations are able to reproduce this feature of the data.
Correlations between the \dstar\ and the jet in the transverse
plane are investigated by measurements
of the difference in azimuthal angle \dphidsj.
A large fraction of the produced \dstarpj\ combinations deviates from 
a back-to-back configuration, indicating the importance of higher order contributions. 
The large tails of the \dphidsj\ distribution can be reproduced reasonably well by 
LO calculations in the collinear factorisation ansatz 
which include parton showers and charm excitation processes and  by
using unintegrated gluon densities in the \kt -factorisation ansatz. 
However, there are differences in shape observed between the data and these
predictions. Especially for \cascade\  the
\dphidsj\ distribution, together with the distributions in \ptds\ and \ptj,
suggests 
that the unintegrated gluon density used for this analysis overestimates
the high \kt\ region.
The available NLO calculations underestimate significantly the 
observed cross sections in the region
$\dphidsj < 120  \grad\ $.

In the {\it \dstardj\ } analysis two jets are required in addition
to the \dstar\ meson and the observable \xgjj\ is studied.
All calculations underestimate the cross section in the region of low \xgjj $<0.6$
where resolved photon processes or other higher order contributions
are expected to be enhanced.


\section*{Acknowledgements}
We are grateful to the HERA machine group whose outstanding
efforts have made this experiment possible. 
We thank
the engineers and technicians for their work in constructing and
maintaining the H1 detector, our funding agencies for 
financial support, the
DESY technical staff for continual assistance
and the DESY directorate for support and for the
hospitality which they extend to the non DESY 
members of the collaboration.
We wish to thank G.~Heinrich and B.~Kniehl for providing us with
the ZMVFNS calculations and G. Kramer and H. Spiesberger for the 
GMVFNS calculations.

\providecommand{\etal}{et al.\xspace}
\providecommand{\href}[2]{#2}
\providecommand{\coll}{Coll.}
\catcode`\@=11
\def\@bibitem#1{%
\ifmc@bstsupport
  \mc@iftail{#1}%
    {;\newline\ignorespaces}%
    {\ifmc@first\else.\fi\orig@bibitem{#1}}
  \mc@firstfalse
\else
  \mc@iftail{#1}%
    {\ignorespaces}%
    {\orig@bibitem{#1}}%
\fi}%
\catcode`\@=12
{\raggedright
}


\clearpage

\begin{table}[tb]
  \begin{center}
    \begin{tabular}{|cc|rrr|}
      \hline
      \multicolumn{5}{|c|}{\bf \boldmath Inclusive \dstar\ Cross Sections} \\
      \hline
      \hline
      \multicolumn{2}{|c|}{\ptds\ range} & \dsdx{\ptds} & stat. & sys. \\  
      \multicolumn{2}{|c|}{[G\eV]} & \multicolumn{3}{c|}{[nb/G\eV]}\\ 
      \hline
      2.0  &  2.5  &  3.37  &  0.63  &  0.36  \\
      2.5  &  3.0  &  3.01  &  0.37  &  0.32  \\
      3.0  &  3.5  &  2.22  &  0.23  &  0.24  \\
      3.5  &  4.25 &  0.97  &  0.11  &  0.10  \\
      4.25 &  5.0  &  0.577 &  0.071 &  0.063 \\
      5.0  &  6.0  &  0.250 &  0.038 &  0.027 \\
      6.0  &  8.5  &  0.063 &  0.013 &  0.0068  \\
      8.5  &  12.0 &  0.0133&  0.0050&  0.0016  \\
      \hline
      \hline
      \multicolumn{2}{|c|}{\etads\ range} & \dsdx{\etads} & stat. & sys. \\  
      \multicolumn{2}{|c|}{            }  & \multicolumn{3}{c|}{[nb]}\\ 
      \hline
      -1.5  &  -1.0  &  3.54 &  0.30 &  0.38 \\
      -1.0  &  -0.5  &  2.93 &  0.25 &  0.31 \\
      -0.5  &  0.0   &  1.71 &  0.26 &  0.19 \\
      0.0   &  0.5   &  2.16 &  0.29 &  0.23 \\
      0.5   &  1.0   &  1.58 &  0.31 &  0.18 \\
      1.0   &  1.5   &  0.74 &  0.41 &  0.08 \\
      \hline
      \hline
      \multicolumn{2}{|c|}{\zDs\ range} & \dsdx{\zDs} & stat. & sys. \\  
      \multicolumn{2}{|c|}{            }  & \multicolumn{3}{c|}{[nb]}\\ 
      \hline
      0  &  0.1  &  10.8 &  2.5 &  1.2  \\
      0.1  &  0.2  &  17.8 &  1.9  &  2.0  \\
      0.2  &  0.3  &  9.9  &  1.2 &  1.1  \\
      0.3  &  0.45  &  8.66 &  0.76 &  0.93 \\
      0.45  &  0.6  &  6.00 &  0.58 &  0.64 \\
      0.6  &  0.8  &  1.48  &  0.24 &  0.19 \\
      \hline
      \hline
      \multicolumn{2}{|c|}{\Wgp\ range} & \dsdx{\Wgp} & stat. & sys. \\  
      \multicolumn{2}{|c|}{ [G\eV]    } & \multicolumn{3}{c|}{[nb/G\eV]}\\ 
      \hline
      172  &  192  &  0.112  &  0.012  &  0.017 \\
      192  &  212  &  0.0784 &  0.0077 &  0.0077  \\
      212  &  232  &  0.0601 &  0.0071 &  0.0058 \\
      232  &  256  &  0.0548 &  0.0078 &  0.0057 \\
      \hline
    \end{tabular}
    \Mycaption{Bin averaged differential cross sections for 
    {\it inclusive \dstar\ } production in bins of \ptds, \etads,
      \zDs\ and \Wgp\ with their statistical and systematical uncertainties.}
    \label{tab:diffXsec1D}
  \end{center}
\end{table}

\begin{table}[tb]
  \begin{center}
    \begin{tabular}{|cc|rrr|}
      \hline
      \multicolumn{5}{|c|}{\bf \boldmath Inclusive \dstar\ Cross Sections} \\
      \hline
      \hline
      \multicolumn{5}{|c|}{$2.0 \le \ptds < 3.0$ G\eV} \\
      \hline
      \multicolumn{2}{|c|}{\etads\ range} & \dsdx{\etads} [nb] & stat. & sys. \\  
      \hline
      -1.5  &  -0.9  &  2.14  &  0.25  &  0.23 \\
      -0.9  &  0.2   &  1.14  &  0.17  &  0.12 \\
      0.2   &  1.5   &  0.63  &  0.20  &  0.07  \\
      \hline
      \hline
      \multicolumn{5}{|c|}{$3.0 \le \ptds < 4.5$ G\eV} \\
      \hline
      \multicolumn{2}{|c|}{\etads\ range} & \dsdx{\etads} [nb] & stat. & sys. \\  
      \hline
      -1.5  &  -0.9  &  1.02  &  0.11  &  0.11  \\
      -0.9  &  0.2   &  0.714 &  0.073 &  0.077  \\
      0.2   &  1.5   &  0.495 &  0.082 &  0.057  \\
      \hline
      \hline
      \multicolumn{5}{|c|}{$4.5 \le \ptds < 8.5$ G\eV} \\
      \hline
      \multicolumn{2}{|c|}{\etads\ range} & \dsdx{\etads} [nb] & stat. & sys. \\  
      \hline
      -1.5  &  -0.9  &  0.263  &  0.043  &  0.028  \\
      -0.9  &  0.2   &  0.240  &  0.037  &  0.027  \\
      0.2   &  1.5   &  0.190  &  0.034  &  0.020  \\
      \hline
    \end{tabular}
    \Mycaption{Bin averaged differential cross sections for 
    {\it inclusive \dstar\ } production in bins of \etads\ for
      three ranges in \ptds\ with their statistical and systematical
      uncertainties.}
    \label{tab:diffXsec1Ddd}
  \end{center}
\end{table}

\begin{table}[tb]
  \begin{center}
    \begin{tabular}{|cc|rrr||c|}
      \hline
      \multicolumn{5}{|c||}{\bf \boldmath \dstarpotherJ\ Cross Sections} & \\
      \hline
      \hline
      \multicolumn{2}{|c|}{\ptds\ range} & \dsdx{\ptds} & stat. & sys. & $f_{had}$ \\  
      \multicolumn{2}{|c|}{[G\eV]} & \multicolumn{3}{c||}{[nb/G\eV]} & \\ 
      \hline
      2.0  &  3.5  &  1.35  &  0.16  &  0.15   & 0.87 \\
      3.5  &  5.0  &  0.407 &  0.051 &  0.044  & 0.92 \\
      5.0  &  7.5  &  0.124 &  0.017 &  0.014  & 0.96 \\
      7.5  &  11.0 &  0.0116&  0.0048&  0.0013 & 0.97 \\
      \hline
      \hline
      \multicolumn{2}{|c|}{\ptj\ range} & \dsdx{\ptj} & stat. & sys. & $f_{had}$ \\  
      \multicolumn{2}{|c|}{[G\eV]} & \multicolumn{3}{c||}{[nb/G\eV]} & \\ 
      \hline
      3.0  &  5.0  &  0.870  &  0.094  &  0.095  & 0.91 \\
      5.0  &  7.5  &  0.334  &  0.049  &  0.037  & 0.85 \\
      7.5  &  11.0 &  0.084  &  0.021  &  0.010  & 0.85 \\
      11.0 &  16.0 &  0.0201 &  0.0071 &  0.0022 & 0.91 \\
      \hline
      \hline
      \multicolumn{2}{|c|}{\etads\ range} & \dsdx{\etads} & stat. & sys. & $f_{had}$ \\  
      \multicolumn{2}{|c|}{} & \multicolumn{3}{c||}{[nb]} & \\ 
      \hline
      -1.5  &  -1.0 &  1.46  &  0.18 &  0.16 & 0.86 \\
      -1.0  &  -0.5 &  1.46  &  0.17 &  0.16 & 0.89 \\
      -0.5  &  0.1  &  0.99  &  0.16 &  0.11 & 0.90 \\
      0.1   &  0.8  &  1.05  &  0.18 &  0.12 & 0.91 \\
      0.8   &  1.5  &  0.33  &  0.22 &  0.04 & 0.90 \\
      \hline
      \hline
      \multicolumn{2}{|c|}{\etaj\ range} & \dsdx{\etaj} & stat. & sys. & $f_{had}$ \\  
      \multicolumn{2}{|c|}{} & \multicolumn{3}{c||}{[nb]} & \\ 
      \hline
      -1.5  &  -1.0  &  0.84  &  0.16  &  0.092 & 0.66 \\
      -1.0  &  -0.5  &  1.15  &  0.19  &  0.13  & 0.90 \\
      -0.5  &  0.1   &  1.15  &  0.18  &  0.13  & 0.94 \\
      0.1   &  0.8   &  0.84  &  0.17  &  0.093 & 0.99 \\
      0.8   &  1.5   &  0.99  &  0.17  &  0.11  & 0.94 \\
      \hline
      \hline
      \multicolumn{2}{|c|}{\detadsj\ range} & \dsdx{(\detadsj)} & stat. & sys. & $f_{had}$ \\  
      \multicolumn{2}{|c|}{} & \multicolumn{3}{c||}{[nb]} & \\ 
      \hline
      -2.8  &  -1.9  &  0.318 &  0.065 &  0.037 & 1.01 \\
      -1.9  &  -1.0  &  0.66  &  0.10  &  0.07  & 0.98 \\
      -1.0  &  0.0   &  1.01  &  0.13  &  0.11  & 0.86 \\
      0.0   &  1.0   &  0.63  &  0.13  &  0.07  & 0.81 \\
      1.0   &  1.9   &  0.44  &  0.11  &  0.05  & 0.88 \\
      1.9   &  2.8   &  0.084 &  0.071 &  0.012 & 0.89 \\
      \hline
    \end{tabular}
    \Mycaption{Bin averaged differential cross sections for 
    the \dstarpotherj\ sample in bins of \ptds, \ptj, \etads, \etaj\
      and \detadsj\ with their statistical and systematical uncertainties.
      The last column shows the hadronisation correction factors applied
      to the NLO calculations.}
    \label{tab:diffXsec2D}
  \end{center}
\end{table}

\begin{table}[tb]
  \begin{center}
    \begin{tabular}{|cc|rrr||c|}
      \hline
      \multicolumn{5}{|c||}{\bf \boldmath \dstarDj\ Cross Sections} & \\
      \hline
      \hline
      \multicolumn{2}{|c|}{\xgjj\ range} & \dsdx{\xgjj} & stat. & sys. & $f_{had}$ \\  
      \multicolumn{2}{|c|}{            } & \multicolumn{3}{c||}{[nb]} & \\ 
      \hline
      0.0  &  0.3  &  0.60  &  0.39  &  0.12  & 0.69 \\
      0.3  &  0.6  &  2.23  &  0.45  &  0.36  & 0.88 \\
      0.6  &  0.8  &  2.12  &  0.51  &  0.27  & 1.40\\
      0.8  &  1.0  &  4.81  &  0.44  &  0.53  & 0.68 \\
      \hline
      \hline
      \multicolumn{5}{|c||}{\bf \boldmath \dstarpotherJ\ Cross Sections} & \\
      \hline
      \hline
      \multicolumn{2}{|c|}{\dphidsj\ range} & \dsdx{\dphidsj} & stat. & sys. & $f_{had}$ \\  
      \multicolumn{2}{|c|}{  [$\grad$]} & \multicolumn{3}{c||}{[nb/$\grad$]}&\\ 
      \hline
      0    &  86   &  0.00167 &  0.00086 &  0.00022 & 0.62 \\
      86   &  114  &  0.0108  &  0.0029  &  0.0012  & 0.91 \\
      114  &  138  &  0.0156  &  0.0040  &  0.0018  & 0.91 \\
      138  &  154  &  0.034   &  0.0062  &  0.0039  & 0.94 \\
      154  &  170  &  0.0561  &  0.0077  &  0.0061  & 0.90 \\
      170  &  180  &  0.073   &  0.010   &  0.008   & 0.91 \\
      154  &  180  &  0.0625  &  0.0061  &  0.0068  & 0.91 \\
      \hline
    \end{tabular}
    \Mycaption{Bin averaged differential cross sections for the \dstardj\ 
      sample in bins of \xgjj\
      and the \dstarpotherj\ cross section as a function of \dphidsj\  
      with their statistical and systematical uncertainties.
      The last column shows the hadronisation correction factors applied
      to the NLO calculations.}
    \label{tab:diffXsecXgammaDphi}
  \end{center}
\end{table}


\clearpage

\begin{mcbibliography}{10}

\bibitem{H1:DstarGluonDens99}
C.~Adloff {\it et al.}, [{H1} Collaboration],
\newblock Nucl. Phys.{} {\bf B545},~21~(1999),
  \href{http://www.arXiv.org/abs/hep-ex/9812023}{{ [hep-ex/9812023]}}\relax
\relax
\bibitem{ZEUS:dstarjetgammap99}
J.~Breitweg {\it et al.}, [{ZEUS} Collaboration],
\newblock Eur. Phys. J.{} {\bf C6},~67~(1999),
  \href{http://www.arXiv.org/abs/hep-ex/9807008}{{ [hep-ex/9807008]}}\relax
\relax
\bibitem{ZEUS:dijetDstar03}
S.~Chekanov {\it et al.}, [{ZEUS} Collaboration],
\newblock Phys. Lett.{} {\bf B565},~87~(2003),
  \href{http://www.arXiv.org/abs/hep-ex/0302025}{{ [hep-ex/0302025]}}\relax
\relax
\bibitem{ZEUS:dstarGammapJetCorr05}
S.~Chekanov {\it et al.}, [{ZEUS} Collaboration],
\newblock Nucl. Phys.{} {\bf B729},~492~(2005),
  \href{http://www.arXiv.org/abs/hep-ex/0507089}{{ [hep-ex/0507089]}}\relax
\relax
\bibitem{H1D*mu:05}
A.~Aktas {\it et al.}, [{H1} Collaboration],
\newblock Phys. Lett.{} {\bf B621},~56~(2005),
  \href{http://www.arXiv.org/abs/hep-ex/0503038}{{ [hep-ex/0503038]}}\relax
\relax
\bibitem{DGLAP:72}
V.~N. Gribov and L.~N. Lipatov,
\newblock Sov. J. Nucl. Phys.{} {\bf 15},~438 and 675~(1972)\relax
\relax
\bibitem{DGLAP:75}
L.~N. Lipatov,
\newblock Sov. J. Nucl. Phys.{} {\bf 20},~94~(1975)\relax
\relax
\bibitem{DGLAP77}
G.~Altarelli and G.~Parisi,
\newblock Nucl. Phys.{} {\bf B126},~298~(1977)\relax
\relax
\bibitem{DGLAP77_2}
Y.~L. Dokshitzer,
\newblock Sov. Phys. JETP{} {\bf 46},~641~(1977)\relax
\relax
\bibitem{GribLevRys:SemihardQcd84}
L.~V. Gribov, E.~M. Levin, and M.~G. Ryskin,
\newblock Phys. Rept.{} {\bf 100},~1~(1983)\relax
\relax
\bibitem{LevRys:HqSemiHard91}
E.~M. Levin, M.~G. Ryskin, Y.~M. Shabelski, and A.~G. Shuvaev,
\newblock Sov. J. Nucl. Phys.{} {\bf 53},~657~(1991)\relax
\relax
\bibitem{CatCiaf:HqSmallX91}
S.~Catani, M.~Ciafaloni, and F.~Hautmann,
\newblock Nucl. Phys.{} {\bf B366},~135~(1991)\relax
\relax
\bibitem{CollEll:HqInHighE91}
J.~C. Collins and R.~K. Ellis,
\newblock Nucl. Phys.{} {\bf B360},~3~(1991)\relax
\relax
\bibitem{H1:Detektor97}
I.~Abt {\it et al.}, [{H1} Collaboration],
\newblock Nucl. Instrum. Meth.{} {\bf A386},~310~(1997)\relax
\relax
\bibitem{H1:Detektor97TrackCalMuon}
I.~Abt {\it et al.}, [{H1} Collaboration],
\newblock Nucl. Instrum. Meth.{} {\bf A386},~348~(1997)\relax
\relax
\bibitem{CST2000}
D.~Pitzl {\it et al.},
\newblock Nucl. Instrum. Meth.{} {\bf A454},~334~(2000),
  \href{http://www.arXiv.org/abs/hep-ex/0002044}{{ [hep-ex/0002044]}}\relax
\relax
\bibitem{H1SPACAL97}
R.~D. Appuhn {\it et al.}, [{H1 SPACAL Group}],
\newblock Nucl. Instrum. Meth.{} {\bf A386},~397~(1997)\relax
\relax
\bibitem{PDG2004}
S.~Eidelman {\it et al.}, [{Particle Data Group}],
\newblock Phys. Lett.{} {\bf B592},~1~(2004)\relax
\relax
\bibitem{diss:geroflucke05}
G.~Flucke,
\newblock {\em Photoproduction of $D^*$ Mesons and $D^*$ Mesons Associated with
  Jets at HERA}.
\newblock Ph.D.\ Thesis, Universit\"at Hamburg, Report
  \mbox{DESY-THESIS-2005-006}, Mar.\ 2005.
\newblock Available at https://www-h1.desy.de/publications/theses\_list.html\relax
\bibitem{jetKT93}
S.~Ellis and D.~Soper,
\newblock Phys. Rev.{} {\bf D48},~3160~(1993),
  \href{http://www.arXiv.org/abs/hep-ph/9305266}{{ [hep-ph/9305266]}}\relax
\relax
\bibitem{Kt++:02}
J.~M. Butterworth, J.~P. Couchman, B.~E. Cox, and B.~M. Waugh,
\newblock Comput. Phys. Commun.{} {\bf 153},~85~(2003),
  \href{http://www.arXiv.org/abs/hep-ph/0210022}{{ [hep-ph/0210022]}}\relax
\relax
\bibitem{Artuso:2004pj}
M.~Artuso {\it et al.}, [{CLEO} Collaboration],
\newblock Phys. Rev.{} {\bf D70},~112001~(2004),
  \href{http://www.arXiv.org/abs/hep-ex/0402040}{{ [hep-ex/0402040]}}\relax
\relax
\bibitem{CollJungNeedUnintPdf:05}
J.~Collins and H.~Jung,
\newblock {\em \mbox{Need for Fully Unintegrated Parton Densities}}, 2005,
\newblock \href{http://www.arXiv.org/abs/hep-ph/0508280}{{
  [hep-ph/0508280]}}\relax
\relax
\bibitem{Frixione:diffHQ95}
S.~Frixione, P.~Nason, and G.~Ridolfi,
\newblock Nucl. Phys.{} {\bf B454},~3~(1995),
  \href{http://www.arXiv.org/abs/hep-ph/9506226}{{ [hep-ph/9506226]}}\relax
\relax
\bibitem{Kniehl:massless02}
B.~A. Kniehl,
\newblock {\em \mbox{Hadron Production in Hadron Hadron and Lepton Hadron
  Collisions}},
\newblock in {\em 14th Topical Conference on Hadron Collider Physics},
  eds.~M.~Erdmann and T.~M\"uller, pp. 161--170.
\newblock Springer, Heidelberg, 2003,
\newblock \href{http://www.arXiv.org/abs/hep-ph/0211008}{{
  [hep-ph/0211008]}}\relax
\relax
\bibitem{heavyQuarkConcept:01}
W.-K. Tung, S.~Kretzer, and C.~Schmidt,
\newblock J. Phys.{} {\bf G28},~983~(2002),
  \href{http://www.arXiv.org/abs/hep-ph/0110247}{{ [hep-ph/0110247]}}\relax
\relax
\bibitem{Kniehl:2004fy}
B.~A. Kniehl, G.~Kramer, I.~Schienbein, and H.~Spiesberger,
\newblock Phys. Rev.{} {\bf D71},~014018~(2005),
  \href{http://www.arXiv.org/abs/hep-ph/0410289}{{ [hep-ph/0410289]}}\relax
\relax
\bibitem{Kniehl:2005mk}
B.~A. Kniehl, G.~Kramer, I.~Schienbein, and H.~Spiesberger,
\newblock Eur. Phys. J.{} {\bf C41},~199~(2005),
  \href{http://www.arXiv.org/abs/hep-ph/0502194}{{ [hep-ph/0502194]}}\relax
\relax
\bibitem{CTEQ6:02}
J.~Pumplin {\it et al.}, [{CTEQ} Collaboration],
\newblock JHEP{} {\bf 07},~012~(2002),
  \href{http://www.arXiv.org/abs/hep-ph/0201195}{{ [hep-ph/0201195]}}\relax
\relax
\bibitem{unintGlu_Jung:04}
H.~Jung,
\newblock {\em \mbox{Unintegrated Parton Density Functions in CCFM}},
\newblock in {\em XII International Workshop on Deep Inelastic Scattering (DIS
  2004)}, eds.~D.~Bruncko, J.~Ferencei, and P.~Str\'{\i}\v{z}enec, Vol.~1, pp.
  299--302.
\newblock SAS, Apr.\ 2004,
\newblock \href{http://www.arXiv.org/abs/hep-ph/0411287}{{
  [hep-ph/0411287]}}\relax
\relax
\bibitem{GRV:92}
M.~Gl\"uck, E.~Reya, and A.~Vogt,
\newblock Phys. Rev.{} {\bf D46},~1973~(1992)\relax
\relax
\bibitem{Nason:99}
P.~Nason and C.~Oleari,
\newblock Nucl. Phys.{} {\bf B565},~245~(2000),
  \href{http://www.arXiv.org/abs/hep-ph/9903541}{{ [hep-ph/9903541]}}\relax
\relax
\bibitem{BKKfrag:98}
J.~Binnewies, B.~A. Kniehl, and G.~Kramer,
\newblock Phys. Rev.{} {\bf D58},~014014~(1998),
  \href{http://www.arXiv.org/abs/hep-ph/9712482}{{ [hep-ph/9712482]}}\relax
\relax
\bibitem{KKSSfrag:05}
B.~A. Kniehl, G.~Kramer, I.~Schienbein, and H.~Spiesberger,
\newblock Phys. Rev. Lett.{} {\bf 96},~012001~(2006),
  \href{http://www.arXiv.org/abs/hep-ph/0508129}{{ [hep-ph/0508129]}}\relax
\relax
\bibitem{Gladilin:c->D*99}
L.~Gladilin,
\newblock {\em \mbox{Charm Hadron Production Fractions}}, 1999,
\newblock \href{http://www.arXiv.org/abs/hep-ex/9912064}{{
  [hep-ex/9912064]}}\relax
\relax
\bibitem{pythia61}
T.~Sj\"ostrand {\it et al.},
\newblock Comput. Phys. Commun.{} {\bf 135},~238~(2001),
  \href{http://www.arXiv.org/abs/hep-ph/0010017}{{ [hep-ph/0010017]}}\relax
\relax
\bibitem{Lund83}
B.~Andersson, G.~Gustafson, G.~Ingelman, and T.~Sj\"ostrand,
\newblock Phys. Rept.{} {\bf 97},~31~(1983)\relax
\relax
\bibitem{Peterson83}
C.~Peterson, D.~Schlatter, I.~Schmitt, and P.~Zerwas,
\newblock Phys. Rev.{} {\bf D27},~105~(1983)\relax
\relax
\bibitem{cascade}
H.~Jung and G.~P. Salam,
\newblock Eur. Phys. J.{} {\bf C19},~351~(2001),
  \href{http://www.arXiv.org/abs/hep-ph/0012143}{{ [hep-ph/0012143]}}\relax
\relax
\bibitem{cascadeII}
H.~Jung,
\newblock Comput. Phys. Commun.{} {\bf 143},~100~(2002),
  \href{http://www.arXiv.org/abs/hep-ph/0109102}{{ [hep-ph/0109102]}}\relax
\relax
\bibitem{CASCADE12:manual}
H.~Jung,
\newblock {\em \mbox{The CCFM Monte Carlo Generator CASCADE Version 1.2010}},
  Nov.\ 2005,
\newblock \mbox{http://www.desy.de/\symbol{126}jung/cascade/}\relax
\relax
\bibitem{CCFM:Ciaf88}
M.~Ciafaloni,
\newblock Nucl. Phys.{} {\bf B296},~49~(1988)\relax
\relax
\bibitem{CCFM:CatFioMarch90a}
S.~Catani, F.~Fiorani, and G.~Marchesini,
\newblock Phys. Lett.{} {\bf B234},~339~(1990)\relax
\relax
\bibitem{CCFM:CatFioMarch90b}
S.~Catani, F.~Fiorani, and G.~Marchesini,
\newblock Nucl. Phys.{} {\bf B336},~18~(1990)\relax
\relax
\bibitem{CCFM:March95}
G.~Marchesini,
\newblock Nucl. Phys.{} {\bf B445},~49~(1995),
  \href{http://www.arXiv.org/abs/hep-ph/9412327}{{ [hep-ph/9412327]}}\relax
\relax
\bibitem{Frixione:totHQ95}
S.~Frixione, M.~L. Mangano, P.~Nason, and G.~Ridolfi,
\newblock Phys. Lett.{} {\bf B348},~633~(1995),
  \href{http://www.arXiv.org/abs/hep-ph/9412348}{{ [hep-ph/9412348]}}\relax
\relax
\bibitem{KniehlHein:D*+jet04}
G.~Heinrich and B.~A. Kniehl,
\newblock Phys. Rev.{} {\bf D70},~094035~(2004),
  \href{http://www.arXiv.org/abs/hep-ph/0409303}{{ [hep-ph/0409303]}}\relax
\relax
\bibitem{H1:PhotonProton95}
S.~Aid {\it et al.}, [{H1} Collaboration],
\newblock Z. Phys.{} {\bf C69},~27~(1995),
  \href{http://www.arXiv.org/abs/hep-ex/9509001}{{ [hep-ex/9509001]}}\relax
\relax
\bibitem{geant3}
R.~Brun, F.~Bruyant, M.~Maire, A.~C. McPherson, and P.~Zanarini,
\newblock {\em GEANT 3 User's Guide}, 1987\relax
\relax
\bibitem{H1dstarDIS:01}
C.~Adloff {\it et al.}, [{H1} Collaboration],
\newblock Phys. Lett.{} {\bf B528},~199~(2002),
  \href{http://www.arXiv.org/abs/hep-ex/0108039}{{ [hep-ex/0108039]}}\relax
\relax
\end{mcbibliography}
\end{document}